\documentclass{aa}  

\bibpunct{(}{)}{;}{a}{}{,} % to follow the A&A style
\usepackage{graphicx}
\usepackage{txfonts}
\usepackage{hyperref}
\makeatletter
\renewcommand*\aa@pageof{, page \thepage{} of \pageref*{LastPage}}
\makeatother
\begin{document}

\title{The temperature dependency of Wolf-Rayet-type mass loss}

\subtitle{An exploratory study for winds launched by the hot iron bump}

\author{
   A. A. C. Sander
   \inst{\ref{inst:ari}}
 \and
   R. R. Lefever
   \inst{\ref{inst:ari}}
 \and
   L. Poniatowski
   \inst{\ref{inst:ari},\ref{inst:kuleuven}}
 \and
   V. Ramachandran
   \inst{\ref{inst:ari}}
 \and
   G. N. Sabhahit
   \inst{\ref{inst:aop}}
 \and 
   J. S. Vink
  \inst{\ref{inst:aop}}
}

% List of institutions
   \institute{%
    Zentrum f{\"u}r Astronomie der Universit{\"a}t Heidelberg, Astronomisches Rechen-Institut, M{\"o}nchhofstr. 12-14, 69120 Heidelberg\\
   \label{inst:ari} \email{andreas.sander@uni-heidelberg.de}
     \and
   Institute for Astronomy (IvS), KU Leuven, Celestijnenlaan 200D, 3000 Leuven, Belgium
   \label{inst:kuleuven}
     \and
   Armagh Observatory and Planetarium, College Hill, BT61 9DG Armagh, Northern Ireland \label{inst:aop}\\
   }

% These dates will be filled out by the publisher
\date{Received 3 October 2022; accepted 27 December 2022}

 \abstract
   {The mass loss of helium-burning stars, which are partially or completely stripped of their outer hydrogen envelope, is a catalyst of the cosmic matter cycle and decisive ingredient of massive star evolution. Yet, its theoretical fundament is only starting to emerge with major dependencies still to be uncovered.}
   {A temperature or radius dependence is usually not included in descriptions for the mass loss of classical Wolf-Rayet (cWR) stars, despite being crucial for other hot star wind domains. We thus aim to determine whether such a dependency will also be necessary for a comprehensive description of mass loss in the cWR regime.}
   {Sequences of dynamically consistent stellar atmosphere models were calculated with the hydrodynamic branch of the PoWR code  along the temperature domain, using different choices for the luminosity, mass, and surface abundances. For the first time, we allowed nonmonotonic velocity fields when solving the hydrodynamic equation of motion. The resulting velocity structures were then interpolated for the comoving-frame radiative transfer, ensuring that the main wind characteristics were preserved.}
  {We find a strong dependence of the mass-loss rate with the temperature of the critical/sonic point which mainly reflects the different radii and resulting gravitational accelerations. Moreover, we obtain a relation between the observed effective temperature and the transformed mass-loss rate $\dot{M}_\text{t}$ which seems to be largely independent of the underlying stellar parameters. The relation is shifted when different density contrasts are assumed for the wind clumping. Below a characteristic value of $\log\,(\dot{M}_\text{t}\,[M_\odot\,\text{yr}^{-1}]) \approx -4.5$, the slope of this relation changes and the winds become transparent for \ion{He}{ii} ionizing photons.}
   {The mass loss of cWR stars is a high-dimensional problem but also shows inherent scalings which can be used to obtain an approximation of the observed effective temperature. For a more realistic treatment of cWR stars and their mass loss in stellar evolution, we recommend the inclusion of a temperature dependency and ideally the calculation of hydrodynamic structure models.}


\maketitle

%
%________________________________________________________________

\section{Introduction}
  \label{sec:intro}

The mass loss of hot, evolved, massive stars plays a critical role on multiple astrophysical scales: strong stellar winds affect the individual appearance of the stars \citep[e.g.,][]{Hamann1985,deKoter+1997,Hillier+2001,Shenar+2020} and consequently also their ionizing and energetic feedback to the environment \citep[e.g.,][]{Smith+2002,CrowtherHadfield2006,Hainich+2015,SanderVink2020}. Although the timescales for all evolutionary stages beyond the main sequence are comparably short, mass loss in these stages still considerably affects the stellar fates \citep[e.g.,][]{Langer+1994,ChieffiLimongi2013,Yusof+2022,MoriyaYoon2022}. In particular for hydrogen-depleted, classical Wolf-Rayet (WR) stars, strong stellar winds provide a major channel for the chemical enrichment of their host environment \citep[e.g.,][]{Maeder1983,Dray+2003,Farmer+2021,Martinet+2022}. 
The strong winds give rise to the emission-line-dominated spectra of WR stars, leaving their imprint even in integrated spectra of whole stellar populations and galaxies \citep[e.g.,][]{Conti1991,Leitherer+1996,Schaerer+1999,Plat+2019}. Since the detectability of gravitational waves \citep[][]{Abbott+2016} and along with it the significant amount of black holes (BHs) above $20\,M_\odot$ \citep[e.g.,][]{LIGO+2021}, the interest in a better understanding of the BH-mass limiting WR mass loss as a function of metallicity ($Z$) has increased even further \citep[e.g.,][]{Woosley+2020,Higgins+2021,Vink+2021}. 

Contrary to their impact, the theoretical understanding of WR-type winds is still rather limited. Despite earlier doubts, partially exacerbated by the high mass-loss rates determined before clumping was incorporated into wind models, the considerations and model efforts of \citet{NugisLamers2002}, \citet{GraefenerHamann2005} and \citet{VdK2005} demonstrated that the winds of WR stars are mainly radiatively driven with iron opacities playing a critical role for the acceleration of the wind and the scaling of the mass-loss rate. Since then, it required a new generation of computers and a considerable update to the modeling techniques to extend these fundamental efforts to a larger parameter space. Only recently did \citet{Sander+2020} and \citet{SanderVink2020} manage to calculate a larger set of dynamically consistent 1D atmosphere models that were able to predict the winds of classical WR (cWR) stars over a wider parameter space, though still covering two dimensions only. A key ingredient of these models is the detailed calculation of the flux-weighted mean opacity $\varkappa_F$ and thus the radiative acceleration $a_\text{rad}$ in an expanding environment without requiring the assumption of local thermodynamic equilibrium (LTE). 
The new generation of computational capabilities has also opened the path toward multidimensional simulations for WR winds \citep[e.g.,][]{Poniatowski+2021,Moens+2022}. In contrast to the 1D models, these 3D calculations are time-dependent, but so far limited to LTE and very few test cases, making the current insights from 1D and 3D modeling quite complementary.

In this work, we mainly follow up on the work of \citet{Sander+2020} and \citet{SanderVink2020}, using 1D stellar atmosphere models to explore an additional dimension that is very important to determine the properties and strength of WR winds. Given the high computational costs of dynamically consistent atmosphere models, all sequences presented in \citet{SanderVink2020} were calculated using a fixed stellar temperature ($T_\ast$) defined at a Rosseland continuum optical depth of $\tau_\text{R,cont} = 20$. The corresponding radii $R_\ast$ for the models were thereby given via the Stefan-Boltzmann law 
\begin{equation}
  \label{eq:stebol}
  L = 4 \pi R_\ast^2 \sigma_\text{SB} T_\ast^4\text{.}    
\end{equation}
While the value of $T_\ast = 141\,$kK in \citet{SanderVink2020} was well motivated by the prototypical solution for a classical WC star \citep{GraefenerHamann2005}, there is a priori no reason to assume that this choice of $T_\ast$ is valid for all He-burning WR stars. In fact, stellar structure models predict a curvature in the zero age main sequence (ZAMS) for He stars \citep[e.g.,][]{Langer1989,Koehler+2015} with lower temperatures obtained for lower masses. Since we could not take this effect into account in \citet{SanderVink2020}, we had to limit the applicability of the derived $\dot{M}$ recipe to He stars of about $10\,M_\odot$ and higher.

The radii of WR stars and -- as a consequence -- also their temperatures are a long-standing topic of active research \citep[e.g.,][]{Hillier1991,HamannGraefener2004,Grassitelli+2018,Sander+2020}. Beside the curvature in the He ZAMS, we are facing a particular challenge for stars with dense winds by the photosphere shifting to highly supersonic velocities. Thereby, the spectral appearance is completely determined in the wind, providing no direct observable (e.g.\ $\log g$) which could be used to determine the (hydrostatic) stellar radius. This raises the problem of connecting the effective temperatures for a Rosseland optical depth of $\tau_\text{R} = 2/3$ ($T_{2/3}$), which can be obtained via quantitative spectroscopy, to the (much) deeper subsonic regime represented by $T_\ast$ for stars with extended envelopes \citep[cf.\ the discussion in][]{Sander+2020}. In principle, the actual hydrostatic radii of the stars could be much larger. This is referred to as (hydrostatic) inflation. Alternatively, a relatively compact star can be cloaked in a wind that is optically thick out to significant radii. Both solutions might actually occur in nature with the realized branch depending on the particular stellar parameters. 

Stellar structure calculations can help to get a handle on $R_\ast$ and $T_\ast$ for helium stars, but their inherent (and computationally necessary) limitation to gray opacities usually prevents a proper estimation of $T_{2/3}$. For a selected evolutionary track of a $60\,M_\odot$ star, \citet{Groh+2014} calculated stellar atmosphere models adopting stellar parameters derived from an evolutionary track. This method led to important revisions on the predicted spectral appearances and their duration during the later evolutionary stages of massive stars. However, despite the more sophisticated method to obtain the improved effective temperatures, their underlying mass-loss rates were taken from a simplified recipe inherent to the evolutionary calculations. 

With the calculation of hydrodynamically-consistent atmosphere models, we can now obtain consistent mass-loss rates $\dot{M}$ and effective temperatures for He-burning stars without requiring any prescription of $\dot{M}$. In this work, we use this technique to investigate the behavior of $T_{2/3}$ and other temperature scales for multiple sequences of models with extended atmospheres. The paper is structured as follows: In Sect.\,\ref{sec:powr}, we briefly introduce the model atmosphere code including its recent updates necessary for our study as well as the calculated model sequences. In Sect.\,\ref{sec:temperatures}, we present the resulting temperature trends, starting with an exemplary discussion of one sequence before exploring the full sample. Afterwards,
we take a closer look at the obtained trends in the terminal wind velocities in Sect.\,\ref{sec:vinf}. Evolutionary implications and resulting scaling relations for the imprint of the WR effective temperatures are discussed in Sect.\,\ref{sec:evol}. The insights on \ion{He}{ii} ionizing fluxes are introduced in Sect.\,\ref{sec:ionflux} before drawing the conclusions in Sect.\,\ref{sec:conclusions}.

\section{Stellar atmosphere models}
  \label{sec:powr}

In this work we employ the PoWR model atmosphere code \citep[][]{Graefener+2002,HamannGraefener2003,Sander+2015} in its hydrodynamical branch (PoWR$^\text{\textsc{HD}}$) to calculate stationary, hydrodynamically-consistent atmosphere models. The implementation concepts for coupling hydrodynamics and radiative transfer are described in \citet{Sander+2017,Sander+2018} and \citet{Sander+2020}. Our hydrodynamic solutions are calculated in a similar manner as described in \citet{SanderVink2020}, that is we keep the stellar parameters $L$ and $M_\ast$ fixed and iteratively adjust $\dot{M}$ and $\varv(r)$ until a consistent solution is obtained.

%------- Figure   ----------------------------------------------------
\begin{figure}
  \includegraphics[width=\columnwidth]{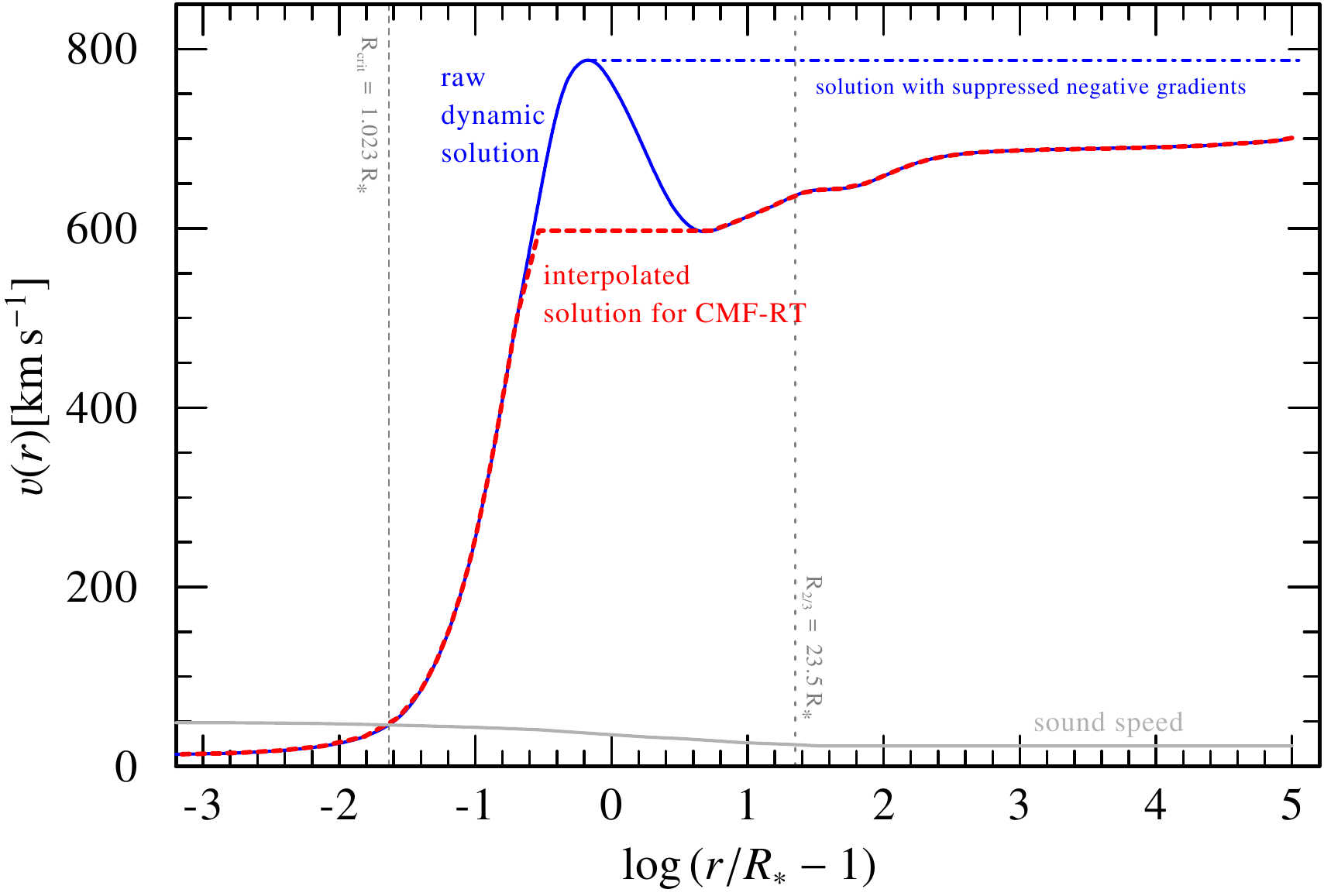}
  \caption{Illustrating example of the possible velocity treatments in case that the integration of the hydrodynamic equation of motions yields a nonmonotonic $\varv(r)$ (solid blue curve): In the simple treatment, negative velocity gradients are suppressed during the integration, yielding the blue, dash-dotted curve. In some cases, such as illustrated in this plot, this can spoil the terminal velocity $\varv_\infty$. In the more sophisticated treatment, negative gradients are therefore taken into account and $\varv(r)$ is modified such that the interpolated solution keeps the obtained $\varv_\infty$.}
  \label{fig:velo-ip-example}
\end{figure}
%------ end Figure ----------------------------------------------------  
In our previous studies, the solutions obtained for the velocity field were always monotonic. In this work, we demonstrate that apart from the high clumping factor ($D_\infty = 50$), this was mainly a result from choosing $T_\ast = 141\,$kK as the anchor point of our model sequences. 

%------- Figure   ----------------------------------------------------
\begin{figure}
  \includegraphics[width=\columnwidth]{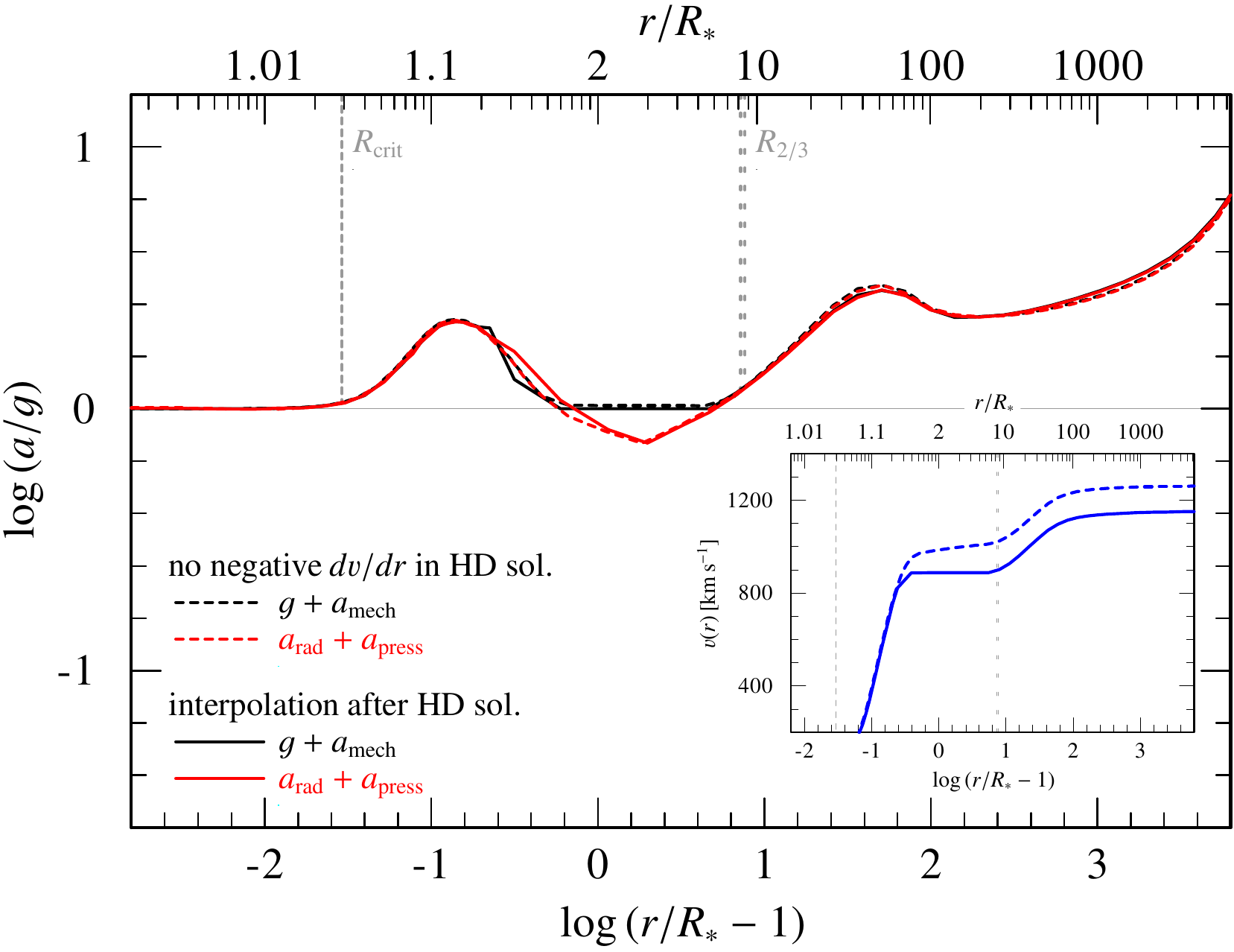}
  \caption{Resulting acceleration stratification of the converged models with two different treatments of nonmonotonic velocity fields: The dashed red and black lines illustrate the two sides of the hydrodynamic equation of motion for a model where negative velocity gradients are ignored in the solution. The corresponding solid lines show the result for the alternative method where the nonmonotonic $\varv(r)$ is interpolated afterwards. The small inlet shows the resulting velocity fields for both models. In this example we show models with $T_\ast = 125\,$kK, $\log L/L_\odot = 5.475$ and $M = 15\,M_\odot$.}
  \label{fig:acc-ip-example}
\end{figure}
%------ end Figure ---------------------------------------------------- 
When extending our modeling efforts to lower $T_\ast$, we now approach a regime where $\Gamma_\text{rad} := a_\text{rad}/g$ can drop below unity after launching the wind. Such a deficiency in the available radiative acceleration is regularly seen in WR atmosphere models including the opacities of the hot iron bump \citep[e.g.,][]{GraefenerHamann2005,Aadland+2022}. When assuming a prescribed velocity field, as traditionally done in spectral analysis, these deficiency regions have no immediate impact on the model calculations. This is different in our case where we solve hydrodynamic equation of motion. Here, a region with $\Gamma_\text{rad} < 1$ in the supersonic regime implies a negative velocity gradient until $\Gamma_\text{rad}$ eventually raises above unity again, resulting in a nonmonotonic velocity $\varv(r)$ \citep[see, e.g., the recent calculations and discussions in][]{Poniatowski+2021}. Given that our atmosphere modeling technique needs to perform the radiative transfer in a co-moving frame, which cannot handle nonmonotonic velocity fields, we therefore have to modify the $\varv(r)$ obtained from the solution of the hydrodynamic equation of motion. 

Two approaches are used in this work, which are illustrated in Fig.\,\ref{fig:velo-ip-example}. In the simple, numerically more robust method, we ignore any negative gradients already during the solution of the equation of motion. With this method, we obtain a locally consistent solution at all depth points except for any supersonic regions where $\Gamma_\text{rad} < 1$. However, this method leads to an over-prediction of $\varv_\infty$ as any reduction in $\varv(r)$ due to parts with negative gradients is ignored. The example with the dashed-dotted curve in Fig.\,\ref{fig:velo-ip-example}  illustrates that this approach can remove all further structure from the outer velocity field. We thus calculate a second type of models where the integration of the hydrodynamic equation of motion is not perturbed and only the resulting velocity field is interpolated afterwards. For the latter interpolation, we use an ``outside-in'' approach, starting at the outer boundary of our model ($R_\text{max}$) and cutting away any parts where $\varv(r)$ increases inward. While the such modified solution actually leads to more local violations of the hydrodynamic equation of motion (cf.\ Fig.\,\ref{fig:acc-ip-example}), we preserve not only the correct $\varv_\infty$ but usually also the whole $\varv(r)$ in the optically thin regime as illustrated with the red-dashed curve in Fig.\,\ref{fig:velo-ip-example}. We therefore use these types of models as the main anchor-point for discussing our results and drawing conclusions.

\begin{table}
  \caption{Input parameters for our hydrodynamically consistent He-ZAMS models. The absolute mass fraction for a particular model sequence can be obtained by obtained by inserting the corresponding value from Table\,\ref{tab:sequence-overview} for $Z/Z_\odot$.}
  \label{tab:inputparams}
  \centering
 \begin{tabular}{lc}
      \hline\hline
       Parameter    &  Value(s) \\
      \hline  
      $T_\ast$\,[kK] &  $80$\dots$220$  \\
      $\log\,(L~[\mathrm{L}_\odot])$ &  $5.35$, $5.475$, $5.7$ \\
      $M_\ast\,[\mathrm{M}_\odot]$ &  $12.9$, $15$, $20$ \\
			$\varv_\text{mic}$\,[km\,s$^{-1}$] & $30$ \\
      \smallskip
      $D_\infty$     &   $10$ or $50$  \\

      \multicolumn{2}{l}{\textit{abundances in mass fractions:}}\\
			$X_\text{H}$  &       $0$  or  $0.2$  \\
  	 $X_\text{He}$  &   $1 - X_\text{H} - 0.014 \cdot Z/Z_\odot$ \\
                $X_\text{C}$   &   $8.7\cdot10^{-5} \cdot Z/Z_\odot$  \\
			$X_\text{N}$   &   $9.1\cdot10^{-3} \cdot Z/Z_\odot$  \\
			$X_\text{O}$   &   $5.5\cdot10^{-5} \cdot Z/Z_\odot$  \\
			$X_\text{Ne}$  &   $1.3\cdot10^{-3} \cdot Z/Z_\odot$  \\
			$X_\text{Na}$  &   $2.7\cdot10^{-6} \cdot Z/Z_\odot$  \\
			$X_\text{Mg}$  &   $6.9\cdot10^{-4} \cdot Z/Z_\odot$  \\
			$X_\text{Al}$  &   $5.3\cdot10^{-5} \cdot Z/Z_\odot$  \\
			$X_\text{Si}$  &   $8.0\cdot10^{-4} \cdot Z/Z_\odot$  \\
			$X_\text{P}$   &   $5.8\cdot10^{-6} \cdot Z/Z_\odot$  \\
			$X_\text{S}$   &   $3.1\cdot10^{-4} \cdot Z/Z_\odot$  \\
			$X_\text{Cl}$  &   $8.2\cdot10^{-6} \cdot Z/Z_\odot$  \\
			$X_\text{Ar}$  &   $7.3\cdot10^{-5} \cdot Z/Z_\odot$  \\
			$X_\text{K}$   &   $3.1\cdot10^{-6} \cdot Z/Z_\odot$  \\
			$X_\text{Ca}$  &   $6.1\cdot10^{-5} \cdot Z/Z_\odot$  \\
      \medskip
			$X_\text{Fe}$  &   $1.6\cdot10^{-3} \cdot Z/Z_\odot$  \\
    \hline
  \end{tabular}
\end{table} 

\begin{table}
  \caption{Overview of the calculated model sequences}
  \label{tab:sequence-overview}
  \centering
 \begin{tabular}{lccccc}
      \hline\hline
        type  & $M\,[M_\odot]$  &  $\log\,(L~[\mathrm{L}_\odot])$ & $X_\text{H}$ & $Z\,[Z_\odot]$ & $D_\infty$ \\\hline
\multicolumn{6}{l}{\textit{Main sequences}} \\
WN & $20$   & $5.7$ & $0.0$ & $1.0$ & $50$  \\
WN & $20$   & $5.7$ & $0.2$ & $0.5$ & $50$  \\
WN & $12.9$ & $5.35$ & $0.2$ & $1.0$ & $50$  \\
WN & $15$   & $5.475$ & $0.0$ & $1.0$ & $50$  \\
WC & $20$   & $5.7$  & $0.0$ & $0.5$ & $50$  \\
WN\medskip & $20$   & $5.7$ & $0.2$ & $1.0$ & $50$  \\
\multicolumn{6}{l}{\textit{Comparison sequences}} \\
WN & $20$   & $5.7$ & $0.2$ & $1.0$ & $10$  \\
WN \smallskip & $12.9$ & $5.35$ & $0.2$ & $1.0$ & $10$ \\
WN & $20$   & $5.7$ & $0.2$ & $1.0$ & $4$  \\\hline
 \end{tabular}
\end{table}

%--------- Figure   -----------------------------------------------
\begin{figure*}
  \includegraphics[width=17cm]{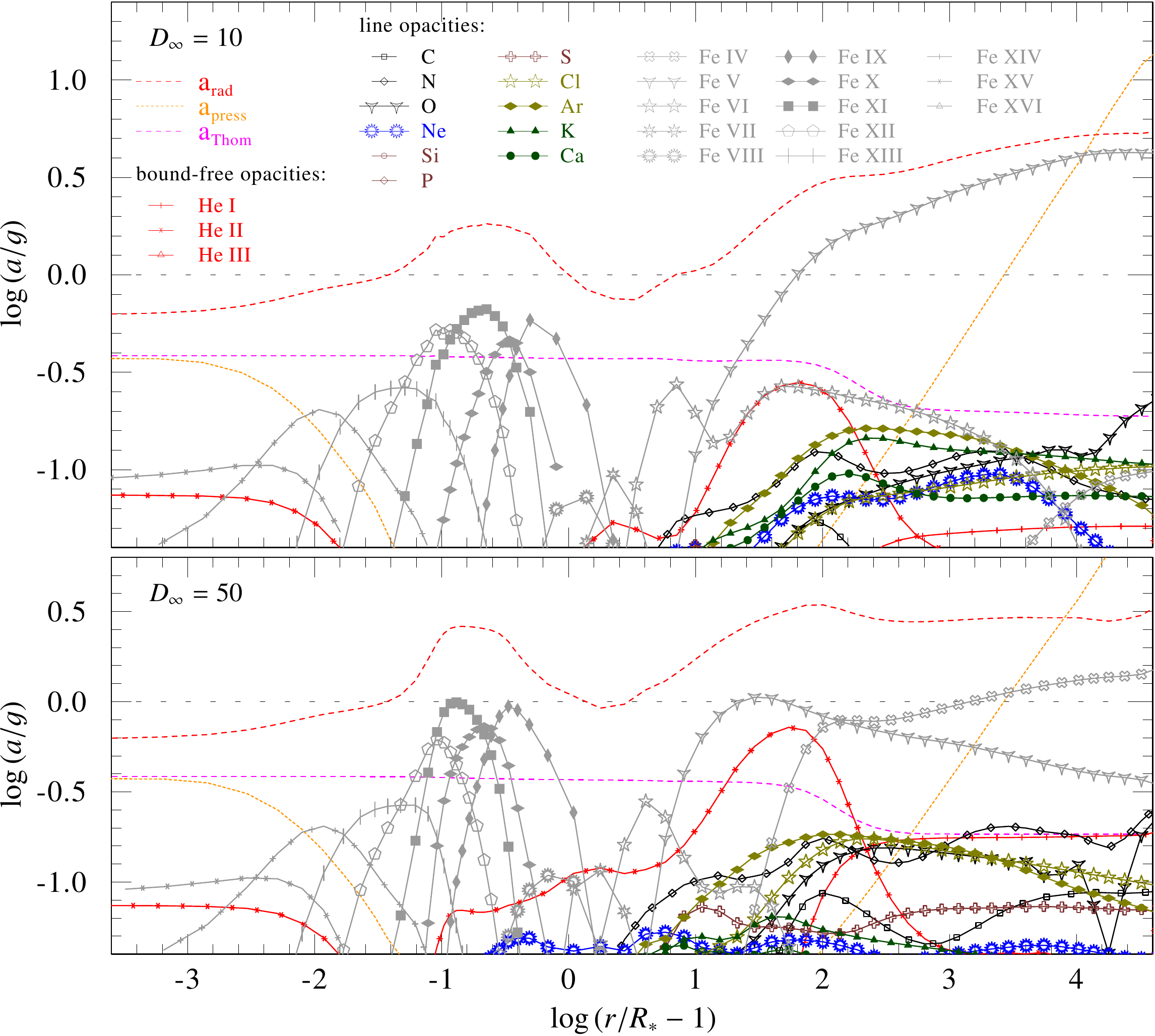}
  \caption{Major contributions to the radiative acceleration for two hydrodynamically consistent, hydrogen-free WN models with $T_\ast = 130\,$kK, $\log L/L_\odot = 5.7$, $M = 20\,M_\odot$, and $D_\infty = 10$ (upper panel) or $D_\infty = 50$ (lower panel): For the line contributions, all elemental contributions except Fe are summed over all ions. The total radiative acceleration ($a_\text{rad}$), the Thomson acceleration from free electrons ($a_\text{Thom} = \Gamma_\text{e} \cdot g$), and the contribution from gas (and turbulence) pressure ($a_\text{press}$) are also shown for comparison. The loosely dashed horizontal line denotes the total Eddington limit that needs to be overcome to launch a wind.}
  \label{fig:driving-dd50-dd10}
\end{figure*}
%--------- end Figure ---------------------------------------------

To be able to compare our results to the large set of calculations performed in \citet{SanderVink2020}, we keep most of our original model input, including the clumping description \citep[$D_\infty = 50$ and $\varv_\text{cl} = 100\,\mathrm{km}\,\mathrm{s}^{-1}$, using the ``Hillier law'' from][]{HillierMiller1999} and the set of considered elements (cf.\,Table\,\ref{tab:inputparams}). However, we have calculated some additional models, including two complete $T_\ast$ sequences for $12.9\,M_\odot$ and $20\,M_\odot$, with $D_\infty = 10$ as well as one sequence with $D_\infty = 4$ to have a comparison sets which turns out to be quite insightful. 

In Fig.\,\ref{fig:driving-dd50-dd10} we display the resulting contributions to the radiative acceleration from two models which only differ in $D_\infty$. The higher $D_\infty$ changes the wind stratification, most notably by stronger recombination from \ion{He}{iii} to \ion{He}{ii} (indicated by the higher \ion{He}{ii} bound-free opacity bump) and an earlier ionization change from \ion{Fe}{vi} to \ion{Fe}{v} in the outer wind, enabling additional line driving from \ion{Fe}{iv}. (A more detailed breakdown with all ionic contributions is provided in appendix Sect.\,\ref{sec:ioncontribclump}. with Figs.\,\ref{fig:leadions-dd50} and \ref{fig:leadions-dd10} and brief discussion about the comparison.)

Furthermore, we calculate a few additional sequences where we add surface hydrogen ($X_\text{H} = 0.2$) and one sequence where we switch to a WC-type ($X_\text{C} = 0.4$, $X_\text{N} = 4 \cdot 10^{-5}$, $X_\text{O} = 0.05$) metal composition. A full overview of the model sequences is given in Table\,\ref{tab:sequence-overview}.
Similar to \citet{SanderVink2020}, we again align $L$ and $M_\ast$ such that they follow the relations for hydrogen-free stars given in \citet{Graefener+2011}. This means that we ignore any potential extra mass (and luminosity) due to surface hydrogen as well as any differences in the $L$-$M$ relation between WN and WC stars. We do so in order to isolate the effects of different chemical compositions on the resulting wind predictions rather than trying to emulate an observed star or a particular evolutionary model. A more detailed investigation of the impact of (surface) hydrogen on the mass loss of WR stars is currently underway and will follow in a separate paper. 

%------- Figure   ----------------------------------------------------
\begin{figure*}
  \includegraphics[width=17cm]{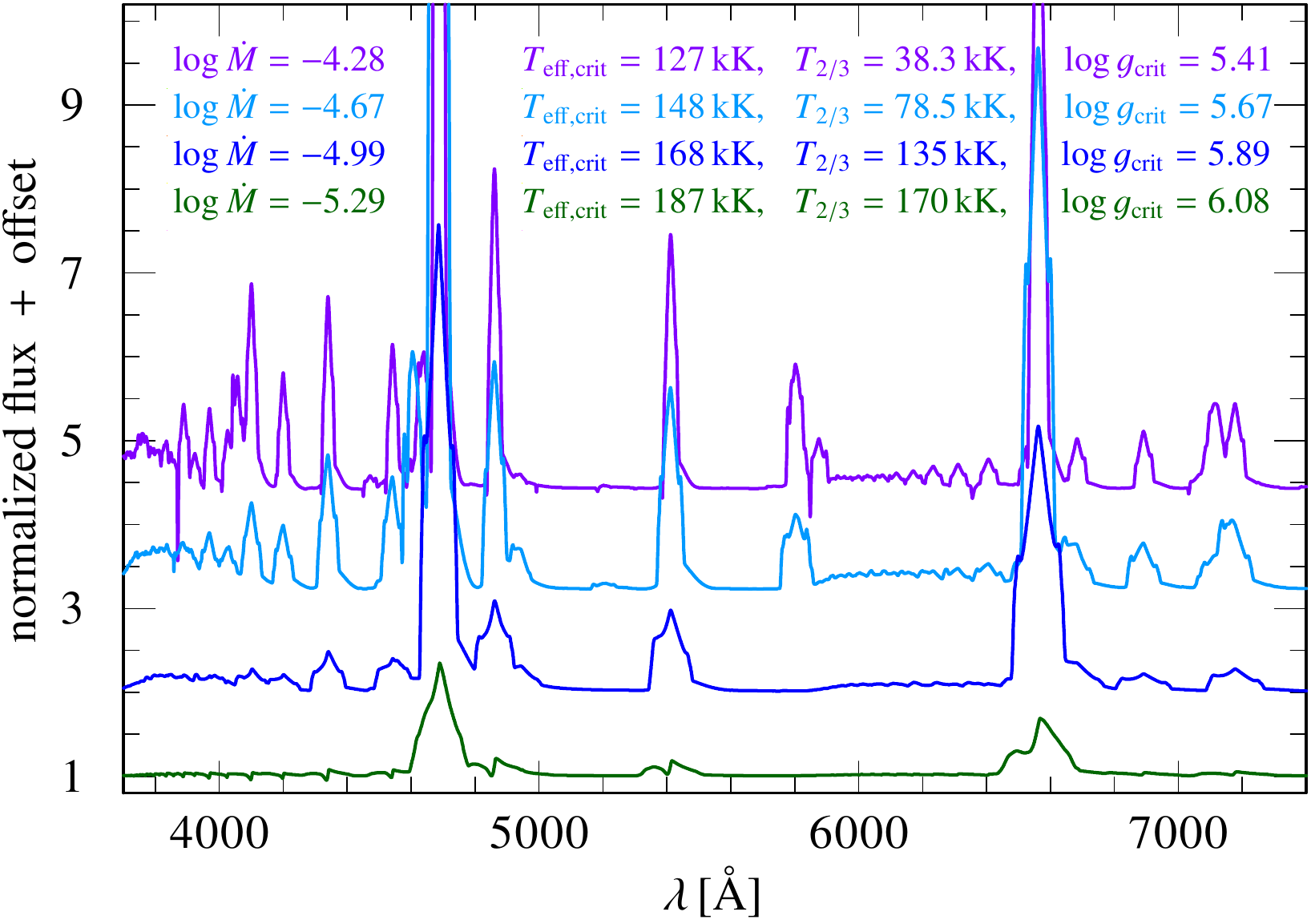}
  \caption{Example spectra from our WN model sequence with $\log L/L_\odot = 5.7$ and $M = 20\,M_\odot$, $X_\mathrm{H} = 0.2$, $Z = Z_\odot$, and $D_\infty = 10$ for the optical wavelength regime with different colors corresponding to the different models}
  \label{fig:spec-example}
\end{figure*}
%------ end Figure ---------------------------------------------------- 

While not tailored to mimic any particular WR star, the spectra resulting from our model sequences display a typical WR-type appearance. As an example, we plot four spectra from the $20\,M_\odot$ WN sequence with $X_\mathrm{H} = 0.2$, $Z = Z_\odot$ and $D_\infty = 10$ in Fig.\,\ref{fig:spec-example}. The hottest model shown mimics typical features of a relative weak-lined, early-type WN with intrinsic absorption lines, e.g., seen in WN3ha stars. Along the sequence, the lines tend to get stronger, but narrower with additional lines from cooler ionization stages appearing in the cooler models that would be classified as later WN types. Fine-tuned model efforts for particular stars will be necessary to further constrain choices of currently free parameters such as microturbulence and clumping. Ideally, one would want to eliminate the necessity for a dedicated input of these parameters completely to get full dynamical consistency, but this would require significant code updates, which is beyond the scope of the present paper.

\section{Temperatures and radii}
  \label{sec:temperatures}

\begin{figure}
  \includegraphics[width=\columnwidth]{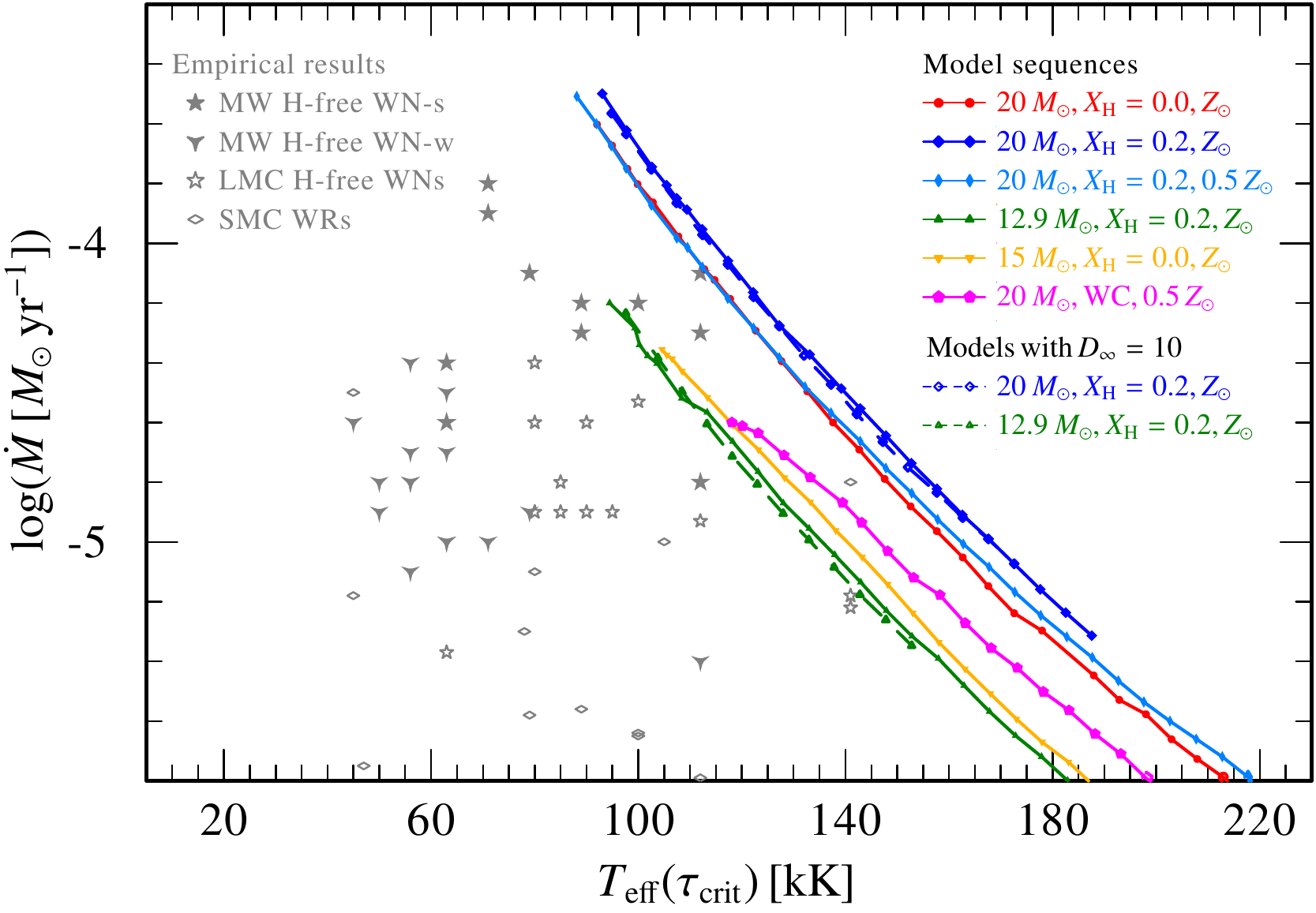}
  \caption{Mass-loss rate $\dot{M}$ as a function of $T_\text{eff}(\tau_\text{crit})$ for our model sequences. For comparison, a set of empirically inferred temperatures $T_\ast$ for different types of WR stars is shown as well (gray symbols), illustrating the well-known ``WR radius problem.'' Since the empirical values are inferred from models without dynamical consistency, the values of $T_\ast$, defined at a Rosseland optical depth of $\tau_\text{R,cont} = 20$, are shown. For our model sequences, the values of $T_\ast$ and $T_\text{eff}(\tau_\text{crit})$ align very closely except for the highest mass-loss rates. A direct comparison with $T_\ast$ from the dynamically-consistent models is provided in Fig.\,\ref{fig:mdot-tstar-all}.}
  \label{fig:mdot-teffcrit-all}
\end{figure}

For our sequences listed in Table\,\ref{tab:sequence-overview}, we study the mass-loss rate as a function of the effective temperature at the critical ($\approx$ sonic) point $T_\text{eff}(\tau_\text{crit})$. The latter value is very close to $T_\ast$ defined at $\tau_\text{R,cont} = 20$, which usually describes the inner boundary of our models, except for models with very high $\dot{M}$ where $\tau_\text{R,cont} = 100$ needs to be chosen as the inner boundary. The results depicted in Fig.\,\ref{fig:mdot-teffcrit-all} reveal a steep, monotonic decrease of $\dot{M}$ with increasing value of $T_\text{eff}(\tau_\text{crit})$ (corresponding to decreasing radii $R_\text{crit}$). This is not unexpected given that larger radii lower the local gravitational acceleration and thus enable an easier escape of material.

%--------- Figure   ----------------------------------------------------
%--------- end Figure ---------------------------------------------------- 

Our sequences with different chemical compositions give us first qualitative insights on the impact of hydrogen on the one hand and carbon and oxygen on the other hand: The direct comparison of the two curves for $20\,M_\odot$ at $Z_\odot$ in Fig.\,\ref{fig:mdot-teffcrit-all} show a systematic shift to slightly higher mass-loss rates in the presence of surface hydrogen (as long as it is negligible for the total stellar mass). Interestingly, a hydrogen surface mass fraction of $X_\text{H} = 0.2$ seems to be sufficient to counter the lower metal abundances in the $0.5\,Z_\odot$ sequence. This result should not yet be generalized given the limited number of sequences, but will be followed up in our dedicated study focusing on surface hydrogen.
Contrary to hydrogen, the inclusion of more carbon and oxygen is not beneficial to $\dot{M}$ as illustrated in Fig.\,\ref{fig:mdot-teffcrit-all} by the sequence with the WC surface composition. In all cases, the local (electron) temperatures at the launching point of the wind ($R_\text{crit}$) are too high to generate any additional line opacity from C or O. Instead, the slightly lower amount of free electrons compared to a WN surface composition decreases the resulting $\dot{M}$ \citep[cf.\ ][]{Sander+2020}. The opposite effect instead happens in the case of WN stars with $X_\text{H} > 0$.

For comparison, Fig.\,\ref{fig:mdot-teffcrit-all} also contains empirical results obtained with standard models using a $\beta$-law or double-$\beta$-law to describe $\varv(r)$ from \citet{Hamann+2006,Hamann+2019,Hainich+2015,Shenar+2016,Shenar+2019}. Illustrating the well-known ``Wolf-Rayet radius problem'' \citep[e.g.,][]{Grassitelli+2018}, most empirically derived temperatures are located at $T_\ast$ values that seem to be too cool for their mass-loss rate. As demonstrated by \citet{GraefenerHamann2005} and \citet{Sander+2020}, the critical radii inferred from dynamically-consistent atmosphere models are much smaller than those obtained by a standard $\beta$-law due to the opacities of the ``hot iron bump'' that enable the launch of a supersonic wind already at deeper layers.
Moreover, the comparison in the $\dot{M}$-$T_\text{eff}(\tau_\text{crit})$ plane is not ideal as the empirically derived values of $\dot{M}$ depend on distances which are still uncertain for some Galactic targets, but as we see below when discussing the transformed mass-loss rate, this is not a major issue here. An inspection of the model sequences indicates a clear shift for different mass regimes. Thus, one could argue that the whole plane in Fig.\,\ref{fig:mdot-teffcrit-all} could be covered if we would calculate further sequences for lower masses. However, lower masses are most likely not the solution and discrepancies remain even when adjusting the mass-loss rates for stars with different luminosities in Sect.\,\ref{sec:t23estimate}. 

\subsection{Mass loss versus different temperature scales}

%--------- Figure   ----------------------------------------------------
\begin{figure}
  \includegraphics[width=\columnwidth]{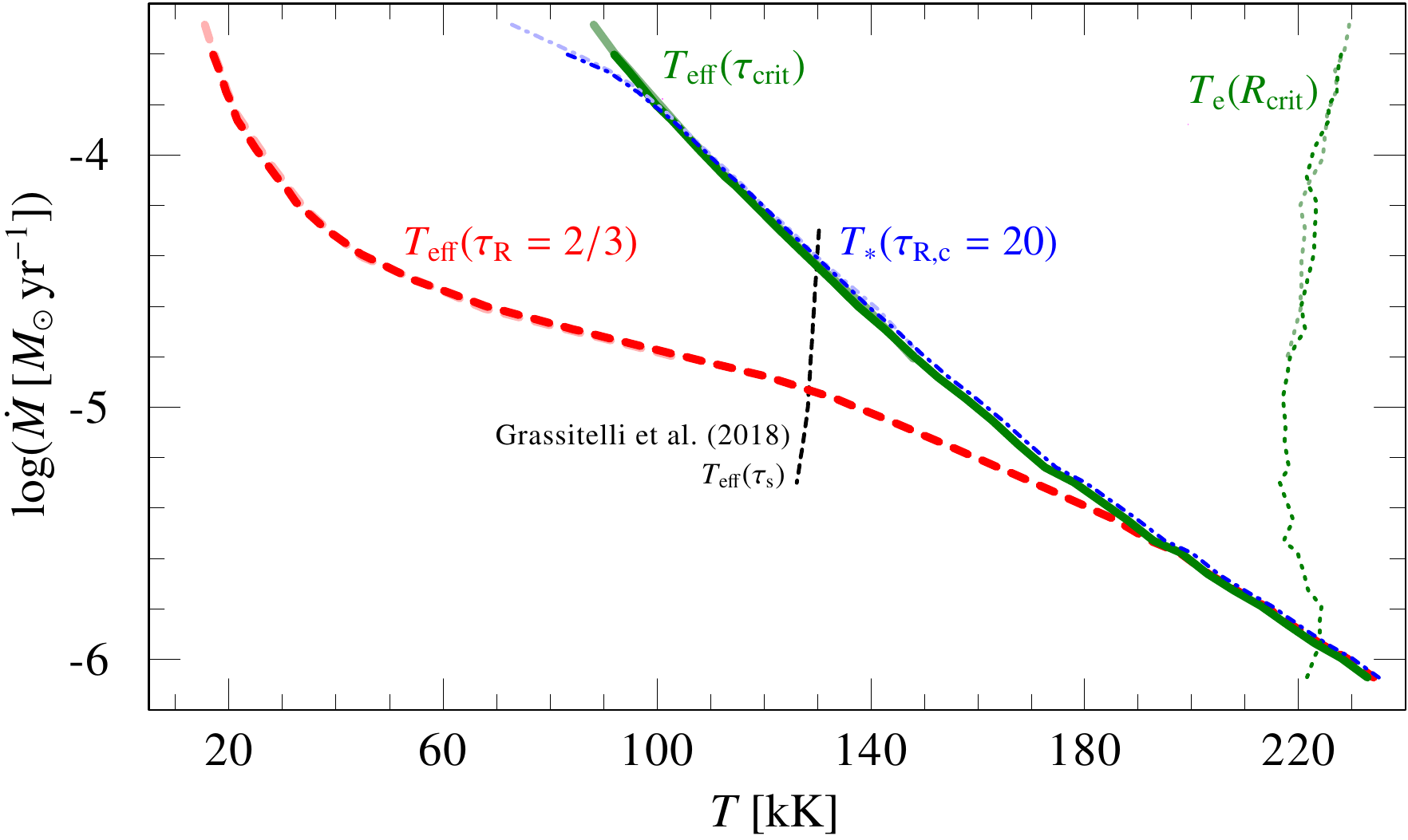}
  \caption{Mass-loss rates versus different temperature scales for a series of dynamically consistent atmosphere models with $\log L/L_\odot = 5.7$, $M = 20\,M_\odot$, and $X_\text{H} = 0$: The thick red dashed line denotes the effective temperatures defined at a Rosseland optical depth of $\tau_\text{R} = 2/3$, while the green solid line and the blue dashed-dotted line denote the effective temperatures referring to $\tau_\text{crit}$ and $\tau_\text{R,cont} = 20$, respectively. The green dotted line on the right denotes the (electron) temperature at the critical point. Curves in lighter colors reflect models using the simple integration treatment suppressing negative velocity gradients (cf.\,Sect.\,\ref{sec:powr}). The black dashed line shows the hydrodynamic structure solutions by \citet{Grassitelli+2018}.}
  \label{fig:mdot-tscales-m20}
\end{figure}
%--------- end Figure ----------------------------------------------------  

%--------- Figure   ----------------------------------------------------
\begin{figure}
  \includegraphics[width=\columnwidth]{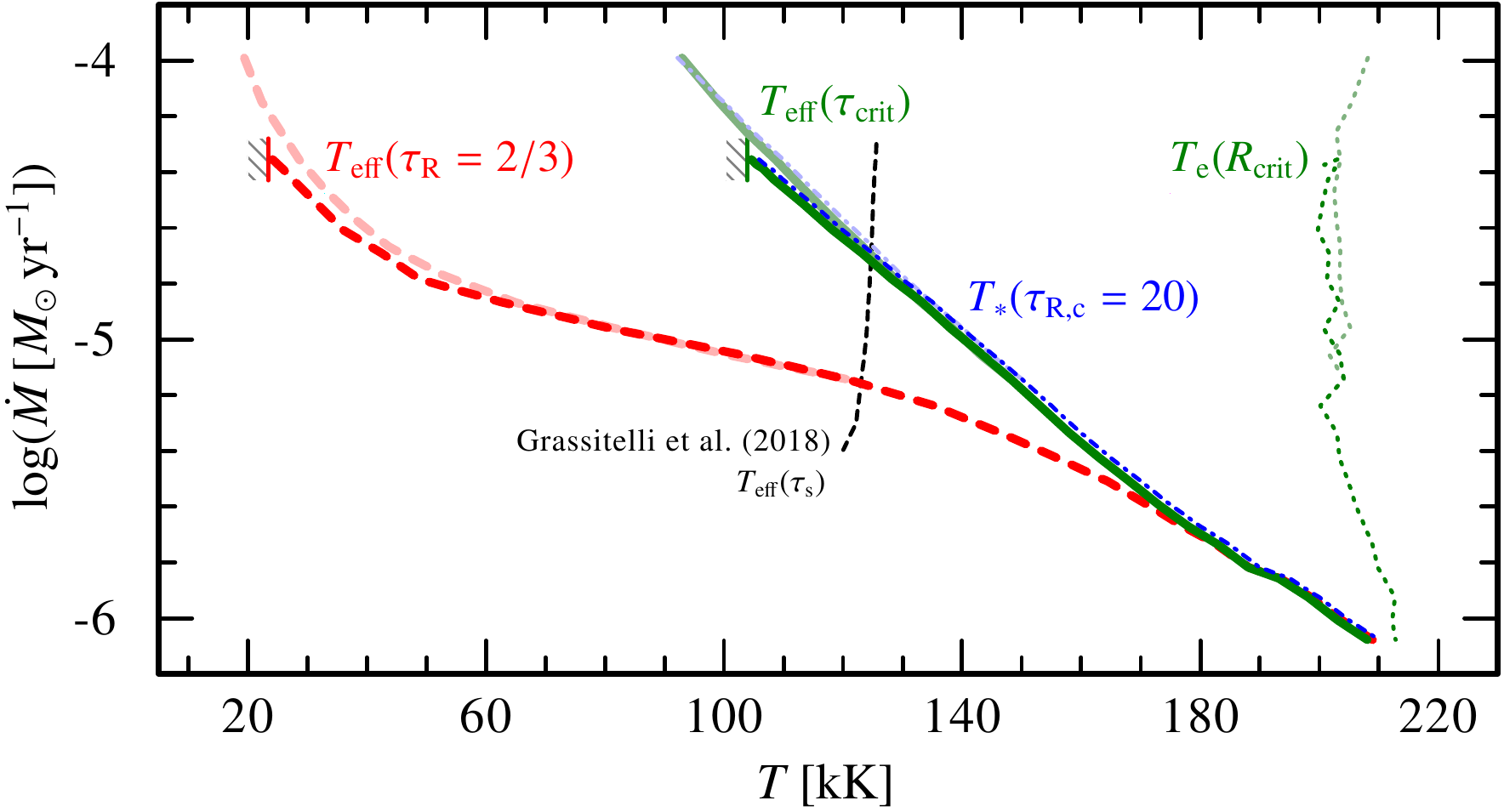}
  \caption{Same as Fig.\,\ref{fig:mdot-tscales-m20}, but for a model sequence with $\log L/L_\odot = 5.475$, $M = 15\,M_\odot$, and $X_\text{H} = 0$. Contrary to the situation in Fig.\,\ref{fig:mdot-tscales-m20} for a $20\,M_\odot$ star, there is an abrupt breakdown of solutions beyond a minimum temperature and a maximum $\dot{M}$. For the $T_{2/3}-$ and $T_\text{eff}(\tau_\text{crit})$-scales these points are marked with a vertical line attached to a gray-hatched area.}
  \label{fig:mdot-tscales-m15}
\end{figure}
%--------- end Figure ----------------------------------------------------  

To discuss the different temperature scales in WR winds and their scaling with $\dot{M}$, we take a closer look at an individual model sequence. In Fig.\,\ref{fig:mdot-tscales-m20}, we plot the mass-loss rate $\dot{M}$ for the sequence of $20\,M_\odot$ models without hydrogen as functions of different temperature definitions, namely (i) the effective temperature $T_{2/3}$ at a Rosseland optical depth of $\tau_\text{R} = 2/3$ (thick red dashed line), often simply denoted as $T_\text{eff}$ in the literature; (ii) the effective temperature $T_\ast$ commonly used in the model setup for PoWR models, defined at a Rosseland continuum optical depth of $\tau_{\text{R},\text{cont}} = 20$; (iii) the effective temperature at the critical point, denoted $T_\text{eff}(\tau_\text{crit})$, $T_\text{eff}(R_\text{crit})$, or simply $T_\text{eff,crit}$; and (iv) the electron temperature $T_\text{e}$ at the critical point (thin green dotted line).
In contrast to the first three temperatures, $T_\text{e}$ is not an effective temperature. In the deeper layers of the atmosphere where the deviations from LTE become negligible, $T_\text{e}$ aligns with the general temperature $T(r)$ defined in (1D) stellar structure models. To further illustrate the effect of the two different nonmonotonic $\varv(r)$-treatments, we plot the more accurate method with posterior interpolation of the velocity field in strong colors while the simple method ignoring negative gradients is drawn in lighter shades of the same line style. Beyond a certain temperature, there are no more deceleration regions and thus both curves agree.

Except for the regimes with highest mass-loss ($\dot{M} > 10^{-4}\,M_\odot\,\mathrm{yr}^{-1}$ in Fig.\,\ref{fig:mdot-tscales-m20}), the values of $T_\ast$ and $T_\text{eff,crit}$ closely align. This does not imply that $\tau_\text{crit}$ has to correspond to $\tau_{R,\text{cont}} \approx 20$, but that the locations of the radii corresponding to $\tau_{R,\text{cont}} = 20$ and $\tau_\text{crit}$ are close enough to yield similar effective temperatures. The alignment between $T_\ast$ and $T_\text{eff,crit}$ at lower $\dot{M}$ is fulfilled in all of our model sequences (see Figs.\,\ref{fig:mdot-tscales-m15} and \ref{fig:mdot-tscales-m12p9-hydrogen} for further examples) and allows us to discuss the physically more meaningful temperatures at the critical point (i.e., the launch of the wind) instead of the slightly more technical $T_\ast$. For higher $\dot{M}$, $\tau_\text{crit}$ moves inward and eventually surpasses $\tau_{R,\text{cont}} = 20$, explaining the deviation of the curves for the highest mass-loss rates. However, these cases require models very close to the Eddington limit.

The growing difference between the effective temperature at the launch of the wind ($T_\text{eff,crit}$) and $T_{2/3}$ with increasing $\dot{M}$ shows the ``extended atmosphere'' of a WR star. For low mass-loss rates, the atmosphere is optically thin and the two temperatures align. Although at usually much lower temperatures, this is similar to what we see for most OB-star winds. With increasing $\dot{M}$, we get a more and more extended optically thick layer. Albeit leading to a much cooler appearance of the star, this kind of layer should not be mixed up with the inflated envelope obtained in various hydrostatic structure models \citep[e.g.,][]{Petrovic+2006,Graefener+2012,RoMatzner2016}. Instead of a subsonic, but still loosely bound extended layer, our models show supersonic (i.e., unbound) layers moving out with hundreds of $\mathrm{km}\,\mathrm{s}^{-1}$. As illustrated in \citet{Sander+2020}, the winds often reach more than $0.5\,\varv_\infty$ before the atmosphere becomes optically thin. In more recent work \citep[e.g.,][]{Poniatowski+2021}, this form of an extended photosphere is termed ``dynamical inflation'' to distinguish it from the hydrostatic inflation. Hydrostatic inflation is not likely to occur if a wind can be launched and maintained, but it could in situations where the latter is not given. In Sect.\,\ref{sec:breakdown}, we discuss the limits of launching a wind from the hot iron bump. While a detailed exploration of wind solutions beyond this limit is not feasible in this study, the numerical results of our ``failed'' models indicate a tendency toward larger sonic radii, potentially indicating some form of hydrostatic inflation.

Finally, we also plot the electron temperature at the critical ($\approx$ sonic) point as a dotted green curve in Fig.\,\ref{fig:mdot-tscales-m20}. The values reflect the expected temperature range of the hot iron bump (around $200\,$kK), although the particular values are slightly higher than predicted in structural studies employing OPAL opacity tables \citep[][]{Grassitelli+2018,NakauchiSaio2018}. The value of $T_\text{e}(R_\text{crit})$ appears relatively constant at first sight, but aside from numerical scatter affecting the results a bit, a subtle trend can be noticed: In the regime of optically thick winds, there is a tendency toward increasing $T_\mathrm{e}(R_\text{crit})$ with higher mass-loss rates. This is qualitatively in line with the predictions by \citet{Grassitelli+2018}, who found that in hydrodynamic stellar structure calculations higher mass-loss rates correspond to higher temperatures at the sonic point. However, this trend is interrupted when the winds become optically more thin and eventually $T_\mathrm{e}(R_\text{crit})$ increases mildly with lower $\dot{M}$ until $T_\text{eff,crit}$ surpasses $T_\mathrm{e}(R_\text{crit})$.

All of the temperature trends described for the exemplary Fig.\,\ref{fig:mdot-tscales-m20} are observed for the other model sequences as well. For comparison, we show similar temperature scale plots for the $15\,M_\odot$ sequence (Fig.\,\ref{fig:mdot-tscales-m15}) and the $12.9\,M_\odot$ sequence with surface hydrogen (Fig.\,\ref{fig:mdot-tscales-m12p9-hydrogen}). While there are shifts in the absolute values, the same general trends are clearly identified.

%--------- Figure ---------------------------------------------------
\begin{figure}
  \includegraphics[width=\columnwidth]{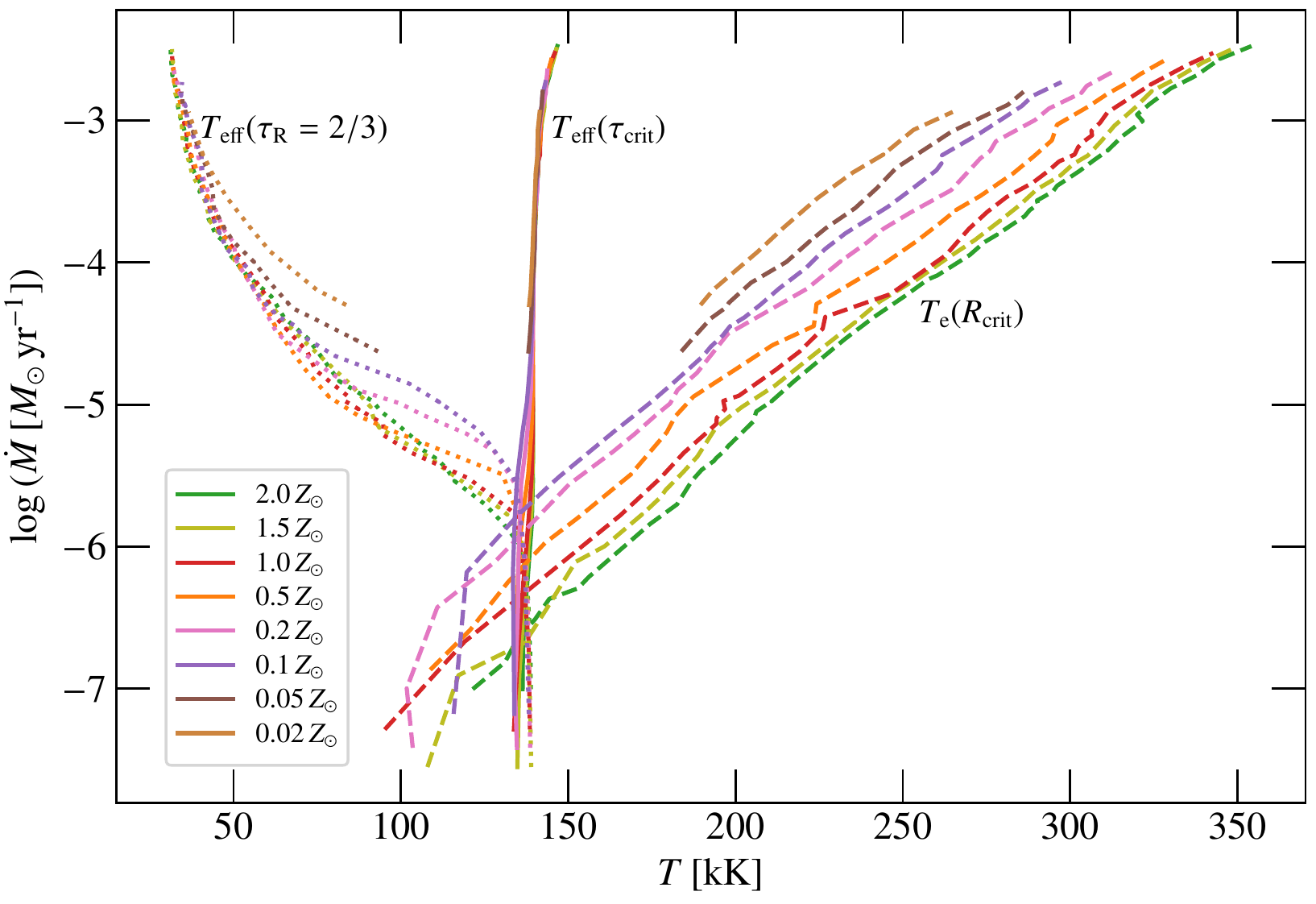}
  \caption{Mass-loss rates as a function of different temperature scales for the $L/M$ model sequences presented in \citet{SanderVink2020}. Higher mass-loss rates correspond to higher $L/M$-ratios in this plot.}
  \label{fig:tscales-sv20}
\end{figure}
%--------------------------------------------

To study whether our findings are more general or limited to our new sample, we check the behavior of the different temperature scales also for the whole set of model sequences from \citet{SanderVink2020}. The resulting curves are presented in Fig.\,\ref{fig:tscales-sv20}. Due to the fixed value of $T_\ast(\tau_\text{R,cont} = 20) = 141\,$kK in \citet{SanderVink2020} and the launching of the winds at high optical depths, the effective temperature referring to the critical point (i.e., the launch of the wind) hardly varies over the whole sample. Still, the resulting $T_{2/3}$-temperatures look very similar to those obtained in our new models with varying $T_\ast$. On the other hand, $T_\text{e}(R_\text{crit})$ varies much more than in any of our new model sequences. As we also see a shift in the $T_\text{e}(R_\text{crit})$-curves between different mass sequences in our new work, we can conclude that $\Gamma_\text{e}$ -- defined by the chemical composition and $L/M$ -- plays a major role in setting the temperature regime of the sonic point. The ratio between the flux and the radius -- which is mapped in $T_\ast$ -- instead only has a minor effect. We do not see the interruption of the $T_\text{e}(R_\text{crit})$-trend in Fig.\,\ref{fig:tscales-sv20} that was apparent in Fig.\,\ref{fig:mdot-tscales-m20} and the other new model sequences. This is likely due to the different dimensionality of the sequences in \citet{SanderVink2020} (fixed $T_\ast$, variable $L/M$ per sequence) and this work (fixed $L/M$, variable $T_\ast$ per sequence). In general, we can conclude that for stars further away from the Eddington Limit, the same $\dot{M}$ can only be reached by shifting the critical point to lower electron temperatures. For the same $L/M$-ratio, however, we see a much lower amplitude of changes in $T_\text{e}(R_\text{crit})$. In a zeroth-order approximation, one could state that $T_\text{e}(R_\text{crit})$ is constant for a given $L/M$ and chemical composition.

\subsection{WR-type mass loss and its breakdown}
  \label{sec:breakdown}

The trend of increasing $\dot{M}$ with lower $T_\text{eff,crit}$ does not automatically continue beyond the plotted values. The comparison between the simpler and the more sophisticated treatment in Fig.\,\ref{fig:mdot-tscales-m15} already suggests that the effect of deceleration regions has to be taken into account for computing a more realistic $\dot{M}$. In some situations, such as the one illustrated in Fig.\,\ref{fig:mdot-tscales-m15}, the deceleration region can become large enough to reduce the wind to subsonic or even negative velocities, making it impossible to launch a wind from the deeper layers of the ``hot iron bump.'' This regime occurs right next to the (theoretical) maximum of $\dot{M}$ along the $T_\text{eff,crit}$-axis which is reached when the deceleration region is just not strong enough to put $\varv(r)$ below the local sound speed in the wind.

%--------- Figure ---------------------------------------------------
\begin{figure}
  \includegraphics[width=\columnwidth]{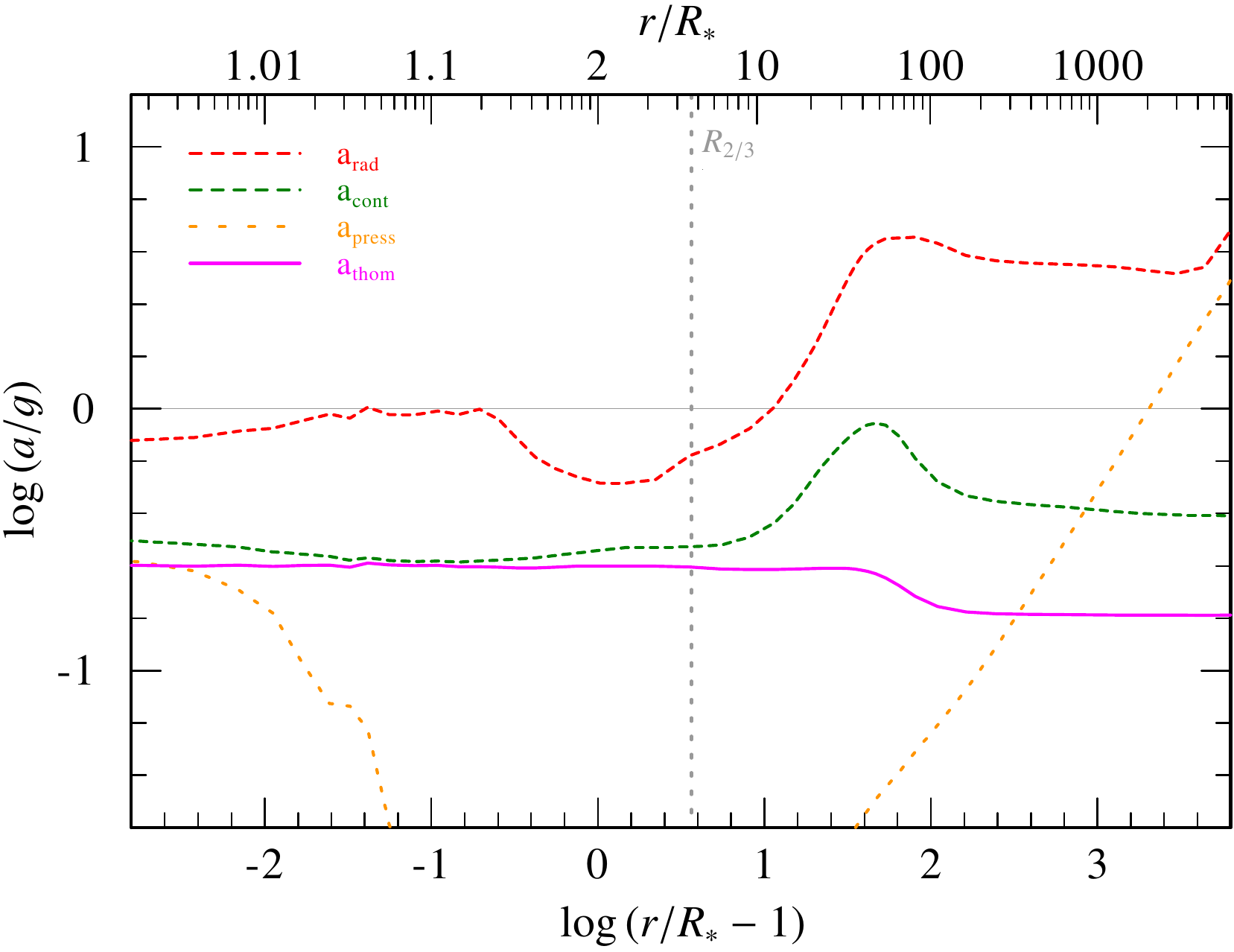}
  \caption{Illustration of the radiative acceleration for a model with $10\,M_\odot$ and $T_\ast \approx 115\,$kK which is not capable of launching a wind from the ``hot iron bump'' and thus is dynamically not converged: In the inner part, the total radiative acceleration (red dashed line) approaches $\Gamma_\text{rad} = 1$, but does not surpass it sufficiently to launch a wind that could be maintained in the following deceleration region.}
  \label{fig:acc-launchlimit}
\end{figure}
%--------------------------------------------

The situation of a failed wind launch is illustrated in Fig.\,\ref{fig:acc-launchlimit}, where the radiative acceleration barely reaches $\Gamma_\text{rad} = 1$ in a model for $10\,M_\odot$. This example also illustrates that for lower $L/M$ values, the regime where no wind can be launched from the hot iron bump gets larger and larger. A hydrogen-rich surface can compensate this to some degree as it helps to get the star closer to the Eddington limit. Still, when getting to lower and lower $L/M$-ratios the regime of WR winds driven by the hot iron bump eventually vanishes. In our models with a fixed $L$-$M$-relation this corresponds to a limit in both luminosity and mass. However, objects of lower masses and luminosity can potentially launch a wind if they have a considerably higher $L/M$-ratios than homogeneous He stars. This might e.g.\ be the case in WR-type central stars of planetary nebulae. \citet{Graefener+2017} and \citet{Ro2019} also pointed out that most of the H-free WN population in the LMC presents a challenge as these stars should not be able to launch a wind from the hot iron bump if their masses would adhere to a typical $L$-$M$ relation for He-burning stars. However, this discrepancy is already reduced if we use the \citet{SanderVink2020} models, likely due to the computed flux-weighted opacities exceeding the OPAL Rosseland opacities assumed as a proxy for $\varkappa_F$ in previous studies. The discrepancy could potentially be reduced even further slightly lower temperatures are considered as well (cf. Table\,\ref{tab:m20cmp}). Nonetheless, the LMC sample remains an interesting test-bed for detailed comparisons with individual objects and the limits of radiation-driven winds from the hot iron bump. 

Summarizing the limits of $\dot{M}$ along the temperature axis, we see two very different behaviors: Toward cooler temperatures, we have an abrupt breakdown of the thick wind regime when the effect of the deceleration region outweighs the initial acceleration by the hot iron bump. This endpoint is reached close to the highest possible mass-loss rate (for the given stellar parameters) in this whole wind regime. On the hot temperature end, we instead proceed rather smoothly into the regime of optically thin winds with lower and lower values of $\dot{M}$. This drop along the $T$-axis is significantly shallower than the strong breakdown of $\dot{M}$ along the $L/M$-axis we obtained in \citet{SanderVink2020}.

\subsection{Scaling with the transformed mass-loss rate}
  \label{sec:scaling}

%--------- Figure ---------------------------------------------------
\begin{figure}
  \includegraphics[width=\columnwidth]{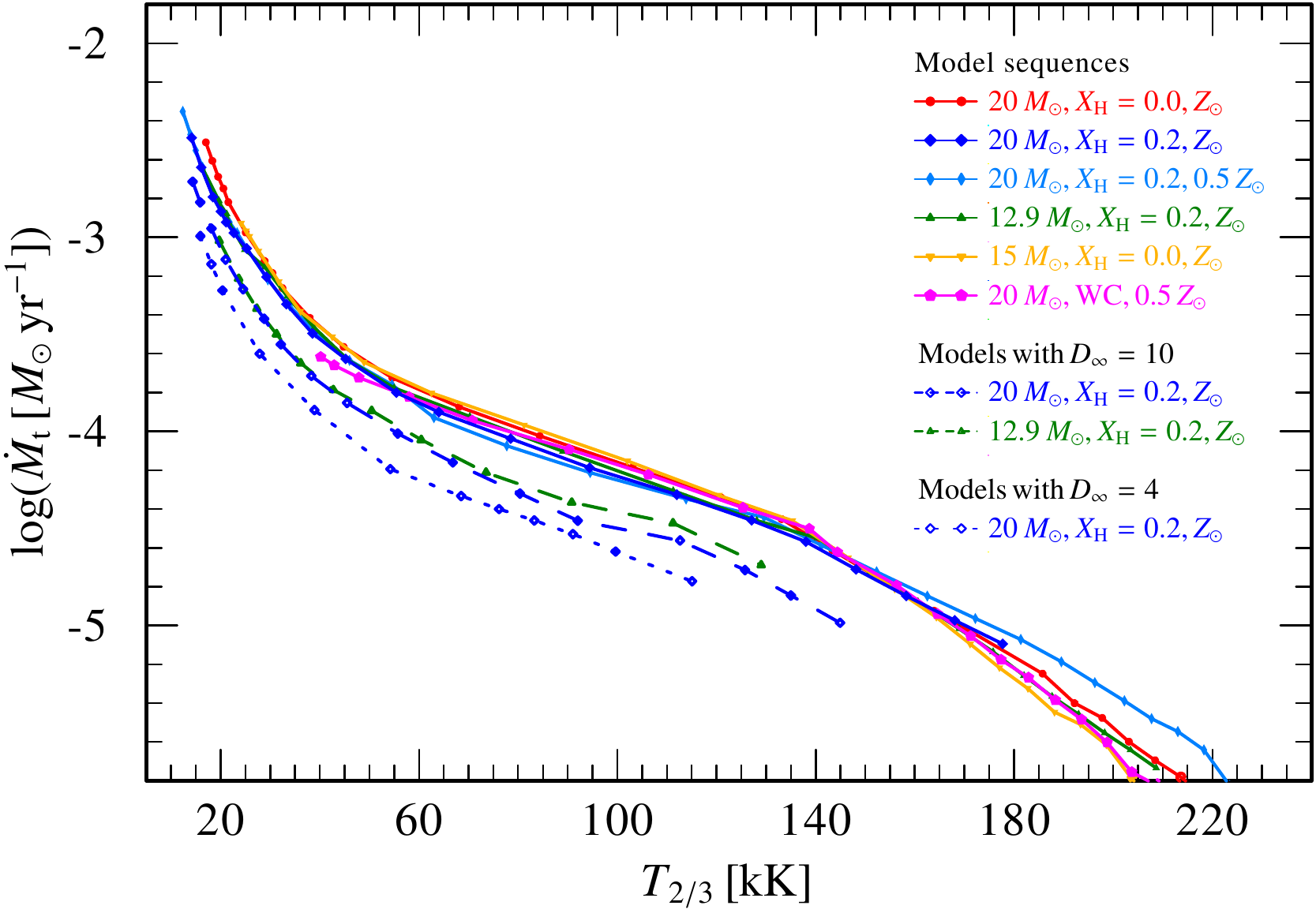}
  \caption{Transformed mass-loss rate as a function of $T_{2/3}$ for our model sequences}
  \label{fig:mdottr-t23}
\end{figure}
%--------------------------------------------

Since the different calculated model sequences show very similar slopes for $\dot{M}(T_{2/3})$, we investigate whether there is a common scaling behind these curves. Given the empirical scaling relations for WR spectra and our findings from \citet{SanderVink2020}, we study the ``transformed mass-loss rate''
\begin{equation}
  \label{eq:mdott}
  \dot{M}_\text{t} = \dot{M} \sqrt{D} \cdot \left( \frac{1000\,\text{km/s}}{\varv_\infty} \right) \left( \frac{10^6 L_\odot}{L} \right)^{3/4}\text{,}
\end{equation}
originally introduced by \citet{GraefenerVink2013}, for our new model sequences as a function of $T_{2/3}$. As depicted in Fig.\,\ref{fig:mdottr-t23}, the resulting curves align extremely well when plotting $\dot{M}_\text{t}$ instead of $\dot{M}$. Major offsets are only introduced when assuming different (maximum) clumping factors $D_\infty$. Given our findings in 
\citet{Sander+2020} and \citet{SanderVink2020}, the latter is not much of a surprise. While the clumping does not directly affect the radiative transfer, the solution of the statistical equations are solved for a higher density $D \cdot \rho$ \citep[][]{HK1998}. This affects the ionization stratification and usually leads to a larger opacity and thus larger terminal velocity $\varv_\infty$. While the mass-loss rate $\dot{M}$ is typically not much affected, the resulting $\dot{M}_\text{t}$ is changed due the increase in $\varv_\infty$ being smaller than the increase in $\sqrt{D_\infty}$. For our $20\,M_\odot$ models with $X_\text{H} = 0.2$, the typical increase was about $20\%$ in $\varv_\infty$ when increasing $D_\infty$ from $4$ to $10$ and about $40\%$ when increasing $D_\infty$ from $10$ to $50$.

%--------- Figure ---------------------------------------------------
\begin{figure}
  \includegraphics[width=\columnwidth]{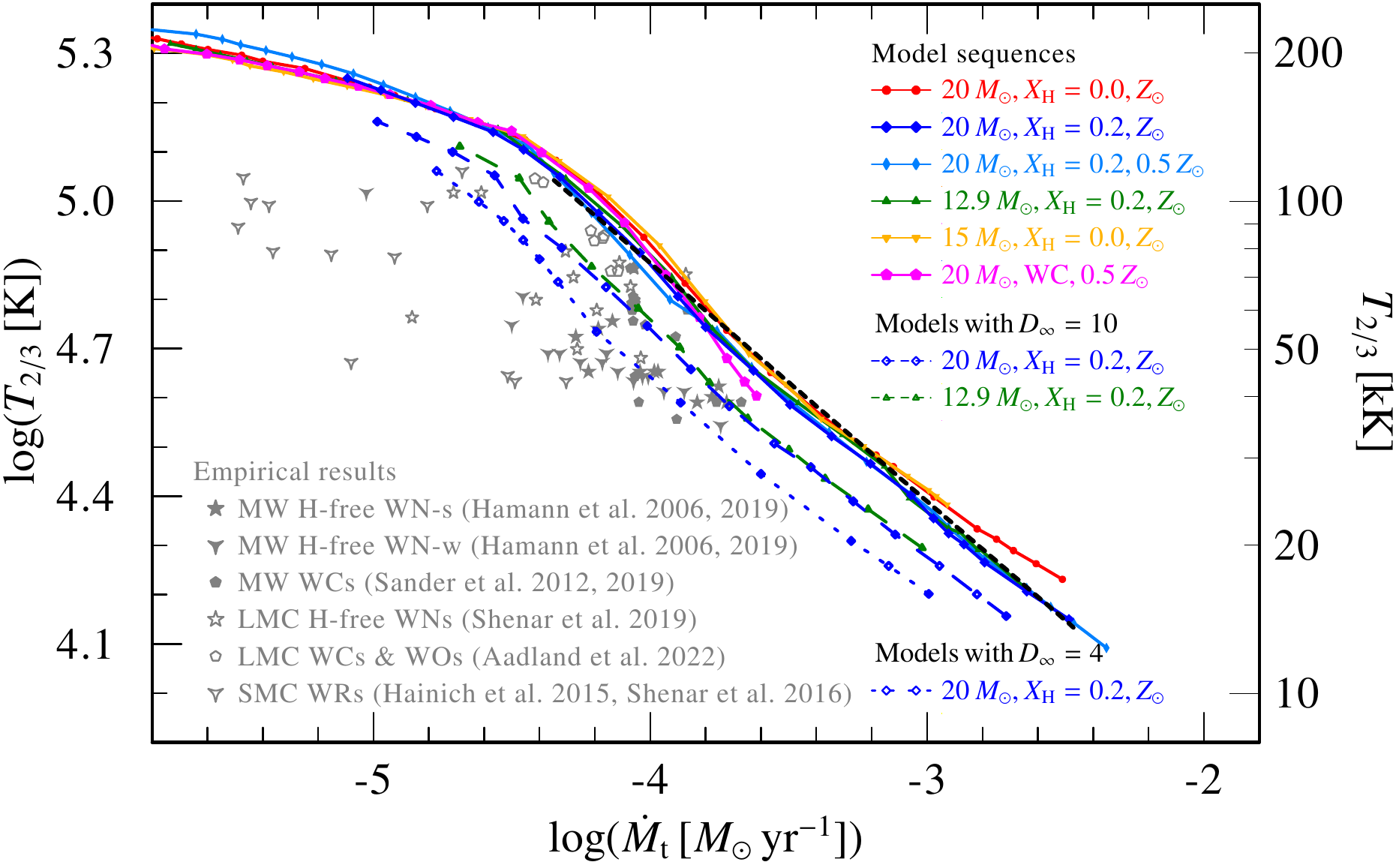}
  \caption{Effective temperature at a Rosseland optical depth of $2/3$ as a function of the transformed mass-loss rate $\dot{M}_\text{t}$ for our new calculated model sequences. The sequences connected by solid lines all employ $D_\infty = 50$, while those with dashed and dotted curves indicate sequences using $D_\infty = 10$ and $4$, respectively, as indicated in the plot.
  For comparison, also various empirical results from the literature are depicted by discrete, gray symbols.}
  \label{fig:tscale-mdottr}
\end{figure}
%--------------------------------------------

To quantify our finding, we flip the axes and show a double-logarithmic plot in Fig.\,\ref{fig:tscale-mdottr}. A clear transition between two regimes is evident with a ``kink'' around $\log \dot{M}_\text{t} \approx -4.5$ that appears to be independent of $D_\infty$. The more dense wind regime ($T_{2/3} < 130\,$kK and $\log \dot{M}_\text{t} > -4.5$) can be reasonably well approximated by a linear fit, yielding
\begin{equation}
  \label{eq:t23mdtrfit}
  \log \frac{T_{2/3}}{\mathrm{K}} = \left(-0.49 \pm 0.01\right) \log \frac{\dot{M}_\mathrm{t}}{M_\odot\,\mathrm{yr}^{-1}} + \left(2.91 \pm 0.02\right)
\end{equation}
for the sequences using $D_\infty = 50$. Given the inherent numerical scatter, in particular in $\varv_\infty$ entering $\dot{M}_\text{t}$, we can conclude that in the limit of dense winds $T_{2/3} \propto \dot{M}_\text{t}^{-1/2}$.

When comparing the relations with empirically obtained values of WN \citep[][]{Hamann+2006,Hamann+2019,Hainich+2015,Shenar+2016,Shenar+2019} and WC stars \citep[][]{Sander+2012,Sander+2019,Aadland+2022}, it is immediately evident that comparing $T_{2/3}$ between empirical and theoretical results yields a much better match than the comparison between the empirical $T_\ast$ and our theoretical $T_\text{eff}(\tau_\text{crit})$ in Fig.\,\ref{fig:mdot-teffcrit-all} (or the direct $T_\ast$-comparison in Fig.\,\ref{fig:mdot-tstar-all}). While the mismatch in Fig.\,\ref{fig:mdot-teffcrit-all} illustrates the ``Wolf-Rayet radius problem'' discussed at the beginning of   Sect.\,\ref{sec:temperatures}, the better alignment of the $T_{2/3}$ values underlines that value of the empirical analysis, despite the dynamical concerns. In empirical studies with fixed velocity fields, models are chosen such that they reproduce the observed spectrum. Although $\tau_{2/3}$ has a significant wavelength dependence in WR winds, the effective temperature corresponding to the Rosseland mean value provides some form of a representative value for the regime that needs to be met when the light eventually escapes from the star. Our dynamically consistent models can generally reproduce these $T_{2/3}$ values, but employing more compact radii that better align with structural predictions.

Despite the generally better match when comparing $T_{2/3}$, it is also evident from Fig.\,\ref{fig:tscale-mdottr} that all symbols are either on or leftward of the derived curve for $D_\infty = 50$. The most striking discrepancies are obtained for the SMC WN stars. The \citet{Aadland+2022} WC and WO results are very close to our obtained relation. As they are the only one assuming $D_\infty = 20$, some discrepancies are likely rooted in different clumping assumptions and treatments. The mismatch of the empirical SMC positions however, cannot be explained with clumping differences alone with most of the stars showing empirical $\dot{M}_\text{t}$ values that are about an order of magnitude lower than our model relations. This could be due to various effects including considerable differences in $L/M$, e.g., due to having significant hydrogen shells and thus not obeying the assumed $L$-$M$-relation in our model sequences, or too low $T_{2/3}$ estimates. Investigating these and other possibilities would add further dimensions to our model sequences and thus we have to postpone a dedicated analysis of individual targets to a separate follow-up paper.

%--------- Figure ---------------------------------------------------
\begin{figure}
  \includegraphics[width=\columnwidth]{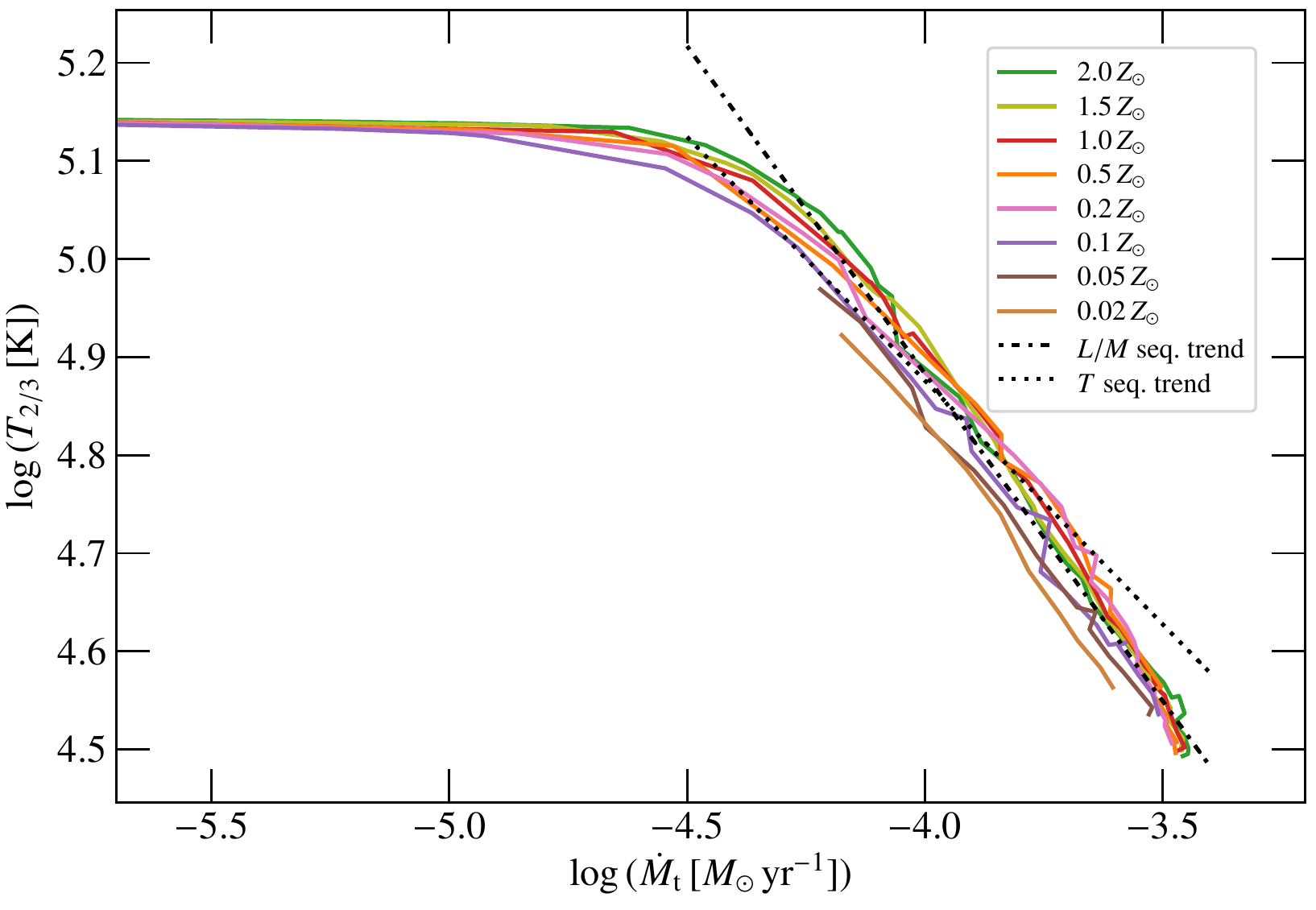}
  \caption{Effective temperature at a Rosseland optical depth of $2/3$ as a function of the transformed mass-loss rate $\dot{M}_\text{t}$ for the whole set of models from \citet{SanderVink2020}. For $\log\,(\dot{M}_\text{t}\,[M_\odot\,\mathrm{yr}^{-1}]) < -5.5$, $T{2/3}$ is effectively independent of $\dot{M}_\text{t}$. In \citet{SanderVink2020}, the value of $T_\ast$ is fixed for all models, but the difference in $L/M$ still yields a wide range of $T_{2/3}$ values. The dashed-dotted line represents a linear fit of the temperature trend for $\log\,(\dot{M}_\text{t}\,[M_\odot\,\mathrm{yr}^{-1}]) > -4.5$ while the dotted curve represents the fit for the new model sequence illustrated in Fig.\,\ref{fig:tscale-mdottr}.}
  \label{fig:zrange-t23-mdott}
\end{figure}
%--------------------------------------------

When considering the obtained curves in the $T_{2/3}$-$\dot{M}_\text{t}$-plane from the current model sequences, it is so far unclear whether the slope and even  the underlying scaling is universal. In Fig.\,\ref{fig:zrange-t23-mdott}, we thus plot the same parameters, now using the sequences from \citet{SanderVink2020}. A first noticeable difference is the upper horizontal cutoff at $T_{2/3} \approx 141\,$kK, but this is expected due to the fixed value of $T_\ast = 141\,$kK in the \citet{SanderVink2020} sample. The behavior at higher values of $\dot{M}_{t}$ looks similar to Fig.\,\ref{fig:tscale-mdottr} at first -- although with considerably more scatter -- but an actual fit of the data reveals a non-negligible difference in the slopes, yielding
\begin{equation}
  \label{eq:t23mdtrfit-SV2020}
  \log \frac{T_{2/3}}{\mathrm{K}} = \left(-0.667 \pm 0.009\right) \log \frac{\dot{M}_\mathrm{t}}{M_\odot\,\mathrm{yr}^{-1}} + \left(2.21 \pm 0.03\right)      
\end{equation}
 for the \citet{SanderVink2020} sample. (For comparison, the derived trend for the new sequences is shown as well in Fig.\,\ref{fig:zrange-t23-mdott}.) We discuss possible origins later in Sect.\,\ref{sec:t23estimate}.

\subsection{Quantitative mass-loss radius-dependence}
  \label{sec:mdot-rcrit-trend}

The similarity of the curves in Fig.\,\ref{fig:mdot-teffcrit-all} and the simplicity of the slopes, indicates a common dependence between $\log\,\dot{M}$ and $\log\,T_\text{eff,crit}$ for our sequences with fixed $L/M$. Performing a linear fit, we obtain a relation in the form of
\begin{equation}
  \label{eq:mdot-teffcrit-trend} 
  \log(\dot{M}\,[M_\odot\,\mathrm{yr}^{-1}]) = -6 \cdot \log (T_\text{eff,crit}\,[\mathrm{K}]) + \text{offset}
\end{equation}
with the detailed fit coefficients being presented in appendix Sect.\,\ref{sec:mdot-teffcrit-fit} and Table\,\ref{tab:mdot-teffcrit-trend}. The factor $-6$ in Eq.\,\eqref{eq:mdot-teffcrit-trend} implies that the obtained temperature dependence essentially reflects the radius change of the stellar models, since $T_\text{eff,crit}^4 \propto R_\text{crit}^{-2}$ and 
\begin{equation}
  \label{eq:mdot-rcrit}
  \dot{M} \propto R_\text{crit}^3   
\end{equation}
for models with $L = \text{const.}$ (as in our model sequences). Our corresponding plot (Fig.\,\ref{fig:mdot-teffcrit-trend}) further shows that deviations from the purely geometrical trend occur when we reach the limit of radiative driving discussed in Sect.\,\ref{sec:breakdown} (and listed explicitly for each sequence in Table \ref{tab:mdot-teffcrit-trend}), as e.g.\ visible at the upper end of the WC sequence (cf.\,Fig.\,\ref{fig:mdot-teffcrit-trend}).

%--------- Figure --------------------------------------
\begin{figure}
  \includegraphics[width=\columnwidth]{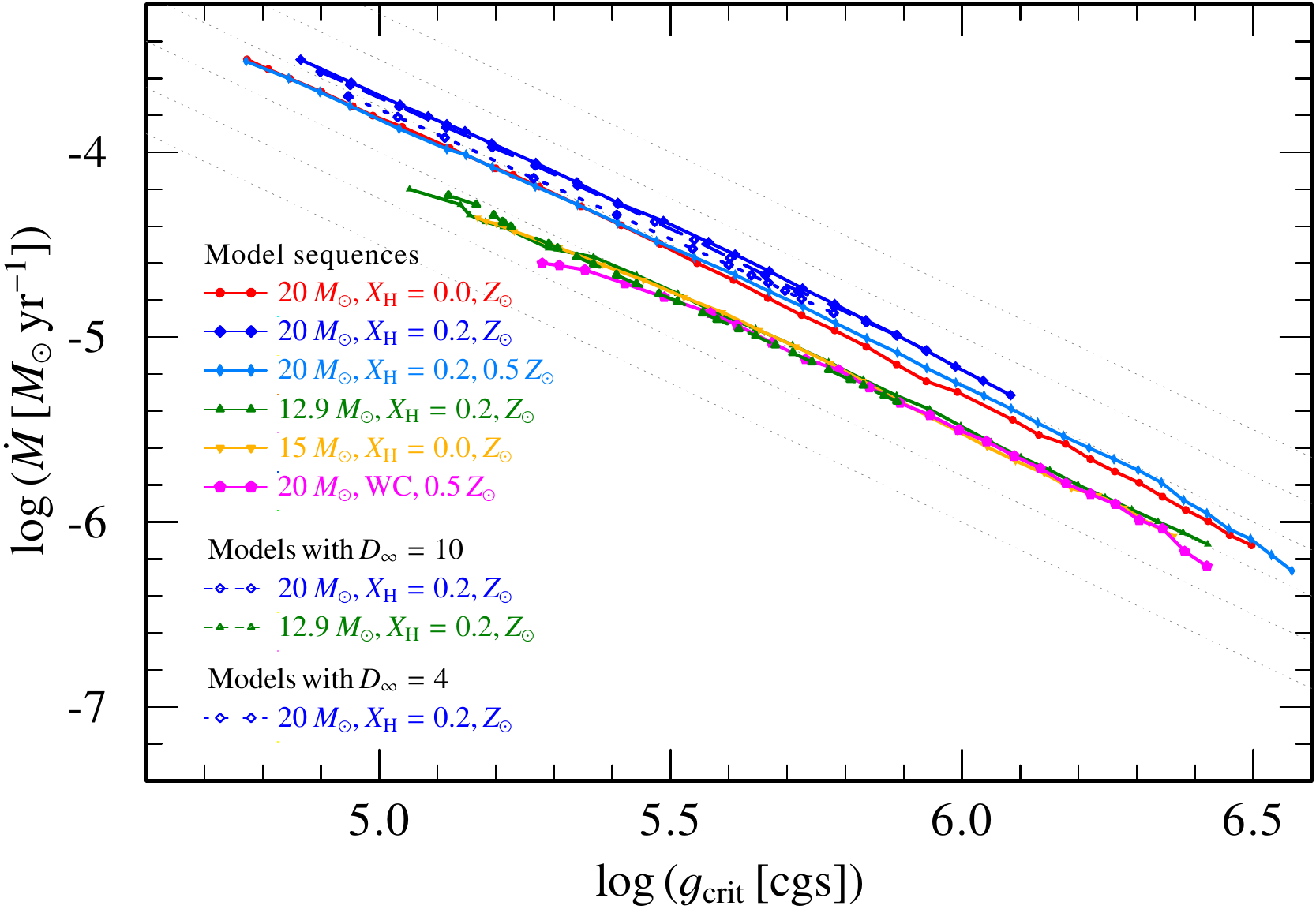}
  \caption{
  Mass-loss rate $\dot{M}$ as a function of the gravitational acceleration at the critical radius $g_\text{crit} = G M R_\text{crit}^{-2}$. Thin, dotted, gray lines indicate curves with $\dot{M} \propto g_\text{crit}^{-3/2}$.}
  \label{fig:mdot-gcrit}
\end{figure}
%--------------------------------------------

From a dynamical perspective, the obtained $R_\text{crit}$-trend for $\dot{M}$ can be understood as a dependence on the gravitational acceleration on the critical point, where the wind is launched. With the straight-forward definition of
\begin{equation}
  \label{eq:gcrit-def}
  g_\text{crit} = g(R_\text{crit}) = \frac{G M}{R_\text{crit}^2}   
\end{equation}
we can rewrite Eq.\,\eqref{eq:mdot-rcrit} as 
\begin{equation}
  \label{eq:mdot-gcrit}
  \dot{M} \propto g_\text{crit}^{-3/2}
\end{equation}
since $M$ is a constant among each of the model sequences.  Trend curves reflecting Eq.\,\eqref{eq:mdot-gcrit} are displayed in Fig.\,\ref{fig:mdot-gcrit} together with the curves from our model sequences. Generally, a decreasing trend of $\dot{M}$ with $\log g_\text{crit}$ is not surprising as an increased gravitational force needs to be overcome. Given the content mass $M$ along the model sequences, the change in $g_\text{crit}$ expected from Eq.\,\eqref{eq:mdot-gcrit} is purely geometrical, that is only from the change in $R_\text{crit}$. The model sequences align well with Eq.\,\eqref{eq:mdot-gcrit}, but there is a notable flattening for the highest mass-loss rate, that is in the case of more dense winds. We cannot rule out that a numerical effect is playing a role here as these high-$\dot{M}$ models often operate on the limits of what the code is capable of. Nonetheless, given that the bending occurs in all sequences, a physical origin seems more likely and we continue our efforts on this assumptions.
For some sequences, there are also notable deviations from Eq.\,\eqref{eq:mdot-gcrit} at the lower $\dot{M}$-end. From the current set of calculations, the apparent kink in the curves approximately coincides with $T_\text{e}(R_\text{rcrit})$ surpassing $T_\text{eff,crit}$, meaning that the electron temperature at the critical point is higher than the effective temperature at this point. However, the low number of models where this trend is clearly observed and the need to include higher ionization stages in these models, which can cause an additional offset in the numerical solutions if not done early enough in the sequence, currently refrain us from concluding whether there is a clear ``kink'' or a more gradual change that might potentially be emphasized by a switch in the numerical setup.
In any case, the model solutions obtained in this thinner wind regime are characterized by (electron) temperature stratification that remain very high, e.g.\ larger than $50\,$kK, until infinity. Their leading acceleration is provided by Fe M-shell ions, which are populated throughout the wind, qualitatively similar to the example shown in Fig.\,16 of \citet{Sander+2020}.
When considering the transformed mass-loss rate $\dot{M}_\text{t}$ instead of $\dot{M}$, the scaling of the velocity with $g_\text{crit}$ has to be considered as well, which we do in the following section.

\section{Terminal velocity trends}
  \label{sec:vinf}

With the intrinsic solution of the hydrodynamic equation of motion, our models automatically predict terminal wind velocities together with $\dot{M}$. While already entering the transformed mass-loss rates, we now take a look at the explicit results for $\varv_\infty$ as a function of $T_{2/3}$ in Fig.\,\ref{fig:vinf-t23}. Although our sequences are not at all adjusted to match any particular observations, we also plot empirical results for WN stars obtained with PoWR for comparison. It is clear that our models match the general regime of the observed sample, but a closer inspection also shows caveats, for example with hydrogen-free WN stars showing values above the $20\,M_\odot$ H-free sequence. Various possibilities could explain this (e.g., higher $L/M$ and/or higher clumping), but a thorough investigation is beyond the scope of this paper.

\subsection{Impact of clumping}
  \label{sec:vinf-clumping}

%--------- Figure ---------------------------------------------------
\begin{figure}
  \includegraphics[width=\columnwidth]{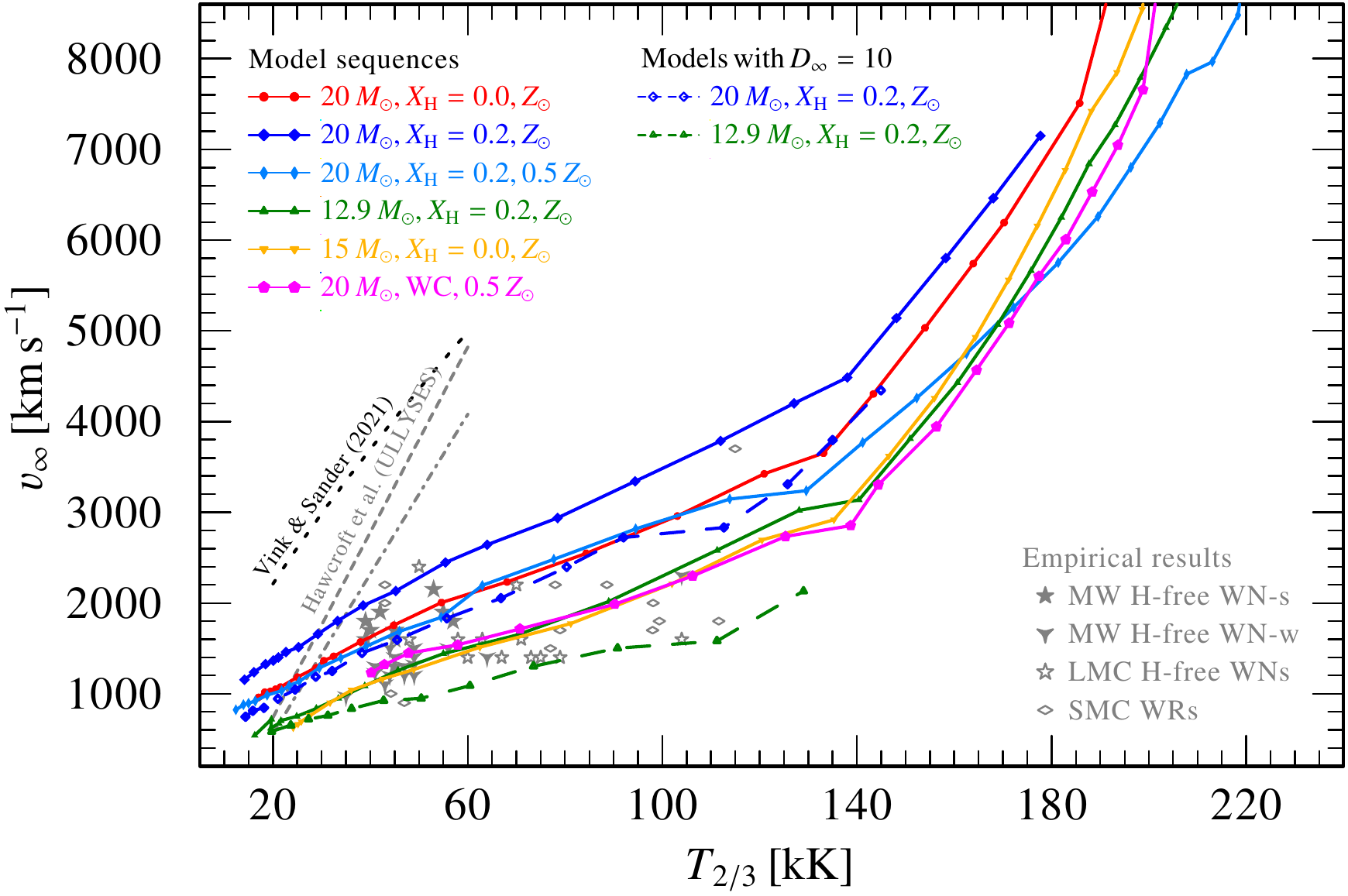}
  \caption{Terminal velocity as a function of $T_{2/3}$ for the different model sets. For comparison, also the derived trend for OB-star winds in the Milky Way (dashed line) and the LMC (dashed-dotted line) from Hawcroft et al. (in prep.) are shown.}
  \label{fig:vinf-t23}
\end{figure}
%--------------------------------------------

In contrast to $\dot{M}$, the values for $\varv_\infty$ tend to scatter a bit more due to being evaluated at the outer boundary of the models. The terminal velocity further strongly depends on the included opacity, so including all ions contributing to the acceleration is necessary in order to avoid underestimating $\varv_\infty$. As apparent from Fig.\,\ref{fig:vinf-t23}, $\varv_\infty$ also reacts on the choice of the clumping factor $D_\infty$. With a depth-dependent onset of the clumping, the response of the mass-loss rate to a change of $D_\infty$ is usually small as the resulting differences in $D(r)$ are small in the subsonic layers \citep[see also Fig.\,1 in][and appendix Sect.\,\ref{sec:ioncontribclump} of this work]{Sander+2020}. In the supersonic layers, however, any differences in $D_\infty$ affect the bulk of the opacities being considered in the hydrodynamic equation of motion. Consequently, the obtained values for $\varv_\infty$ are notably higher for higher $D_\infty$, typically on the order of $20\%$ when calculating a model for the same $L$, $M$, $T_\ast$, and chemical composition with $D_\infty = 50$ instead of $10$. In particular cases, the effects can be much larger, e.g.\ when a model has a deceleration regime in the case of a lower $D_\infty$, while there is no such regime for higher $D_\infty$. Significant changes of the wind density regime due to a switch of $D_\infty$ can then lead to a stronger change in the derived $\dot{M}$. Moreover, the assumption of little to no clumping in the deeper layers would become invalid in case of a porous, optically thick medium, which can result from a subsonic, super-Eddington situation \citep[e.g.,][]{Shaviv1998,Shaviv2000}.

\subsection{Scaling with $T_{2/3}$}
  \label{sec:vinf-t23}

Beside the differences due to clumping, Fig.\,\ref{fig:vinf-t23} demonstrates that any significant change of the chemical composition usually affects the derived $\varv_\infty$-values. In the figure, we see higher terminal velocities for the $20\,M_\odot$ model sequence with surface hydrogen ($X_\text{H} = 0.2$) compared to the corresponding hydrogen-free sequence. This result might be counter-intuitive at first as hydrogen does not provide significant line opacity that could be used to increase $\varv_\infty$. However, the additional hydrogen is able to boost the mass-loss rate of the star as the hydrogen atoms in the atmosphere provide a higher budget of free electrons compared to a hydrogen-free atmosphere\footnote{A small hydrogen-layer on the surface has also the structural consequence of an increased stellar radius, which would again affect the wind parameters. Here, we discuss only the immediate atmospheric consequences for a fixed set of stellar parameters.}. With more acceleration available already in the deeper layers, the critical point of the wind moves inward to higher optical depths. Line opacities which were subsonic in the hydrogen-free case can now be used to further boost the terminal wind speed. Due to the higher mass loss of the hydrogen-containing model, the value of $T_{2/3}$ decreases when comparing models with the same $T_\ast$. Interestingly, as we saw in Fig.\,\ref{fig:mdot-tstar-all}, the value of $\dot{M}_\mathrm{t}$ remains the same when comparing against $T_{2/3}$. In a follow-up study, we will test whether this behavior is universal when considering surface hydrogen or whether the chosen fraction of $X_\text{H} = 0.2$ coincidentally balances out other effects for a $20\,M_\odot$ He-burning star as we e.g.\ saw with the metallicity reduction being balanced by the surface hydrogen when considering only $\dot{M}$ in Fig.\,\ref{fig:mdot-teffcrit-all}.

To get some insights on the general scaling of WR-type winds with 
effective temperature (here: $T_{2/3}$), we also compare our sequences to the trends for OB-type stars. In Fig.\,\ref{fig:vinf-t23}, we plot the trends obtained for the ULLYSES OB stars for the SMC and a Galactic comparison sample by Hawcroft et al. (in prep.) as well as the predictions from \citet{VinkSander2021}. We see that generally the terminal velocity increases much steeper with the temperature in OB-type winds than in WR-type winds. Only at higher temperatures ($T_{2/3} > 130$\,kK), when the winds become more optically thin as the mass-loss rates decrease, the steepness of the $\varv_\infty(T_{2/3})$-curves increases to a value more comparable to those obtained for OB-star winds.

\subsection{Critical-point dependencies}
  \label{sec:vinf-rcrit}

In addition to studying the behavior of $\varv_\infty$ as a function of the observable quantity $T_{2/3}$, we further investigate the behavior of $\varv_\infty$ as a function of $T_\ast$ or the physically more meaningful $T_\text{eff,crit}$. As noted above, the values of $\varv_\infty$ tend to scatter a bit more, but if we restrict the linear fitting to the optically thick wind regime, we find
\begin{equation}
  \log \left(\varv_\infty\,[\mathrm{km}\,\mathrm{s}^{-1}]\right) = 1.2 \log \left( T_\text{eff}(\tau_\text{crit})\,[\mathrm{K}]\right) + \text{offset}\text{.}
\end{equation}
Given that the number of models in the optically thick regime is restricted and not all sequences reach far enough to see the flattening of the trend, this coefficient has to be considered as rather uncertain. Nonetheless, the corresponding scaling of $\varv_\infty \propto R_\text{crit}^{-0.6}$ can also be obtained from directly fitting $\varv_\infty(R_\text{crit})$ in the optically thick limit. Using the critical radius to define the escape velocity
\begin{equation}
  \varv_\text{esc} = \sqrt{\frac{2 G M}{R_\text{crit}}}  
\end{equation}
we obtain
\begin{equation}
  \varv_\infty \propto \varv_\text{esc}^{1.2}\text{.}
\end{equation}
This relation also holds for the effective escape velocity 
\begin{equation}
  \varv_\text{esc} = \sqrt{\frac{2 G M}{R_\text{crit}}\left(1-\Gamma_\text{e}\right)}  
\end{equation}
since all of the model sequences have a constant $L/M$ and $\Gamma_\text{e}$ is approximately constant. The latter is consequence of the unchanged free electron budget below $R_\text{crit}$ in the considered temperature range. Thus, in the optically thick wind regime we have a slight difference with $\varv_\infty \propto \varv_\text{esc,eff}^{1.2}$ along the $T_\ast$-dimension compared to the well-known $\varv_\infty \propto \varv_\text{esc,eff}$ in the well-known CAK theory \citep*[named after][]{Castor+1975}. For the sequences along the $L/M$-dimension from \citet{SanderVink2020}, we instead obtain a negative trend of $\log \varv_\infty \approx -4.6 \log \varv_\text{esc,eff} + \text{const.}$ in the optically thick regime with a flattening of the trend for the highest mass-loss rates. In both cases, the scaling of $\varv_\infty$ with $\varv_\text{esc,eff}$ remains complicated with no straight-forward prediction as offsets remain in all scalings. This is in sharp contrast to the classical (m)CAK result, where $\varv_\infty$ follows as an offset-free value from $\varv_\text{esc}$.

For lower mass-loss rates ($\log\,\dot{M}_\mathrm{t} < -4.5$), corresponding usually to $T_\text{eff,crit} > 150\,$kK, we reach the regime where winds are mostly optically thin. Above, we could show that when reaching this regime, there seems to be an alignment of $\varv_\infty(T_{2/3})$, with the slopes known from OB-type winds. With considerable scatter in the exponent of up to $\pm 0.5$, we find
\begin{equation}
  \varv_\infty \propto T_\text{eff,crit}^4\text{,}
\end{equation}
corresponding to $\varv_\infty \propto R_\text{crit}^{-2}$ or $\varv_\infty \propto \varv_\text{esc,eff}^4$, that is a steeper relation, contrary to the expected flattening of the slope.
However, there is growing evidence from both observations \citep[e.g.,][]{Garcia+2014} as well as theoretical CMF-based and Monte Carlo calculations \citep[e.g.,][]{Bjoerklund+2021,VinkSander2021} that even in the typical OB-type regime the scaling of $\varv_\infty$ with $\varv_\text{esc,eff}$ is likely more complicated.

\subsection{Influence on $\dot{M}_\text{t}$}
  \label{sec:vinf-mdott}

%--------- Figure --------------------------------------
\begin{figure}
  \includegraphics[width=\columnwidth]{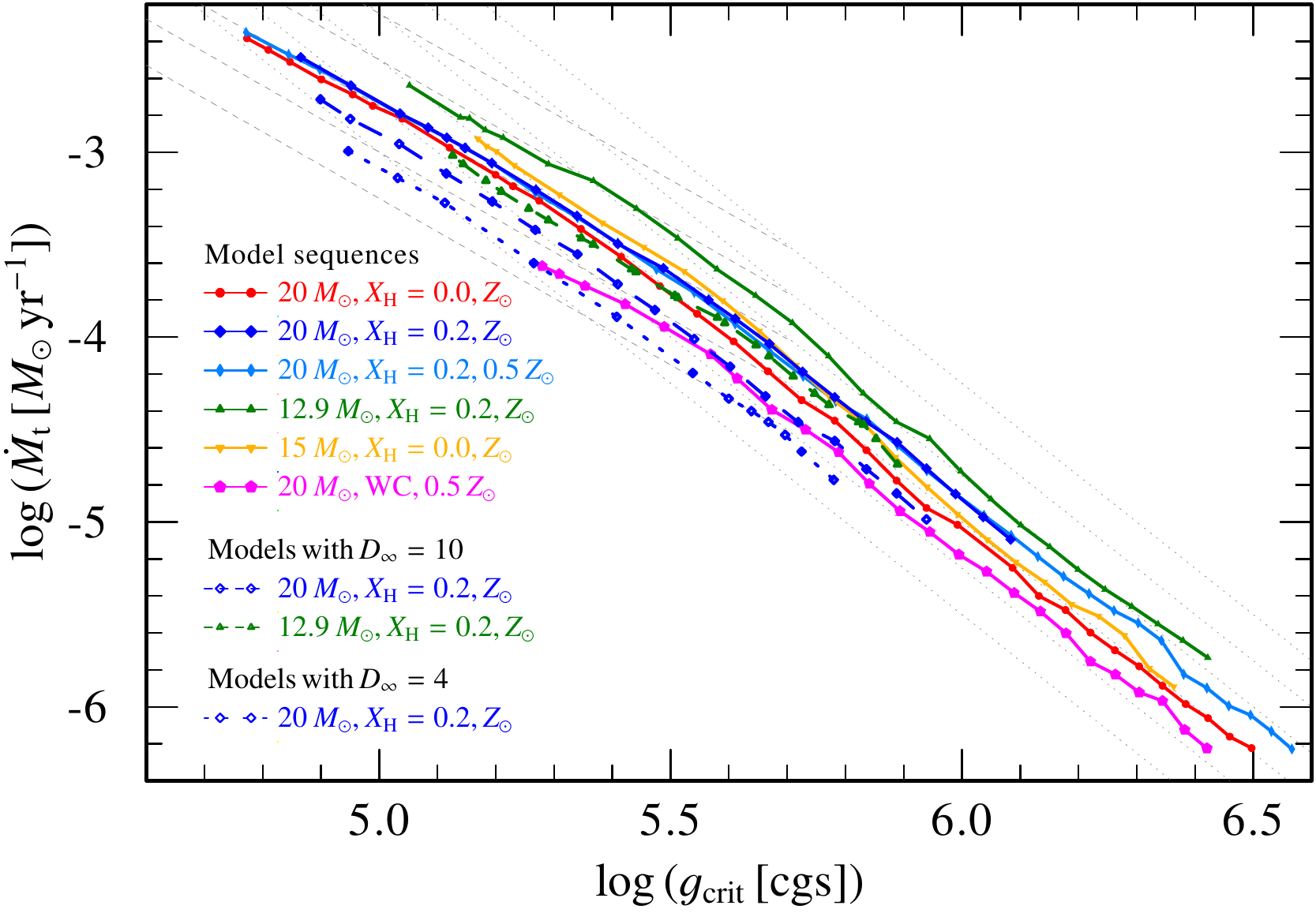}
  \caption{
  Transformed mass-loss rate $\dot{M}_\text{t}$ as a function of the gravitational acceleration at the critical radius $g_\text{crit} = G M R_\text{crit}^{-2}$. 
  To reflect the expected trends from the $R_\text{crit}$-fits, thin gray lines are plotted in the background. The
  dotted, gray lines indicate $\dot{M} \propto g_\text{crit}^{-2.5}$ (optically thin regime), while the dashed, gray lines correspond to $\dot{M} \propto g_\text{crit}^{-1.8}$ (optically thick regime).}
  \label{fig:mdott-gcrit}
\end{figure}
%--------------------------------------------

The scaling of $\varv_\infty$ with $R_\text{crit}$ introduces an additional dependency when considering the transformed mass-loss rate $\dot{M}_\text{t}$ as a function of $R_\text{crit}$ or $g_\text{crit}$.

From Eqs.\,\eqref{eq:mdot-teffcrit-trend} and \eqref{eq:mdot-rcrit}, we know that $\dot{M} \propto T_\text{eff,crit}^{-6}$ or $\dot{M} \propto R_\text{crit}^{3}$. From the definition of $\dot{M}_\text{t}$ (Eq.\,\ref{eq:mdott}) we get
\begin{equation}
  \log \dot{M}_\text{t}(R_\text{crit}) = \log \dot{M}(R_\text{crit}) - \log \varv_\infty(R_\text{crit}) + \text{offset.}
\end{equation}
With the different trends derived for the optically thick and thin limit in Sect.\,\ref{sec:vinf-rcrit}, we obtain $\dot{M}_\text{t} \propto R_\text{crit}^{3.6}$ and $\dot{M}_\text{t} \propto R_\text{crit}^{5}$ respectively.
Using $g_\text{crit}$ as defined in Eq.\,\eqref{eq:gcrit-def} and Eq.\,\eqref{eq:mdot-gcrit}, this yields
\begin{equation}
  \log \dot{M}_\text{t} = 1.8 \log g_\text{crit} + \text{offset} 
\end{equation}
for the optically thick limit and
\begin{equation}
  \log \dot{M}_\text{t} = 2.5 \log g_\text{crit} + \text{offset}  
\end{equation}
in the optically thin limit. These trends are depicted as sets of gray lines in Fig.\,\ref{fig:mdott-gcrit}, where the curves from the model sequences are shown as well. In contrast to the $\dot{M}(g_\text{crit})$-behavior discussed in Sect.\,\ref{sec:mdot-rcrit-trend}, the representation of the slope in the optically thin regime is less precise. In the optically thick limit, some curves align well, but others appear to be slightly steeper or shallower. Hence, the overall results for $\dot{M}_\text{t}(g_\text{crit})$ should be considered less robust than the $\dot{M}(g_\text{crit})$-trends. Interestingly, we do not see the ``kink'' or clear bending for some sequences at lowest (transformed) mass-loss rates that we see in Fig.\,\ref{fig:mdot-gcrit} for $\dot{M}(g_\text{crit})$. While it is hard to draw strong conclusions, there at least appears to be one continuous slope for $\dot{M}_\text{t}(g_\text{crit})$ in the thinner wind regime, regardless of whether the critical point is located at temperatures below or above $T_\text{eff,crit}$. Since we find a change for $\dot{M}$ alone, this would imply that $\varv_\infty$ outweighs this effect. While the inspection of the corresponding sequences in Fig.\,\ref{fig:vinf-t23} is only indicative here, indeed the $\varv_\infty$-curves of these sequences bend again toward shallower slopes then plotting them as functions of $T_\text{eff,crit}$. 

\section{Potential consequences for stellar evolution}
  \label{sec:evol}

Our study presents the very first sequences of hydrodynamically consistent atmosphere models in the cWR regime, where we vary the input parameter $T_\ast$ -- corresponding roughly to $T_\text{eff,crit}$ for most models. WR mass-loss recipes commonly do not incorporate any temperature/radius dependency, which can be seen as a consequence of the optically dense winds of WR stars. When reproducing their spectra with prescribed velocity fields, there is a degeneracy of solutions making it impossible to find a unique value of $T_\ast$ for more dense winds \citep[e.g.,][]{Hillier1991,HamannGraefener2004,Lefever+2022}. 

From an evolutionary standpoint, one could justify the omission of a $T_\ast$-dependence arguing that hydrogen-free WN stars -- and to some extend also WC stars -- may form a 1D sequence as they represent He-burning stars that do not contain any further shell structure which could skew the relation between the luminosity and mass. In reality, effects such as inflation, convection, or rotation augment the physical conditions of wind launching and mass loss, especially when considering their multidimensional nature. However, in the currently typical 1D spherical approach ignoring such issues, no further parameter would be necessary if the He star evolution could be perfectly mapped to one of the fundamental stellar parameters.
For He stars above $10\,M_\odot$, the intrinsic curvature of the HeZAMS indeed gets relatively small \citep[e.g.,][]{Langer1989} and the obtained tracks of WR evolution in different codes yield very similar temperatures around $\log (T\,[\mathrm{K}]) = 5.1$\footnote{In this discussion, we do not consider structure models that show hydrostatic envelope inflation for more massive He stars \citep[see, e.g., Fig.\,19 in][]{Koehler+2015} as \citet{Grassitelli+2018} demonstrated that such an inflation likely does not occur if a strong wind can be launched.}, regardless whether these have been calculated from pure He stars or including all prior evolution from the ZAMS \citep[e.g.,][]{Georgy+2012,LimongiChieffi2018, Higgins+2021}.

\subsection{Mass-loss comparison for a representative model}

In the models from \citet{SanderVink2020}, we thus ignored the width of $\approx 0.1\,$dex in $\log T_\ast$ and fixed $T_\ast$ in order to keep the total amount of models manageable. In this work, we now calculated a number of model sequences where we vary $T_\ast$ in order to investigate the effect of a wider range of $T_\ast$, which probes not only the curvature of the He ZAMS, but also gives a glimpse of how $\dot{M}$ might be affected for stars which are not yet or no longer (exactly) on the He ZAMS. We find that despite the narrow range in temperature, the effect on $\dot{M}$ is quite noticeable. For our $20\,M_\odot$ model sequence (at $Z_\odot$) even a narrow range of only $0.05\,$dex (i.e., $T_\ast = 125\dots140$) results in a factor of two in $\dot{M}$. Whether such a significant correction is really necessary depends on the difference between the most realistic choice of $T_\ast$ and the fixed value ($141\,$kK) in \citet{SanderVink2020}. Combining the structural constraints by \citet{Grassitelli+2018} with our model sequence, we find an ideal value of $T_\text{eff,crit} \approx T_\ast \approx 130\,$kK for a $20\,M_\odot$ at $Z_\odot$ model without any hydrogen. 

\begin{table}
  \caption{Comparison of mass-loss rates obtained with different methods for a hydrogen-free WN star with $\log L/L_\odot = 5.7$ and $20\,M_\odot$}
  \label{tab:m20cmp}
  \centering
 \begin{tabular}{lcc}
      \hline\hline
       Paper    &  \multicolumn{2}{c}{$\log (\dot{M}\,[M_\odot\,\mathrm{yr}^{-1}])$}  \\
                &     $Z_\odot$    &   $0.5\,Z_\odot$  \\
    \hline
   \citet{Graefener+2017}, s.-analytic\tablefootmark{(a)}  &   $-4.72$  &  no sol. \\
   \citet{Graefener+2017}, num.\tablefootmark{(b)}  &   $-4.65$   &  no sol. \\[0.2em]
   \citet{SanderVink2020} ($D_\infty = 50$)    &  $-4.61$  &  $-4.75$  \\
   \citet{SanderVink2020} ($D_\infty = 10$)    &  $-4.64$  &   $-4.83$\tablefootmark{(c)} \\[0.2em]
   this work, $T_\ast = 130\,$kK\tablefootmark{(d)} ($D_\infty = 50$) &  $\mathbf{-4.40}$ & $\mathbf{-4.56}$ \\ 
   this work, $T_\ast = 130\,$kK\tablefootmark{(d)} ($D_\infty = 10$) &  $\mathbf{-4.42}$ & $\mathit{-4.83}$\tablefootmark{(e)} \\[0.5em]
   \citet{NugisLamers2000} recipe & $-4.52$ & $-4.66$ \\ 
   Hamann95+\tablefootmark{(f)} recipe & $-4.40$ & $-4.65$ \\ 
   \citet{Yoon2017} recipe ($f_\text{WR} = 1$) & $-4.59$ & $-4.77$ \\
   \citet{Yoon2017} recipe ($f_\text{WR} = 1.6$) & $-4.39$ & $-4.57$ \\
      \hline  
 \end{tabular}
 \tablefoot{
    The $\dot{M}$ determinations by \citet{Graefener+2017} employ the $P_\text{rad}$-$P_\text{gas}$-plane with the sonic point conditions using \tablefoottext{a}{their Eq.\,(27) and assuming $\varv_\infty = 1800\,\mathrm{km}\,\mathrm{s}^{-1}$} or \tablefoottext{b}{a numerically integrated $\mathrm{d}P_\text{rad}/\mathrm{d}r$}.\\
    \tablefoottext{c}{New calculation, but with $T_\ast = 141\,$kK as in \citet{SanderVink2020}.}\\
    \tablefoottext{d}{Choice of $T_\ast$ based on matched $T_\text{eff}(\tau_\text{crit})$ with  \citet{Grassitelli+2018}}\\
    \tablefoottext{e}{Unstable solution close to driving breakdown, see Sect.\,\ref{sec:breakdown}}\\
    \tablefoottext{f}{Mass-loss rates from \citet{Hamann+1995} divided by a factor 10 and scaled with the $Z$-dependence from \citet{VdK2005}, as suggested by \citet{Yoon+2006}.}
 }
\end{table}

In Table\,\ref{tab:m20cmp}, we provide a comparison of the resulting mass-loss rates for a $20\,M_\odot$ star. Beside the values employing the new estimate of $T_\ast$ and the solutions for $T_\ast = 141\,$kK from \citet{SanderVink2020}, we also list the resulting $\dot{M}$-values from \citet{Graefener+2017} and commonly used (semi-)empirical recipes.
For $Z_\odot$ we find a difference of $\sim$$0.2\,$dex in $\dot{M}$, unaffected by the choice of $D_\infty$. Using the values of Table\,\ref{tab:m20cmp} as an average mass-loss rate during the typical He burning lifetime ($300\,$kyr), this corresponds to a difference between $12.5\,M_\odot$ and $8.1\,M_\odot$ at the end of core He-burning. This calculation is of course only a rough estimate and does not take any change of the stellar parameters or surface abundances into account. Nonetheless, the value using the $\dot{M}$ from \citet{SanderVink2020} is close to what we obtain with actual stellar evolution calculations in \citet{Higgins+2021}.

At $0.5\,Z_\odot$, a value roughly corresponding to the LMC, the $0.2\,$dex shift holds as well when adopting $D_\infty = 50$. For the hydrogen-free $130\,$kK model at $0.5\,Z_\odot$ with $D_\infty = 10$, however, we only find a solution if we relax the stability criterion on $\dot{M}$ between consecutive updates that we otherwise enforce. For the $141\,$kK model, we already see a notable difference in $\dot{M}$ when reducing from $D_\infty = 50$ down to $10$. The reason is that we are already close to the regime where we can no longer obtain a wind solution driven by the hot iron bump (see Sect.\,\ref{sec:breakdown}). For the $130\,$kK we have reached already a meta-stable situation with respect to the solution stability. Thus, the obtained value of $\dot{M}$ for $D_\infty = 10$ is much lower than expected. The matching of the absolute values for $130\,$kK and $141\,$kK is a pure coincidence with higher, also meta-stable solutions up to $\approx -4.7$ for $130\,$kK being possible as well.

\subsection{Structural limits and the role of hydrogen}
  \label{sec:struclimits}

The presence of hydrogen at the surface can considerably change the limits of the wind onset derived above. In contrast to our hydrogen-free results shown in Table\,\ref{tab:m20cmp} and depicted in Fig.\,\ref{fig:mdot-tscales-m20}, our model sequence with $X_\text{H} = 0.2$ and $0.5\,Z_\odot$ extends to much cooler temperatures ($T_\text{eff,crit} < 100\,$kK) as the additional acceleration from free electrons helps to compensate the effect of the deceleration regions. The choice of $D_\infty$ has some impact on the results, but in both cases the effect of surface hydrogen as such is much larger.

While we do not aim at a detailed comparison with observations in this work -- which would require new analyses with dynamically-consistent models -- it is striking that all hydrogen-free WN stars in the LMC are of the subtype WN4 or earlier \citep[][]{Hainich+2014,Shenar+2019}. Moreover, all WC stars in the LMC show early subtypes as well. While one has to be careful drawing absolute conclusions, our temperature study now indicates that beside the metallicity limiting the lower luminosity of the observed WN population, the observed restriction of the subtype regime might be a direct consequence of the inability to launch WR-type winds below a certain (sonic point) temperature. From the perspective of fixed stellar parameters, the lower mass-loss rate reached at a lower metallicity corresponds to a shift to earlier subtypes (cf.\,Fig.\,\ref{fig:spec-example}), thereby confirming the suggestion by \citet{Crowther+2002}.

Coming from a different angle, but addressing essentially the same problem, \citet{Grassitelli+2018} and \citet{Ro2019} used hydrodynamic stellar structure models and semi-analytic approaches to predict the existence of a ``minimum mass-loss rate'' for launching a stellar wind from the hot iron bump. For values of $\dot{M}$ below this limit, extended low-density regions were predicted. In this work, we do not aim to obtain the latter type of solutions, but the calculations failing to launch a wind show a tendency toward trying to launch a wind further out with a (much) lower $\dot{M}$. We can thus qualitatively confirm the structural predictions by \citet{Grassitelli+2018} and \citet{Ro2019}, assuming that we are bound to a choice of $T_\ast$ following the -- ideally hydrodynamical -- structure calculations for the HeZAMS. This underlines once more that unifying structural and atmosphere models remains a challenge that requires a new generation of both atmosphere and structure models.

\subsection{An approximated handling of the temperature shift}

In light of the structural considerations above, it appears likely that any future description of WR-type mass loss needs a temperature or radius-dependency. Simpler treatments might be reasonable in a time-averaged situation, but cannot predict a realistic mass loss for individual points along an evolutionary track. Hence, a detailed update of the \citet{SanderVink2020} formula will eventually be necessary, but the current amount of models does not allow a wide-space parameter investigation. The applicability to lower masses also turns out to be nontrivial: 
One the one hand, the curvature of the HeZAMS toward cooler temperatures should soften the sharp drop obtained in \citet{SanderVink2020} of $\dot{M}$ toward lower He star masses. On the other hand, we reach lower limits of $T_\text{eff,crit}$ for driving winds by the hot iron bump (cf.\ Sect.\,\ref{sec:breakdown}). In our $12.9\,M_\odot$ sequence with $X_\text{H} = 0.2$, this limit is at $\approx 97\,$kK. Given that this limit seems to increase to slightly higher temperatures for lower $L/M$ values, it appears unlikely that one can find any solution for a wind driven by the hot iron bump for stars with $M \leq 10\,M_\odot$ fulfilling the $L$-$M$ relation from \citet{Graefener+2011}.

While a full coverage of the driving limit at cooler temperatures will require its own tailored study, we can use Eq.\,\eqref{eq:mdot-teffcrit-trend} from Sect.\,\ref{sec:mdot-rcrit-trend} to derive a decent temperature description up to discontinuity in $\dot{M}$. Since Eq.\,\eqref{eq:mdot-teffcrit-trend} seems to be valid across both the optically thick and thin regime, we can approximate $\dot{M}$ for WN winds driven by the hot iron bump via
\begin{equation}
  \label{eq:mdot-tadj-rcrit}
   \log \left(\frac{\dot{M}}{M_\odot\,\mathrm{yr}^{-1}}\right) = \log \left(\frac{\dot{M}_\text{SV2020}}{M_\odot\,\mathrm{yr}^{-1}}\right) + 3 \log \left( \frac{R_\text{crit}}{R_{\text{crit},T141}} \right)
\end{equation}
with $\dot{M}_\text{SV2020}$ denoting the mass-loss rate from \citet{SanderVink2020} and $R_{\text{crit},T141} = R_\text{crit}(T_\ast = 141\,\mathrm{kK})$ being the critical radius (in $R_\odot$) of their corresponding model (e.g., $1.217\,R_\odot$ for the $20\,M_\odot$ He star without hydrogen). Although  $R_{\text{crit},T141}$ could be obtained from $L/M$ or $\Gamma_\text{e}$ via a nonlinear fit of the \citet{SanderVink2020} data, it is much more convenient to reformulate Eq.\,\eqref{eq:mdot-tadj-rcrit} in terms of the effective temperature at the critical ($\approx$ sonic) point $T_\text{eff,crit}$, yielding
\begin{equation}
  \label{eq:mdot-tadj-teffcrit}
   \log \left(\frac{\dot{M}}{M_\odot\,\mathrm{yr}^{-1}}\right) = \log \left(\frac{\dot{M}_\text{SV2020}}{M_\odot\,\mathrm{yr}^{-1}}\right) - 6 \log \left( \frac{T_\text{eff,crit}}{141\,\mathrm{kK}} \right)\text{.}
\end{equation}
Apart from small deviations for the highest mass-loss rates ($\dot{M} \gg 10^{-4}\,M_\odot\,\mathrm{yr}^{-1}$), the fixed value of $141\,$kK accurately represents the value of $T_\text{eff,crit}$ in \citet{SanderVink2020}, as illustrated previously in Fig.\,\ref{fig:tscales-sv20}.

This adjusted $\dot{M}$-recipe requires the knowledge of either $T_\text{eff,crit}$ or $R_\text{crit}$. As we did include only a small microturbulent velocity in our modeling efforts ($30\,\mathrm{km}\,\mathrm{s}^{-1}$), the quantities can be replaced by the sonic point values without introducing a considerable error.
Still, the accurate use of Eq.\,\eqref{eq:mdot-tadj-rcrit} and \eqref{eq:mdot-tadj-teffcrit} requires models with a meaningful sonic point in a hydrodynamical sense to prevent reintroducing any further radius/temperature discrepancies. Stellar atmosphere analyses typically employ predefined velocity fields (usually $\beta$-laws) and thus do not have a sonic point that is hydrodynamically consistent. Purely hydrostatic stellar structure calculations are problematic as well as they do not yield a sonic point by construction. This underlines that in order to obtain a really insight- and meaningful comparison between theory and observation for optically thick winds, a new generation of both atmosphere and stellar structure models will be necessary.

The results obtained in our study could be helpful to eventually obtain realistic predictions for the effective temperatures ($T_{2/3}$) of WR stars in stellar evolution models. A route toward such a recipe based on our findings is given in appendix Sect.\,\ref{sec:t23estimate}.

\section{Transparency to He II ionizing photons}
  \label{sec:ionflux}

Despite having intrinsically quite hot temperatures, classical WR stars do not necessarily emit a significant number of ionizing photons beyond the \ion{He}{ii} ionization edge, that is below $227\,$\AA\ or above $54\,$eV. As first described in \citet{Schmutz+1992}, the transparency of the wind for photons with energies above $54\,$eV depends on the mass-loss rate, and thus the density of the wind. In more dense winds, \ion{He}{iii} recombines to \ion{He}{ii}, making the atmosphere opaque to \ion{He}{ii} ionizing photons out to very large radii. The presence of line blanketing further affects the absolute ionizing fluxes significantly \citep[cf.][]{Smith+2002}. Beside usually leading to a reduction of the $\ion{He}{i}$ ionizing flux, it can also affect the $\ion{He}{ii}$ ionizing flux transition by a few orders of magnitude as we will see in our model calculations.
To cover the region where the (continuum) optical depth drops below unity in this wavelength region, we extended the outer boundary radius $R_\text{max}$ of our atmosphere models to extremely large values, often up to $100\,000\,R_\ast$. (Typical atmosphere models for spectral fitting require only $R_\text{max} = 100\dots1000\,R_\ast$.)

%--------- Figure ---------------------------------------------------
\begin{figure}
  \includegraphics[width=\columnwidth]{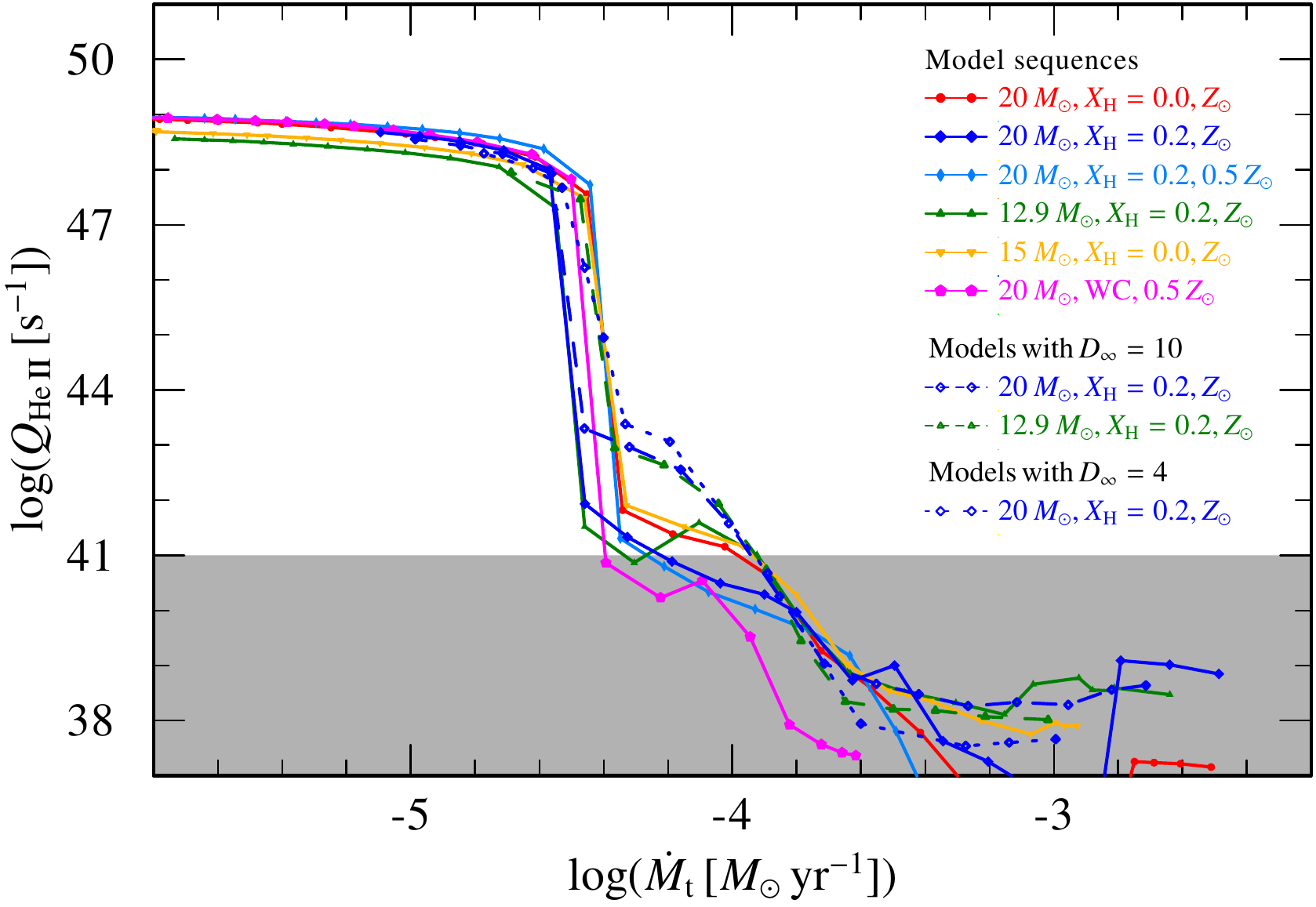}
  \caption{Number of helium ionizing photons per second $Q_\ion{He}{ii}$ as a function of the transformed mass-loss rate $\dot{M}_\text{t}$ for the new model sequences calculated in this work.}
  \label{fig:qheii-mdott}
\end{figure}
%--------------------------------------------

In Fig.\,\ref{fig:qheii-mdott}, we plot the rate of ionizing photons per second $Q_\ion{He}{ii}$ as a function of the transformed mass-loss rate $\dot{M}_\text{t}$. 
The absolute numbers in the regime with low $Q_\ion{He}{ii}$ are more uncertain as they depend strongly on the precise boundary treatment. If the wind becomes transparent around and below $227\,$\AA\ only very close to the model boundary or even remains optically thick at $R_\text{max}$, the value for $Q_\ion{He}{ii}$ can be underestimated, but should never exceed $10^{41}\,\mathrm{s}^{-1}$. Given that this is many orders of magnitude below the actually strong $Q_\ion{He}{ii}$ emitters with rates $> 10^{47}\,\mathrm{s}^{-1}$, the values of the shaded regime in Fig.\,\ref{fig:qheii-mdott} should have no practical consequences.

%--------- Figure ---------------------------------------------------
\begin{figure}
  \includegraphics[width=\columnwidth]{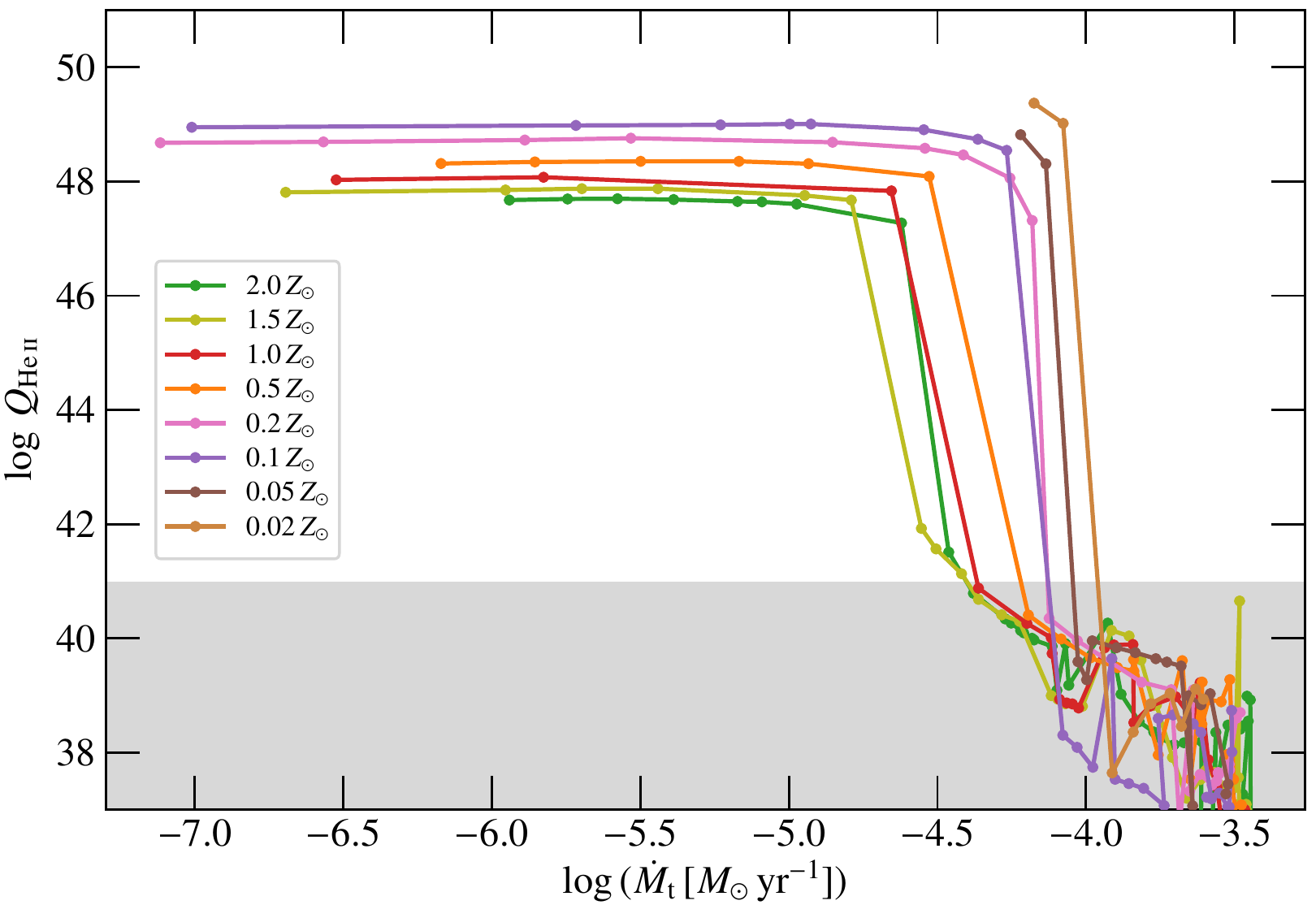}
  \caption{Number of helium ionizing photons per second $Q_\ion{He}{ii}$ as a function of the transformed mass-loss rate $\dot{M}_\text{t}$ for the model sequences computed in \citet{SanderVink2020}.}
  \label{fig:qheii-mdott-sv20}
\end{figure}
%--------------------------------------------

Displaying the \ion{He}{ii} ionizing flux as a function of $\dot{M}_\text{t}$ in Fig.\,\ref{fig:qheii-mdott} confirms that similar to other quantities, the switch in transparency is caused by the lower wind density. In fact, there seems to be a critical lower boundary around $\log (\dot{M}_\text{t}\,[M_\odot\,\mathrm{yr}^{-1}]) = -4.6$ to $-4.4$ where all model sequences switch abruptly. To study whether this transition value might be more universal, we created the same plot for the model sequences from \citet{SanderVink2020}. Their model set, shown in Fig.\,\ref{fig:qheii-mdott-sv20}, clearly hints at a $Z$-dependency for the transition, which we do only sparsely map in our new model set. In fact, our new sequences are quite complementary to the datasets from \citet{SanderVink2020}, indicating that the transition does not only depend on $Z$ as a total value, but likely on the detailed composition and -- notably especially at the lower end of the transition in Fig.\,\ref{fig:qheii-mdott} -- on the choice of the clumping factor. Nonetheless, we can conclude that all stars in our large model sample with $\log (\dot{M}_\text{t}\,[M_\odot\,\mathrm{yr}^{-1}]) < -4.6$ are strong emitters of \ion{He}{ii} ionizing flux. Thus, we propose to take this value as an upper limit of whether to consider WR stars as notable contributors to the \ion{He}{ii} ionizing photon budget.

\section{Summary and conclusions}
  \label{sec:conclusions}

In this work, we presented an exploratory study for the temperature-dependency of radiation-driven winds launched by the so-called hot iron opacity bump. For the first time, we calculated temperature-dependent sequences of hydrodynamically consistent stellar atmosphere models in the cWR regime. To achieve our results, we had to allow for nonmonotonic velocity field solutions when solving the hydrodynamic equation of motion. In order to perform the necessary radiative transfer in the comoving frame, we afterwards interpolated the obtained velocity fields such that the main wind properties ($\dot{M}, \varv_\infty$) as well as the characteristics in the outer wind were maintained. We draw the following conclusions:
\begin{itemize}
  \item The mass-loss rates $\dot{M}$ depend significantly on the critical radius $R_\text{crit}$ and thus also on the assumed model temperature setting $T_\text{eff}(R_\text{crit})$. For model sequences with constant luminosity $L$ and stellar mass $M$, we obtain $\dot{M} \propto R_\text{crit}^3$ over a wide range with moderate deviations from this purely geometrical effect occurring at the lower and upper end of our sequences. This finding can also be expressed in the form of $\dot{M} \propto g_\text{crit}^{-3/2}$, reflecting that larger radii for the critical point imply a lower gravitational force. Our findings underline that WR-type mass-loss depends on multiple parameters and the 2D description from \citet{SanderVink2020} needs to be extended further to describe all relevant effects.

  \item Except for very dense winds -- corresponding to $\log (\dot{M}_\text{t}\,[M_\odot\,\mathrm{yr}^{-1}]) \approx -3.0$ and above -- the effective temperature at the critical point $T_\text{eff}(\tau_\text{crit})$ is close to the effective temperature at a Rosseland continuum optical depth of $\tau_\text{R,c} = 20$. For WN-type models $\tau_\text{R,c} = 20$ typically corresponds to $\tau_\text{Thom} \approx 17$, albeit with considerable scatter along the model sequences.
  
  \item We find a characteristic value of $\log (\dot{M}_\text{t}\,[M_\odot\,\mathrm{yr}^{-1}]) \approx -4.5$ for the transition between the optically thin and thick regime. While there is some scatter between different model sequences, this characteristic value of $\dot{M}_\text{t}$ (plus some error margin) provides a very convenient tool to distinguish between the regimes as $\dot{M}_{t}$ can also be determined with empirical models. Known WC stars show values well above this \citep[e.g.,][]{GraefenerVink2013,Sander+2019} while WO stars might be found on both sides of the transition.
  Whether the characteristic value of $\dot{M}_{t}$ also holds for winds that might not be driven by the hot iron bump is currently unclear. We calculated the transformed mass-loss rates for stars at the spectral transition from Of to WNh, which likely happens at a cooler temperature regime than studied in this work\footnote{The transformed mass-loss rate $\dot{M}_\text{t}$ as such should not be confused with the \textit{transition} mass-loss rate $\dot{M}_\text{trans}$ from \citet{VinkGraefener2012}. Nonetheless, if the other necessary parameters are known, one can estimate the corresponding transformed mass-loss rates for stars defining the transition mass-loss rate, denoted as $\dot{M}_\text{t,trans}$}. Their corresponding transformed mass-loss rates $\dot{M}_\text{t,trans}$ appear to be below $-4.5$, e.g.\ at $\log (\dot{M}_\text{t,trans}\,[M_\odot\,\mathrm{yr}^{-1}]) \approx -5.0$ in the Arches cluster \citep[][]{Martins+2008,VinkGraefener2012} and even lower in R136 \citep[][]{Bestenlehner2020,Bestenlehner+2020}. 
  
  \item The choice of the maximum clumping factor $D_\infty$ does not affect our derived $\dot{M}(T_{2/3})$ trends, but leads to an additional shift in the obtained relations with higher clumping factors corresponding to higher $\dot{M}$ for the same $T_{2/3}$. In contrast to all shifts introduced by varying fundamental stellar parameters or abundances, the shift due to $D_\infty$ does not vanish when considering $\dot{M}_\text{t}(T_{2/3})$ instead of $\dot{M}(T_{2/3})$.

  \item In the limit of optically thick winds, we obtain a linear relation between $\log T_{2/3}$ and $\log \dot{M}_\text{t}$, independent of chemical composition (but for a fixed clumping factor). Combined with the also $Z$-independent result from \citet{SanderVink2020} that $\log \dot{M}_\text{t} \propto \log (L/M)$, this could provide an easy-to-use prediction for WR effective temperatures in stellar structure and evolution models.

  \item Classical WR stars and non-WR helium stars with $\log (\dot{M}_\text{t}\,[M_\odot\,\mathrm{yr}^{-1}]) < -4.6$ are strong emitters of \ion{He}{ii} ionizing flux (with $Q_\ion{He}{ii} > 10^{48}\,\mathrm{s}^{-1}$). Helium stars with stronger winds are (mostly) opaque to radiation above $54\,$eV and thus should not be considered as sources of hard ionizing radiation.

  \item Albeit being limited in comparability to particular observed targets, our findings indicate that high clumping factors ($D \approx 50$) might be necessary to reproduce the observed combinations of $\dot{M}$ and $\varv_\infty$. This is in sharp contrast to the first results obtained from 3D wind modeling by \citet{Moens+2022} arguing for much lower clumping factors of $D \approx 2$. At present, the reason for this discrepancy is unclear. Various solutions are possible, e.g., missing opacities in our wind models -- where the presently assumed high clumping would act as a ``fudge factor'' to make up for that. Alternatively, the sharp contrast in the clumping factor might simply be the result of a mismatch between the considered regimes. Currently, the 3D models from \citet{Moens+2022} probe only the wind onset where even in our 1D models we assume $D(r) \ll D_\infty$ with $D(\tau_\text{crit})$ typically ranging between $1.5$ and $4$.

  \item When comparing empirically obtained results in the $T_{2/3}$-$\dot{M}_\text{t}$-plane to our derived curves, we find a significant fraction of stars to have lower values of $\dot{M}_\text{t}$ than predicted by our curves using hydrodynamic models. It is currently unclear whether this is due to a deviation from the theoretical setup in this work (e.g., different clumping stratification, other $L$-$M$ combinations) or inherent simplifications in the empirical analyses (e.g., the use of a $\beta$-type velocity law). A dedicated analysis of individual objects with hydrodynamical model atmospheres will be necessary to uncover the origin of this discrepancy.

  \item The limits of driving optically thick winds crucially depend on our knowledge of opacities. In case of considerable changes -- e.g., a higher iron opacity as reported by \citet{Bailey+2015} for the so-called ``deep iron bump'' at $T_\text{e} \approx 2 \cdot 10^{6}\,$K -- wind quantity predictions such as $\dot{M}$ and $\varv_\infty$ could shift significantly. Moreover, our understanding of the limits of radiative driving would be affected as well, e.g., due to strengthening or weakening the bumpy radius dependency of the flux-weighted mean opacity $\varkappa_F$. Beside the impact of multi-D effects, higher (Fe) opacities could play an important role to resolve current discrepancies, such as the lower luminosity end of the LMC WN population or the aforementioned need for higher $D_\infty$ to reach the observed terminal velocities.
\end{itemize}

With these conclusions, our study underlines the complexity of radiation-driven mass loss, revealing both parameter regimes with a clear scalings and characteristic transitions as well as more obscure parameter regions where $\dot{M}$ appears to break down suddenly. We provide an adjustment of the recent $\dot{M}$-description from \citet{SanderVink2020} to account for different radii (or effective temperatures respectively) and emphasize that the model efforts presented there as well as in this work were limited to the regime where winds are launched by the hot iron bump. We thus consider our work as an intermediate step on the way toward a more comprehensive understanding of WR-type mass loss and will expand to other regimes in future studies.

\begin{acknowledgements}
The authors would like to thank the anonymous referee for their careful and constructive comments and suggestions.
AACS and VR acknowledge support by the Deutsche Forschungsgemeinschaft (DFG, German Research Foundation) in the form of an Emmy Noether Research Group -- Project-ID 445674056 (SA4064/1-1, PI Sander). RRL is funded by the Deutsche Forschungsgemeinschaft (DFG, German Research Foundation) – Project-ID 138713538 -- SFB 881 (“The Milky Way System”, subproject P04). LP acknowledges support by the Deutsche Forschungsgemeinschaft -- Project-ID 496854903 (SA 4046/2-1, PI Sander). JSV is supported by STFC funding under grant number ST/V000233/1.
This publication has benefited from discussions in a team meeting (PI: Oskinova) sponsored by the International Space Science Institute (ISSI) at Bern, Switzerland.
A significant number of figures in this work were created with \textsc{WRplot}, developed by W.-R.\ Hamann.
\end{acknowledgements}

%%%%%%%%%%%%%%%%%%%%%%%%%%%%%%%%%%%%%%%%%%%%%%%%%%



\bibliographystyle{aa}
\bibliography{literatur} 

\begin{thebibliography}{80}
\expandafter\ifx\csname natexlab\endcsname\relax\def\natexlab#1{#1}\fi

\bibitem[{{Aadland} {et~al.}(2022){Aadland}, {Massey}, {John Hillier},
  {Morrell}, {Neugent}, \& {Eldridge}}]{Aadland+2022}
{Aadland}, E., {Massey}, P., {John Hillier}, D., {et~al.} 2022, \apj, 931, 157

\bibitem[{{Abbott} {et~al.}(2016){Abbott}, {Abbott}, {Abbott}, {Abernathy},
  {Acernese}, {Ackley}, {Adams}, {Adams}, {Addesso}, {Adhikari}, \&
  et~al.}]{Abbott+2016}
{Abbott}, B.~P., {Abbott}, R., {Abbott}, T.~D., {et~al.} 2016, \apjl, 818, L22

\bibitem[{{Bailey} {et~al.}(2015){Bailey}, {Nagayama}, {Loisel}, {Rochau},
  {Blancard}, {Colgan}, {Cosse}, {Faussurier}, {Fontes}, {Gilleron},
  {Golovkin}, {Hansen}, {Iglesias}, {Kilcrease}, {Macfarlane}, {Mancini},
  {Nahar}, {Orban}, {Pain}, {Pradhan}, {Sherrill}, \& {Wilson}}]{Bailey+2015}
{Bailey}, J.~E., {Nagayama}, T., {Loisel}, G.~P., {et~al.} 2015, \nat, 517, 56

\bibitem[{{Bestenlehner}(2020)}]{Bestenlehner2020}
{Bestenlehner}, J.~M. 2020, \mnras, 493, 3938

\bibitem[{{Bestenlehner} {et~al.}(2020){Bestenlehner}, {Crowther},
  {Caballero-Nieves}, {Schneider}, {Sim{\'o}n-D{\'\i}az}, {Brands}, {de Koter},
  {Gr{\"a}fener}, {Herrero}, {Langer}, {Lennon}, {Ma{\'\i}z Apell{\'a}niz},
  {Puls}, \& {Vink}}]{Bestenlehner+2020}
{Bestenlehner}, J.~M., {Crowther}, P.~A., {Caballero-Nieves}, S.~M., {et~al.}
  2020, \mnras, 499, 1918

\bibitem[{{Bj{\"o}rklund} {et~al.}(2021){Bj{\"o}rklund}, {Sundqvist}, {Puls},
  \& {Najarro}}]{Bjoerklund+2021}
{Bj{\"o}rklund}, R., {Sundqvist}, J.~O., {Puls}, J., \& {Najarro}, F. 2021,
  \aap, 648, A36

\bibitem[{{Castor} {et~al.}(1975){Castor}, {Abbott}, \& {Klein}}]{Castor+1975}
{Castor}, J.~I., {Abbott}, D.~C., \& {Klein}, R.~I. 1975, \apj, 195, 157

\bibitem[{{Chieffi} \& {Limongi}(2013)}]{ChieffiLimongi2013}
{Chieffi}, A. \& {Limongi}, M. 2013, \apj, 764, 21

\bibitem[{{Conti}(1991)}]{Conti1991}
{Conti}, P.~S. 1991, \apj, 377, 115

\bibitem[{{Crowther} {et~al.}(2002){Crowther}, {Dessart}, {Hillier}, {Abbott},
  \& {Fullerton}}]{Crowther+2002}
{Crowther}, P.~A., {Dessart}, L., {Hillier}, D.~J., {Abbott}, J.~B., \&
  {Fullerton}, A.~W. 2002, \aap, 392, 653

\bibitem[{{Crowther} \& {Hadfield}(2006)}]{CrowtherHadfield2006}
{Crowther}, P.~A. \& {Hadfield}, L.~J. 2006, \aap, 449, 711

\bibitem[{{de Koter} {et~al.}(1997){de Koter}, {Heap}, \&
  {Hubeny}}]{deKoter+1997}
{de Koter}, A., {Heap}, S.~R., \& {Hubeny}, I. 1997, \apj, 477, 792

\bibitem[{{Dray} {et~al.}(2003){Dray}, {Tout}, {Karakas}, \&
  {Lattanzio}}]{Dray+2003}
{Dray}, L.~M., {Tout}, C.~A., {Karakas}, A.~I., \& {Lattanzio}, J.~C. 2003,
  \mnras, 338, 973

\bibitem[{{Farmer} {et~al.}(2021){Farmer}, {Laplace}, {de Mink}, \&
  {Justham}}]{Farmer+2021}
{Farmer}, R., {Laplace}, E., {de Mink}, S.~E., \& {Justham}, S. 2021, \apj,
  923, 214

\bibitem[{{Garcia} {et~al.}(2014){Garcia}, {Herrero}, {Najarro}, {Lennon}, \&
  {Alejandro Urbaneja}}]{Garcia+2014}
{Garcia}, M., {Herrero}, A., {Najarro}, F., {Lennon}, D.~J., \& {Alejandro
  Urbaneja}, M. 2014, \apj, 788, 64

\bibitem[{{Georgy} {et~al.}(2012){Georgy}, {Ekstr{\"o}m}, {Meynet}, {Massey},
  {Levesque}, {Hirschi}, {Eggenberger}, \& {Maeder}}]{Georgy+2012}
{Georgy}, C., {Ekstr{\"o}m}, S., {Meynet}, G., {et~al.} 2012, \aap, 542, A29

\bibitem[{{Gr{\"a}fener} \& {Hamann}(2005)}]{GraefenerHamann2005}
{Gr{\"a}fener}, G. \& {Hamann}, W.-R. 2005, \aap, 432, 633

\bibitem[{{Gr{\"a}fener} {et~al.}(2002){Gr{\"a}fener}, {Koesterke}, \&
  {Hamann}}]{Graefener+2002}
{Gr{\"a}fener}, G., {Koesterke}, L., \& {Hamann}, W.-R. 2002, \aap, 387, 244

\bibitem[{{Gr{\"a}fener} {et~al.}(2017){Gr{\"a}fener}, {Owocki}, {Grassitelli},
  \& {Langer}}]{Graefener+2017}
{Gr{\"a}fener}, G., {Owocki}, S.~P., {Grassitelli}, L., \& {Langer}, N. 2017,
  \aap, 608, A34

\bibitem[{{Gr{\"a}fener} \& {Vink}(2013)}]{GraefenerVink2013}
{Gr{\"a}fener}, G. \& {Vink}, J.~S. 2013, \aap, 560, A6

\bibitem[{{Gr{\"a}fener} {et~al.}(2011){Gr{\"a}fener}, {Vink}, {de Koter}, \&
  {Langer}}]{Graefener+2011}
{Gr{\"a}fener}, G., {Vink}, J.~S., {de Koter}, A., \& {Langer}, N. 2011, \aap,
  535, A56

\bibitem[{{Gr{\"a}fener} {et~al.}(2012){Gr{\"a}fener}, {Vink}, {Harries}, \&
  {Langer}}]{Graefener+2012}
{Gr{\"a}fener}, G., {Vink}, J.~S., {Harries}, T.~J., \& {Langer}, N. 2012,
  \aap, 547, A83

\bibitem[{{Grassitelli} {et~al.}(2018){Grassitelli}, {Langer}, {Grin},
  {Mackey}, {Bestenlehner}, \& {Gr{\"a}fener}}]{Grassitelli+2018}
{Grassitelli}, L., {Langer}, N., {Grin}, N.~J., {et~al.} 2018, \aap, 614, A86

\bibitem[{{Groh} {et~al.}(2014){Groh}, {Meynet}, {Ekstr{\"o}m}, \&
  {Georgy}}]{Groh+2014}
{Groh}, J.~H., {Meynet}, G., {Ekstr{\"o}m}, S., \& {Georgy}, C. 2014, \aap,
  564, A30

\bibitem[{{Hainich} {et~al.}(2015){Hainich}, {Pasemann}, {Todt}, {Shenar},
  {Sander}, \& {Hamann}}]{Hainich+2015}
{Hainich}, R., {Pasemann}, D., {Todt}, H., {et~al.} 2015, \aap, 581, A21

\bibitem[{{Hainich} {et~al.}(2014){Hainich}, {R{\"u}hling}, {Todt}, {Oskinova},
  {Liermann}, {Gr{\"a}fener}, {Foellmi}, {Schnurr}, \& {Hamann}}]{Hainich+2014}
{Hainich}, R., {R{\"u}hling}, U., {Todt}, H., {et~al.} 2014, \aap, 565, A27

\bibitem[{{Hamann} {et~al.}(2006){Hamann}, {Gr{\"a}fener}, \&
  {Liermann}}]{Hamann+2006}
{Hamann}, W., {Gr{\"a}fener}, G., \& {Liermann}, A. 2006, \aap, 457, 1015

\bibitem[{{Hamann} \& {Koesterke}(1998)}]{HK1998}
{Hamann}, W. \& {Koesterke}, L. 1998, \aap, 335, 1003

\bibitem[{{Hamann}(1985)}]{Hamann1985}
{Hamann}, W.~R. 1985, \aap, 145, 443

\bibitem[{{Hamann} \& {Gr{\"a}fener}(2003)}]{HamannGraefener2003}
{Hamann}, W.-R. \& {Gr{\"a}fener}, G. 2003, \aap, 410, 993

\bibitem[{{Hamann} \& {Gr{\"a}fener}(2004)}]{HamannGraefener2004}
{Hamann}, W.-R. \& {Gr{\"a}fener}, G. 2004, \aap, 427, 697

\bibitem[{{Hamann} {et~al.}(2019){Hamann}, {Gr{\"a}fener}, {Liermann},
  {Hainich}, {Sander}, {Shenar}, {Ramachandran}, {Todt}, \&
  {Oskinova}}]{Hamann+2019}
{Hamann}, W.-R., {Gr{\"a}fener}, G., {Liermann}, A., {et~al.} 2019, \aap, 625,
  A57

\bibitem[{{Hamann} {et~al.}(1995){Hamann}, {Koesterke}, \&
  {Wessolowski}}]{Hamann+1995}
{Hamann}, W.~R., {Koesterke}, L., \& {Wessolowski}, U. 1995, \aap, 299, 151

\bibitem[{{Higgins} {et~al.}(2021){Higgins}, {Sander}, {Vink}, \&
  {Hirschi}}]{Higgins+2021}
{Higgins}, E.~R., {Sander}, A.~A.~C., {Vink}, J.~S., \& {Hirschi}, R. 2021,
  \mnras, 505, 4874

\bibitem[{{Hillier}(1991)}]{Hillier1991}
{Hillier}, D.~J. 1991, in Wolf-Rayet Stars and Interrelations with Other
  Massive Stars in Galaxies, ed. K.~A. {van der Hucht} \& B.~{Hidayat}, Vol.
  143, 59

\bibitem[{{Hillier} {et~al.}(2001){Hillier}, {Davidson}, {Ishibashi}, \&
  {Gull}}]{Hillier+2001}
{Hillier}, D.~J., {Davidson}, K., {Ishibashi}, K., \& {Gull}, T. 2001, \apj,
  553, 837

\bibitem[{{Hillier} \& {Miller}(1999)}]{HillierMiller1999}
{Hillier}, D.~J. \& {Miller}, D.~L. 1999, \apj, 519, 354

\bibitem[{{K{\"o}hler} {et~al.}(2015){K{\"o}hler}, {Langer}, {de Koter}, {de
  Mink}, {Crowther}, {Evans}, {Gr{\"a}fener}, {Sana}, {Sanyal}, {Schneider}, \&
  {Vink}}]{Koehler+2015}
{K{\"o}hler}, K., {Langer}, N., {de Koter}, A., {et~al.} 2015, \aap, 573, A71

\bibitem[{{Langer}(1989)}]{Langer1989}
{Langer}, N. 1989, \aap, 210, 93

\bibitem[{{Langer} {et~al.}(1994){Langer}, {Hamann}, {Lennon}, {Najarro},
  {Pauldrach}, \& {Puls}}]{Langer+1994}
{Langer}, N., {Hamann}, W.~R., {Lennon}, M., {et~al.} 1994, \aap, 290, 819

\bibitem[{{Lefever} {et~al.}(2022){Lefever}, {Shenar}, {Sander}, {Poniatowski},
  {Dsilva}, \& {Todt}}]{Lefever+2022}
{Lefever}, R.~R., {Shenar}, T., {Sander}, A. A.~C., {et~al.} 2022, arXiv
  e-prints, arXiv:2209.06043

\bibitem[{{Leitherer} {et~al.}(1996){Leitherer}, {Vacca}, {Conti},
  {Filippenko}, {Robert}, \& {Sargent}}]{Leitherer+1996}
{Leitherer}, C., {Vacca}, W.~D., {Conti}, P.~S., {et~al.} 1996, \apj, 465, 717

\bibitem[{{Limongi} \& {Chieffi}(2018)}]{LimongiChieffi2018}
{Limongi}, M. \& {Chieffi}, A. 2018, \apjs, 237, 13

\bibitem[{{Maeder}(1983)}]{Maeder1983}
{Maeder}, A. 1983, \aap, 120, 113

\bibitem[{{Martinet} {et~al.}(2022){Martinet}, {Meynet}, {Nandal},
  {Ekstr{\"o}m}, {Georgy}, {Haemmerl{\'e}}, {Hirschi}, {Yusof}, {Gounelle}, \&
  {Dwarkadas}}]{Martinet+2022}
{Martinet}, S., {Meynet}, G., {Nandal}, D., {et~al.} 2022, \aap, 664, A181

\bibitem[{{Martins} {et~al.}(2008){Martins}, {Hillier}, {Paumard},
  {Eisenhauer}, {Ott}, \& {Genzel}}]{Martins+2008}
{Martins}, F., {Hillier}, D.~J., {Paumard}, T., {et~al.} 2008, \aap, 478, 219

\bibitem[{{Moens} {et~al.}(2022){Moens}, {Poniatowski}, {Hennicker},
  {Sundqvist}, {El Mellah}, \& {Kee}}]{Moens+2022}
{Moens}, N., {Poniatowski}, L.~G., {Hennicker}, L., {et~al.} 2022, \aap, 665,
  A42

\bibitem[{{Moriya} \& {Yoon}(2022)}]{MoriyaYoon2022}
{Moriya}, T.~J. \& {Yoon}, S.-C. 2022, \mnras, 513, 5606

\bibitem[{{Nakauchi} \& {Saio}(2018)}]{NakauchiSaio2018}
{Nakauchi}, D. \& {Saio}, H. 2018, \apj, 852, 126

\bibitem[{{Nugis} \& {Lamers}(2000)}]{NugisLamers2000}
{Nugis}, T. \& {Lamers}, H.~J.~G.~L.~M. 2000, \aap, 360, 227

\bibitem[{{Nugis} \& {Lamers}(2002)}]{NugisLamers2002}
{Nugis}, T. \& {Lamers}, H.~J.~G.~L.~M. 2002, \aap, 389, 162

\bibitem[{{Petrovic} {et~al.}(2006){Petrovic}, {Pols}, \&
  {Langer}}]{Petrovic+2006}
{Petrovic}, J., {Pols}, O., \& {Langer}, N. 2006, \aap, 450, 219

\bibitem[{{Plat} {et~al.}(2019){Plat}, {Charlot}, {Bruzual}, {Feltre},
  {Vidal-Garc{\'\i}a}, {Morisset}, {Chevallard}, \& {Todt}}]{Plat+2019}
{Plat}, A., {Charlot}, S., {Bruzual}, G., {et~al.} 2019, \mnras, 490, 978

\bibitem[{{Poniatowski} {et~al.}(2021){Poniatowski}, {Sundqvist}, {Kee},
  {Owocki}, {Marchant}, {Decin}, {de Koter}, {Mahy}, \&
  {Sana}}]{Poniatowski+2021}
{Poniatowski}, L.~G., {Sundqvist}, J.~O., {Kee}, N.~D., {et~al.} 2021, \aap,
  647, A151

\bibitem[{{Ro}(2019)}]{Ro2019}
{Ro}, S. 2019, \apj, 873, 76

\bibitem[{{Ro} \& {Matzner}(2016)}]{RoMatzner2016}
{Ro}, S. \& {Matzner}, C.~D. 2016, \apj, 821, 109

\bibitem[{{Sander} {et~al.}(2012){Sander}, {Hamann}, \& {Todt}}]{Sander+2012}
{Sander}, A., {Hamann}, W.-R., \& {Todt}, H. 2012, \aap, 540, A144

\bibitem[{{Sander} {et~al.}(2015){Sander}, {Shenar}, {Hainich},
  {G{\'{\i}}menez-Garc{\'{\i}}a}, {Todt}, \& {Hamann}}]{Sander+2015}
{Sander}, A., {Shenar}, T., {Hainich}, R., {et~al.} 2015, \aap, 577, A13

\bibitem[{{Sander} {et~al.}(2018){Sander}, {F{\"u}rst}, {Kretschmar},
  {Oskinova}, {Todt}, {Hainich}, {Shenar}, \& {Hamann}}]{Sander+2018}
{Sander}, A.~A.~C., {F{\"u}rst}, F., {Kretschmar}, P., {et~al.} 2018, \aap,
  610, A60

\bibitem[{{Sander} {et~al.}(2017){Sander}, {Hamann}, {Todt}, {Hainich}, \&
  {Shenar}}]{Sander+2017}
{Sander}, A.~A.~C., {Hamann}, W.-R., {Todt}, H., {Hainich}, R., \& {Shenar}, T.
  2017, \aap, 603, A86

\bibitem[{{Sander} {et~al.}(2019){Sander}, {Hamann}, {Todt}, {Hainich},
  {Shenar}, {Ramachandran}, \& {Oskinova}}]{Sander+2019}
{Sander}, A.~A.~C., {Hamann}, W.-R., {Todt}, H., {et~al.} 2019, \aap, 621, A92

\bibitem[{{Sander} \& {Vink}(2020)}]{SanderVink2020}
{Sander}, A. A.~C. \& {Vink}, J.~S. 2020, \mnras, 499, 873

\bibitem[{{Sander} {et~al.}(2020){Sander}, {Vink}, \& {Hamann}}]{Sander+2020}
{Sander}, A. A.~C., {Vink}, J.~S., \& {Hamann}, W.~R. 2020, \mnras, 491, 4406

\bibitem[{{Schaerer} {et~al.}(1999){Schaerer}, {Contini}, \&
  {Kunth}}]{Schaerer+1999}
{Schaerer}, D., {Contini}, T., \& {Kunth}, D. 1999, \aap, 341, 399

\bibitem[{{Schmutz} {et~al.}(1992){Schmutz}, {Leitherer}, \&
  {Gruenwald}}]{Schmutz+1992}
{Schmutz}, W., {Leitherer}, C., \& {Gruenwald}, R. 1992, \pasp, 104, 1164

\bibitem[{{Shaviv}(1998)}]{Shaviv1998}
{Shaviv}, N.~J. 1998, \apjl, 494, L193

\bibitem[{{Shaviv}(2000)}]{Shaviv2000}
{Shaviv}, N.~J. 2000, \apjl, 532, L137

\bibitem[{{Shenar} {et~al.}(2020){Shenar}, {Gilkis}, {Vink}, {Sana}, \& {Sand
  er}}]{Shenar+2020}
{Shenar}, T., {Gilkis}, A., {Vink}, J.~S., {Sana}, H., \& {Sand er}, A.~A.~C.
  2020, \aap, 634, A79

\bibitem[{{Shenar} {et~al.}(2016){Shenar}, {Hainich}, {Todt}, {Sander},
  {Hamann}, {Moffat}, {Eldridge}, {Pablo}, {Oskinova}, \&
  {Richardson}}]{Shenar+2016}
{Shenar}, T., {Hainich}, R., {Todt}, H., {et~al.} 2016, \aap, 591, A22

\bibitem[{{Shenar} {et~al.}(2019){Shenar}, {Sablowski}, {Hainich}, {Todt},
  {Moffat}, {Oskinova}, {Ramachandran}, {Sana}, {Sander}, {Schnurr},
  {St-Louis}, {Vanbeveren}, {G{\"o}tberg}, \& {Hamann}}]{Shenar+2019}
{Shenar}, T., {Sablowski}, D.~P., {Hainich}, R., {et~al.} 2019, \aap, 627, A151

\bibitem[{{Smith} {et~al.}(2002){Smith}, {Norris}, \& {Crowther}}]{Smith+2002}
{Smith}, L.~J., {Norris}, R. P.~F., \& {Crowther}, P.~A. 2002, \mnras, 337,
  1309

\bibitem[{{The LIGO Scientific Collaboration} {et~al.}(2021){The LIGO
  Scientific Collaboration}, {the Virgo Collaboration}, {the KAGRA
  Collaboration}, {Abbott}, {Abbott}, {Acernese}, {Ackley}, {Adams},
  {Adhikari}, {Adhikari}, \& et~al.}]{LIGO+2021}
{The LIGO Scientific Collaboration}, {the Virgo Collaboration}, {the KAGRA
  Collaboration}, {et~al.} 2021, arXiv e-prints, arXiv:2111.03634

\bibitem[{{Vink} \& {de Koter}(2005)}]{VdK2005}
{Vink}, J.~S. \& {de Koter}, A. 2005, \aap, 442, 587

\bibitem[{{Vink} \& {Gr{\"a}fener}(2012)}]{VinkGraefener2012}
{Vink}, J.~S. \& {Gr{\"a}fener}, G. 2012, \apjl, 751, L34

\bibitem[{{Vink} {et~al.}(2021){Vink}, {Higgins}, {Sander}, \&
  {Sabhahit}}]{Vink+2021}
{Vink}, J.~S., {Higgins}, E.~R., {Sander}, A. A.~C., \& {Sabhahit}, G.~N. 2021,
  \mnras, 504, 146

\bibitem[{{Vink} \& {Sander}(2021)}]{VinkSander2021}
{Vink}, J.~S. \& {Sander}, A. A.~C. 2021, \mnras, 504, 2051

\bibitem[{{Woosley} {et~al.}(2020){Woosley}, {Sukhbold}, \&
  {Janka}}]{Woosley+2020}
{Woosley}, S.~E., {Sukhbold}, T., \& {Janka}, H.~T. 2020, \apj, 896, 56

\bibitem[{{Yoon}(2017)}]{Yoon2017}
{Yoon}, S.-C. 2017, \mnras, 470, 3970

\bibitem[{{Yoon} {et~al.}(2006){Yoon}, {Langer}, \& {Norman}}]{Yoon+2006}
{Yoon}, S.~C., {Langer}, N., \& {Norman}, C. 2006, \aap, 460, 199

\bibitem[{{Yusof} {et~al.}(2022){Yusof}, {Hirschi}, {Eggenberger},
  {Ekstr{\"o}m}, {Georgy}, {Sibony}, {Crowther}, {Meynet}, {Kassim}, {Harun},
  {Maeder}, {Groh}, {Farrell}, \& {Murphy}}]{Yusof+2022}
{Yusof}, N., {Hirschi}, R., {Eggenberger}, P., {et~al.} 2022, \mnras, 511, 2814

\end{thebibliography}

%%%%%%%%%%%%%%%%%%%%%%%%%%%%%%%%%%%%%%%%%%%%%%%%%%

%%%%%%%%%%%%%%%%% APPENDICES %%%%%%%%%%%%%%%%%%%%%
%

\begin{appendix}

\section{$\dot{M}$-temperature Fit}
  \label{sec:mdot-teffcrit-fit}

\begin{figure}
  \includegraphics[width=\columnwidth]{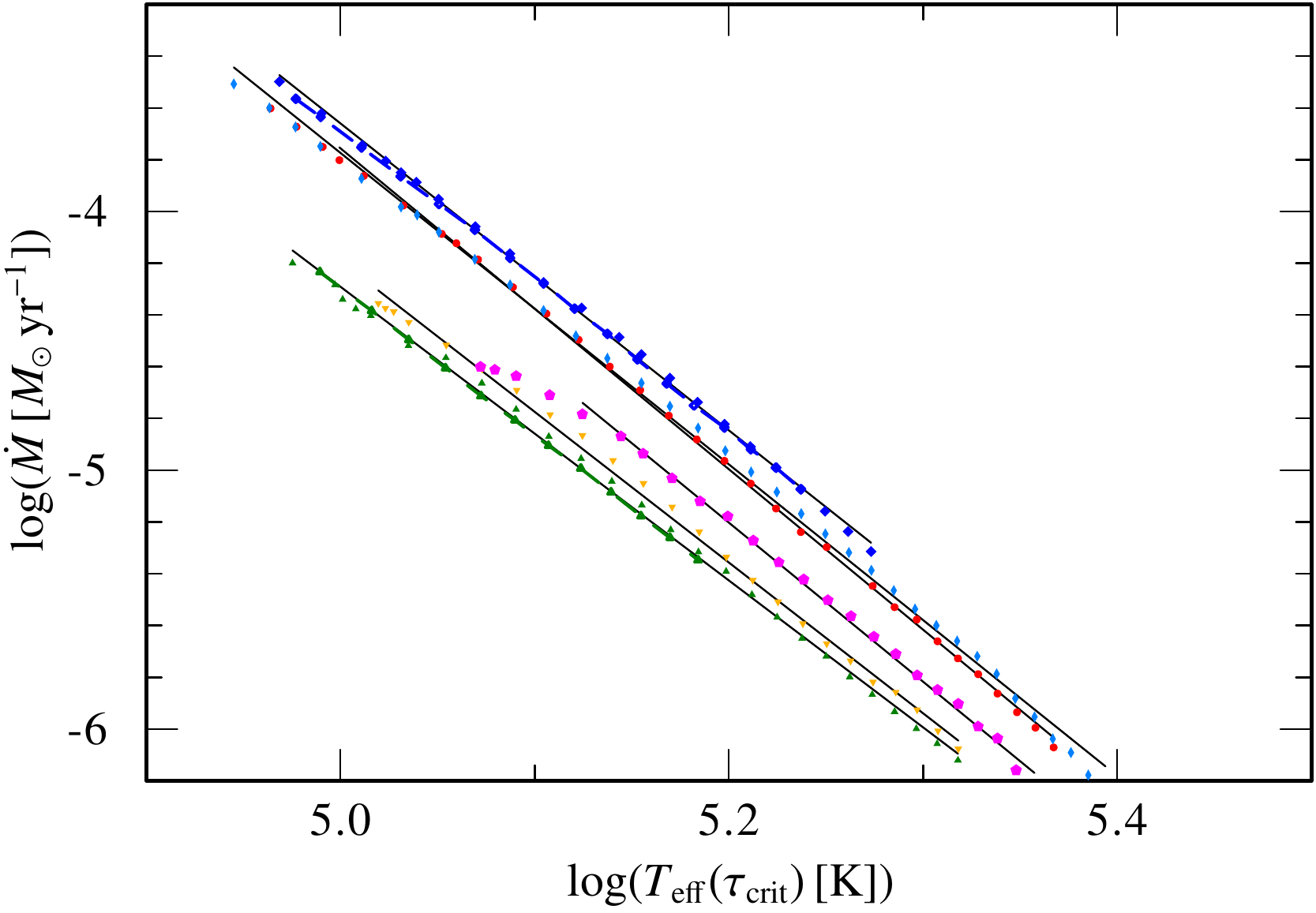}
  \caption{Linear fits (solid lines) to the Mass-loss rate $\dot{M}$ as a function of $T_\text{eff}(\tau_\text{crit})$ in a double-logarithmic-plane. The fit coefficients for the different datasets are given in Table\,\ref{tab:mdot-teffcrit-trend}.}
  \label{fig:mdot-teffcrit-trend}
\end{figure}

For all of our model sequences, the data points in the $\log \dot{M}$-$\log T_\text{eff}(\tau_\text{crit})$-plane suggest a linear relation between the two quantities with deviations occurring only close to the wind driving limit (corresponding to the maximum $\dot{M}$ in Fig.\,\ref{fig:mdot-teffcrit-trend}). In the fits, we thus exclude the uppermost $0.15\,$dex in $\log \dot{M}$. The resulting fit coefficients for the slope including their error margins are given in Table\,\ref{tab:mdot-teffcrit-trend}.
For each individual sequence, the mass loss can be well described for $T_\text{eff}(\tau_\text{crit}) > T_\text{eff,crit,min}$ and $\dot{M} < \dot{M}_\text{max}$ by 
\begin{equation}
   \label{eq:mdot-tdep-fit-seq}
    \log \left(\frac{\dot{M}}{M_\odot\,\mathrm{yr}^{-1}}\right) =  - 6 \log \left( \frac{T_\text{eff,crit}}{141\,\mathrm{kK}} \right) + \log \left( \frac{\dot{M}_\text{offset}}{M_\odot\,\mathrm{yr}^{-1}} \right)
\end{equation}
with the value for $\dot{M}_\text{offset}$ being different for each model sequence. Table\,\ref{tab:mdot-teffcrit-trend} also lists these coefficients together with the corresponding validity limits $ T_\text{eff,crit,min}$ and $\dot{M}_\text{max}$.

\begin{table*}
  \caption{Linear fit results for $\log\,\dot{M}$ versus $\log\,T_\text{eff,crit}$ plus offsets and limitations for the temperature-dependent mass loss of our model sequences described by Eq.\,\eqref{eq:mdot-tdep-fit-seq}.}
  \label{tab:mdot-teffcrit-trend}
  \centering
 \begin{tabular}{lccccccccc}
      \hline\hline
       \multicolumn{5}{c}{Sequence}     &   slope   & formal  &
          $\log\,(\dot{M}_\text{offset}\,[M_\odot\,\mathrm{yr}^{-1}])$ & $T_\text{eff,crit,min}\,[\mathrm{kK}]$ & $\log\,(\dot{M}_\text{max}\,[M_\odot\,\mathrm{yr}^{-1}])$ \\
          & $M\,[M_\odot]$ & $X_\text{H}$ & $Z\,[Z_\odot]$ & $D_\infty$ & & error \\\hline
WN & $20$   & $0.0$ & $1.0$ & $50$ & $-6.02$  &  $0.03$ & $26.33$  &  $92$ & $-3.60$  \\
WN & $20$   & $0.2$ & $0.5$ & $50$ & $-6.02$  &  $0.05$ & $26.31$  &  $88$ & $-3.51$ \\
WN & $12.9$ & $0.2$ & $1.0$ & $50$ & $-5.67$  &  $0.06$ & $24.09$  &  $94$\tablefootmark{(br)} & $-4.20$ \\
WN & $15$   & $0.0$ & $1.0$ & $50$ & $-5.82$  &  $0.07$ & $24.92$  &  $105$\tablefootmark{(br)} & $-4.36$ \\
WC & $20$   & $0.0$ & $0.5$ & $50$ & $-5.96$  &  $0.05$ & $25.81$  &  $118$\tablefootmark{(br)} & $-4.60$\\
WN & $20$   & $0.2$ & $1.0$ & $50$ & $-5.96$  &  $0.04$ & $26.13$  &  $98$ & $-3.62$\\\hline
 \end{tabular}
 \tablefoot{
   \tablefoottext{br}{For marked sequences, $T_\text{eff,crit,min}$  reflects the lower limit for winds driven by the hot iron bump.  In all other sequences breakdown, this values just refers to the minimum explored value.}
 }
\end{table*}

%--------- Figure   ----------------------------------------------------
%--------

\section{$T_{2/3}$ temperatures for the \citet{SanderVink2020} sample}
  \label{sec:t23sv2020}

%--------- Figure   ----------------------------------------------------
\begin{figure}
  \includegraphics[width=\columnwidth]{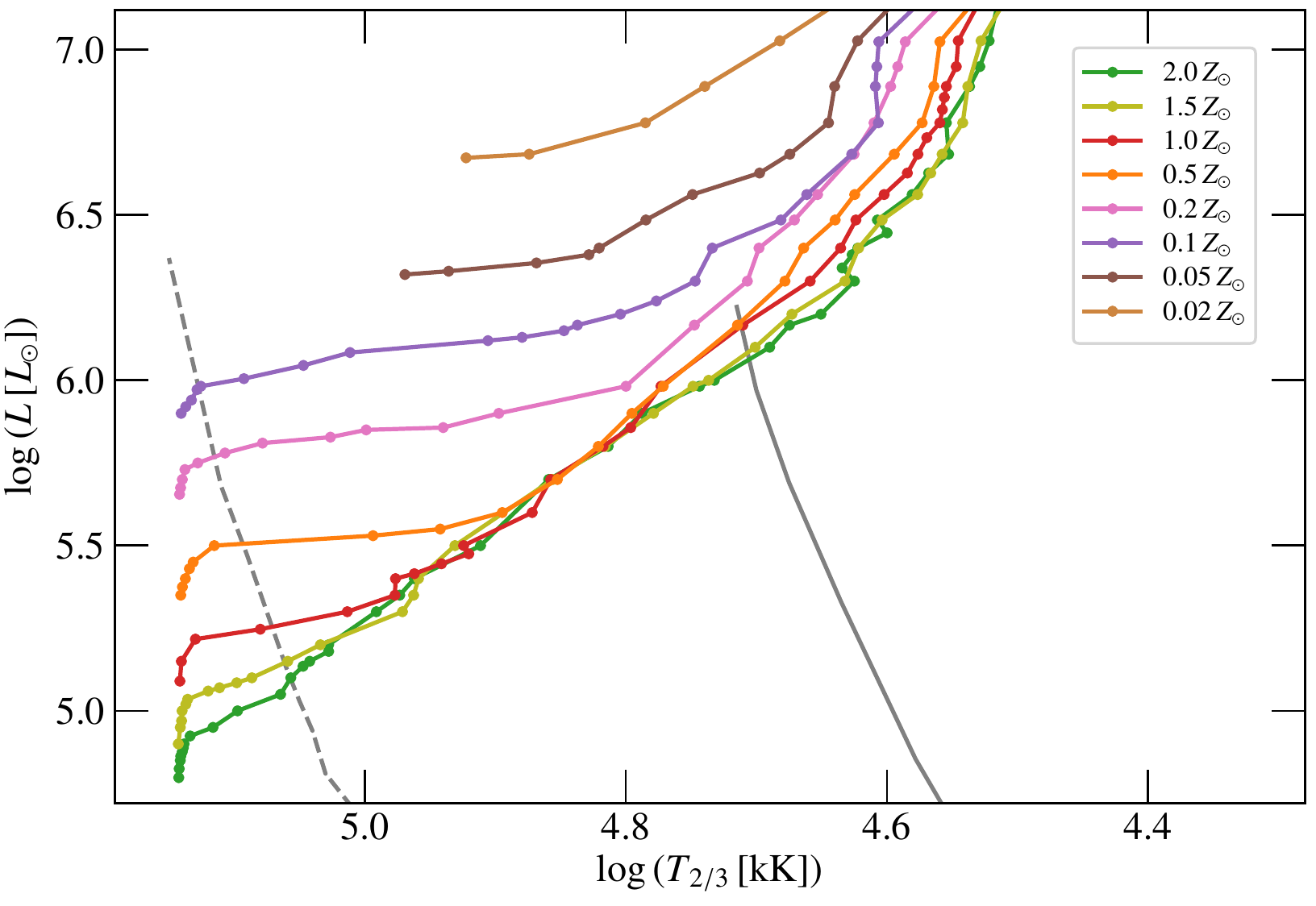}
  \caption{HRD with the effective temperature $T_{2/3}$ defined at a Rosseland optical depth of $\tau_\text{Ross} = 2/3$  and the model luminosity $L$ for our sets of He ZAMS models at different $Z$. For comparison, the HeZAMS (gray, dashed) and ZAMS (gray, solid) are shown as well.}
  \label{fig:t23l}
\end{figure}
%--------- end Figure ----------------------------------------------------  

The effective temperatures $T_{2/3}$ at $\tau_\text{Ross} = 2/3$ resulting from the model sequences in \citet{SanderVink2020} are depicted in Fig.\,\ref{fig:t23l}. Similar to what we obtain when varying $T_\ast$, cooler values of $T_{2/3}$ require higher mass-loss rates. As $T_\ast$ is fixed to $\approx 141\,$kK in \citet{SanderVink2020}, higher luminosities or $L/M$-ratios are required to reach higher mass-loss rates for the same $Z$. At lower metallicity, stars have to get closer to the Eddington Limit to reach sufficient mass loss, shifting the onset of the drop in $T_{2/3}$ to higher $L$ and steepening in particular this drop.
The differences in $T_{2/3}$ and the range of luminosities covered has quite some interesting implications on the spectral appearance of the stars and consequently also which WR subtypes one would expect in a certain environment. Assuming that at least all the winds of early-type WR stars are launched at the hot iron bump, the temperatures of the lowest luminosity WN stars should get hotter at low $Z$. This seems to be the case when comparing the WN populations in the Milky Way and the LMC \citep[e.g.,][]{Hamann+2019,Shenar+2020}, but further -- ideally hydrogen-free -- WN populations in other Galaxies need to be studied to draw any firm conclusions. In the SMC, apart from the small sample size, all WN stars contain hydrogen and might not align with the L-M relation we assume in \citet{SanderVink2020} and this work.

\section{The effect of enhanced clumping on the radiative acceleration}
  \label{sec:ioncontribclump}

\begin{table}
  \caption{Derived wind parameters for WN models with different $D_\infty$}
    \label{tab:clumping-comparison}
    \centering
 \begin{tabular}{cccc}
   \hline\hline
   $D_\infty$   &  $\log\,(\dot{M}\,[M_\odot\,\mathrm{yr}^{-1}])$ & $\varv_\infty\,[\mathrm{km}\,\mathrm{s}^{-1}])$ & $\log\,(\dot{M}_\text{t}\,[M_\odot\,\mathrm{yr}^{-1}])$ \\\hline
   \multicolumn{4}{l}{\textit{WN, $20\,M_\odot$, $X_\mathrm{H} = 0$, $Z_\odot$}} \\
    $10$ &  $-4.42$  &  $1186$  & $-3.77$  \\
    $50$ &  $-4.40$  &  $1754$  & $-3.57$  \\
   \multicolumn{4}{l}{\textit{WN, $20\,M_\odot$, $X_\mathrm{H} = 0.2$, $Z_\odot$}} \\
    $4$   &  $-4.34$  &  $1194$  & $-3.89$  \\
    $10$  &  $-4.28$  &  $1448$  & $-3.71$  \\
    $50$  &  $-4.28$  &  $1970$  & $-3.50$  \\
   \multicolumn{4}{l}{\textit{WN, $20\,M_\odot$, $X_\mathrm{H} = 0.2$, $0.5\,Z_\odot$}} \\
    $4$    &   $-4.44$  &  $609$    &   $-3.70$   \\
    $10$   &   $-4.35$  &  $848$    &   $-3.56$   \\
    $50$   &   $-4.28$  &  $1394$   &   $-3.35$   \\
   \hline   
\end{tabular}

\end{table}

The effect of clumping on our hydrodynamic wind solutions is not trivial. Given that we only use the so-called microclumping approximation assuming optically thin clumps and the solution of radiative transfer in the comoving frame is performed with the average density and not the clumped density, one might expect that the choice of $D_\infty$ could have no effect at all. However, this is not the case. Different values of $D_\infty$ affect the population numbers which in turn affect the radiative transfer. In particular, higher choices of $D_\infty$ favor recombination in the wind. For most elements, lower ionization stages provide more opacity and thus more radiative acceleration. 

\begin{figure*}
  \includegraphics[width=17cm]{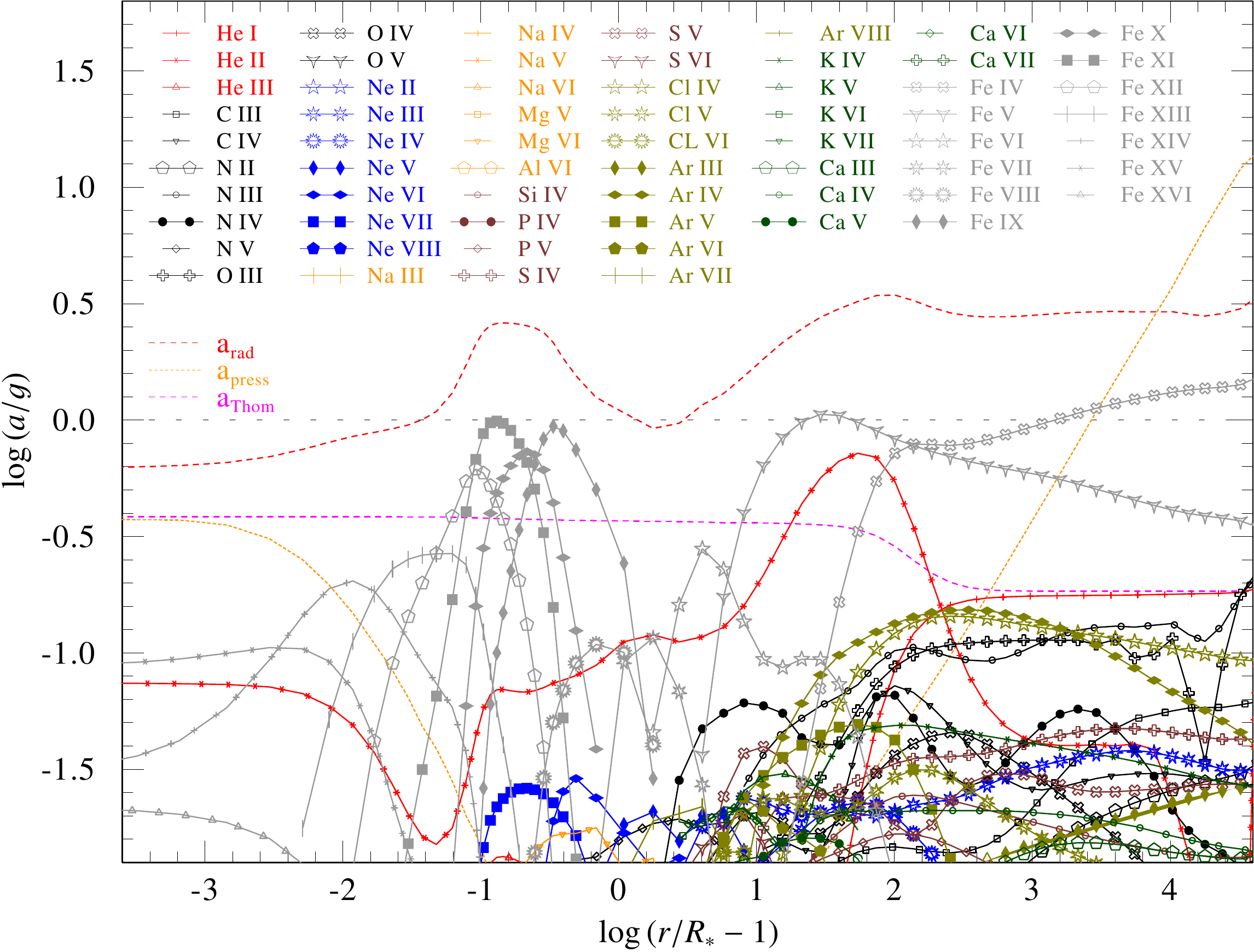}
  \caption{Contributions of the different ions to the radiative acceleration of a hydrodynamically consistent, hydrogen-free WN model with $T_\ast = 130\,$kK, $\log L/L_\odot = 5.7$, $M = 20\,M_\odot$, and $D_\infty = 50$: Different ions are denoted by a combination of different color and symbol. The total radiative acceleration ($a_\text{rad}$), the Thomson acceleration from free electrons ($a_\text{Thom} = \Gamma_\text{e} \cdot g$), and the contribution from gas (and turbulence) pressure ($a_\text{press}$) are also shown for comparison. The loosely dashed horizontal line denotes the total Eddington limit that needs to be overcome to launch a wind.}
  \label{fig:leadions-dd50}
\end{figure*}

\begin{figure*}
  \includegraphics[width=17cm]{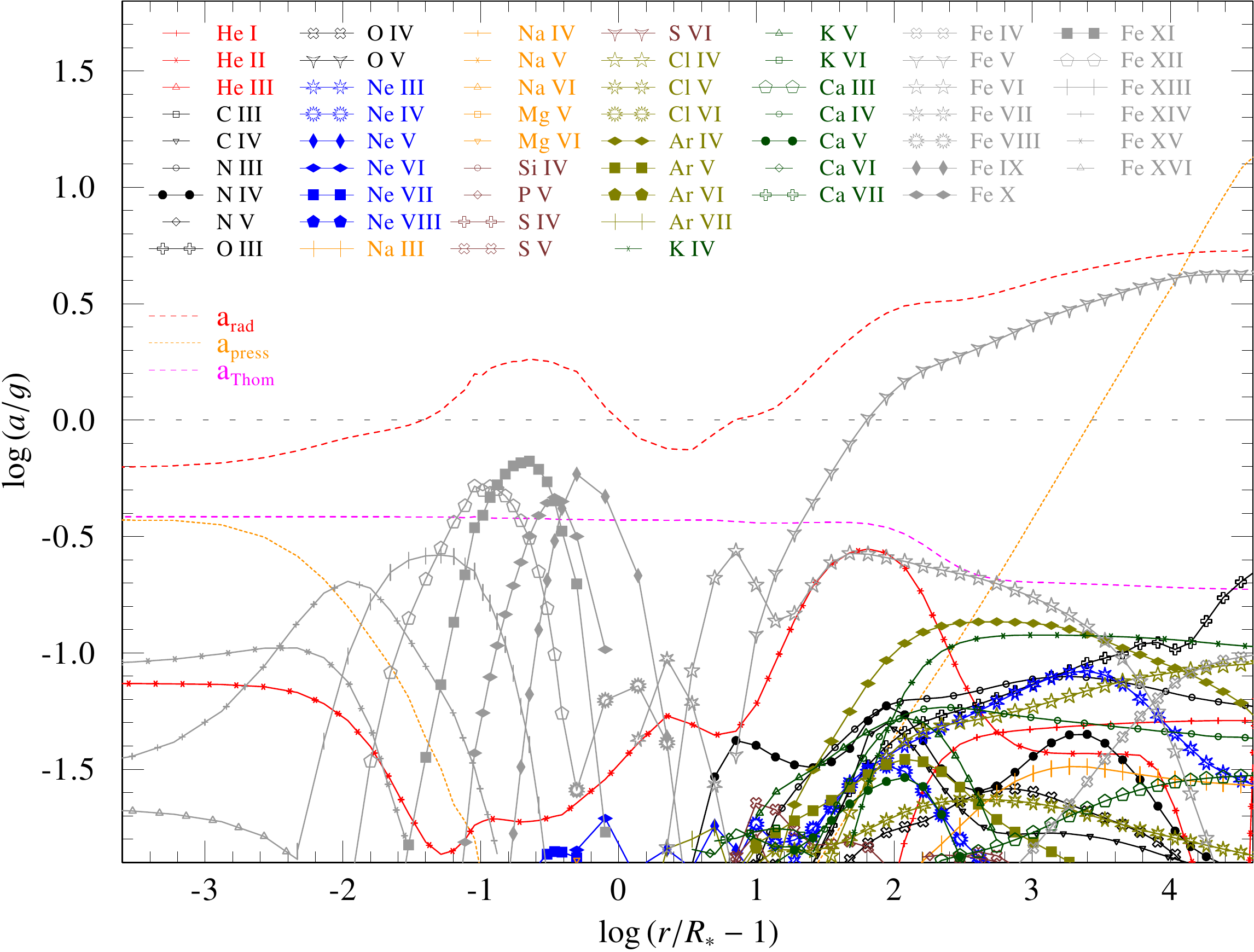}
  \caption{Contributions of different ions to the radiative acceleration, plotted similar to Fig.\,\ref{fig:leadions-dd50}, but now for a model employing $D_\infty = 10$. The wind reaches a lower terminal velocity and the supersonic region with $\Gamma_\text{rad} = a_\text{rad}/g < 1$ is more pronounced than in the model with $D_\infty = 50$.}
  \label{fig:leadions-dd10}
\end{figure*}

In Figs.\,\ref{fig:leadions-dd50} and \ref{fig:leadions-dd10} we present the resulting acceleration contributions for the hydrogen-free $20\,M_\odot$ WN models with $D_\infty = 50$ and $10$, respectively. 
To obtain these curves, the individual opacities resulting from the different ions are stored in addition to the total opacity. Beside the total calculation of 
\begin{equation}
  \label{eq:arad}
  a_\text{rad}(r) = \frac{4\pi}{c} \int\limits_0^{\infty} \varkappa_\nu(r)\,H_\nu(r)\,\mathrm{d}\nu
\end{equation}
similar integrals to Eq.\,\eqref{eq:arad} are calculated using only the ion-specific opacities (e.g.\ $\varkappa_\nu^\text{Fe\,V}$) instead of the total $\varkappa_\nu$, yielding the specific acceleration contribution for each ion.
The corresponding wind parameters of the two displayed models and two other sets with varying $D_\infty$ are listed in Table\,\ref{tab:clumping-comparison}. With out fixed characteristic velocity for the clumping onset of $\varv_\text{cl} = 100\,\mathrm{km}\,\mathrm{s}^{-1}$, the depicted models increase from almost no clumping to $D_\infty$ within the hot iron bump. Thus, the mass-loss rates are barely affected, but the terminal velocity increases significantly from $1186\,\mathrm{km}\,\mathrm{s}^{-1}$ to $1754\,\mathrm{km}\,\mathrm{s}^{-1}$. 

The comparison of Fig.\,\ref{fig:leadions-dd10} with Fig.\,\ref{fig:leadions-dd50} confirms that the additional opacity to reach the higher terminal velocity is provided by lower ions, most notably \ion{Fe}{iv}, which is the leading accelerator in the outer wind for the model with $D_\infty = 50$, while \ion{Fe}{v} remains in the lead for the model with $D_\infty = 10$. In both cases \ion{Fe}{v} is the most populated Fe ion in the outermost wind, while the ``fresh'' opacity provided by the lesser populated \ion{Fe}{iv} is most efficient for the line acceleration in the case of $D_\infty = 50$. In the case of $D_\infty = 10$, the population of \ion{Fe}{iv} instead is too low to contribute significantly. The change in $\varv_\infty$ is further enlarged by the significantly smaller deceleration zone in the  $D_\infty = 50$ model. In the deeper wind layers, the higher clumping boosts the contribution from the iron M-shell opacities and leads to an increased bound-free contribution (i.e.\ recombination) from \ion{He}{ii}.

%--------- Figure   -----------------------------------------------

%--------- end Figure ---------------------------------------------  
The two other examples in Table\,\ref{tab:clumping-comparison} illustrate that in some cases also the mass-loss rate can be notably affected by changes of $D_\infty$. In our models, this is a consequence of the fixed value of $\varv_\text{cl}$. For regimes where generally lower values of $\varv_\infty$ are reached, e.g.\ in lower metallicity model set, often the whole amount of acceleration is reduced, shifting also the region with $\varv \approx 100\,\mathrm{km}\,\mathrm{s}^{-1}$. This can then have two effects leading to a lower $\dot{M}$ for lower values of $D_\infty$: first, a direct reduction of opacities in the region that determines $\dot{M}$. In addition, the reduced wind density could push the star out of the regime where the critical point is in a totally optically thick region \citep[cf.][]{SanderVink2020}, which would lead to a further reduction in $\dot{M}$. 

In the last column of Table\,\ref{tab:clumping-comparison}, we provide the resulting transformed mass-loss rates $\dot{M}_\text{t}$. While there is already a scaling with $\dot{M} \sqrt{D_\infty}$ in these, it does not compensate the clumping changes as the square root of the ${D_\infty}$-ratios is much larger than the changes in $\varv_\infty$ (and $\dot{M}$). For example, the hydrogen free models differ by $\sqrt{50/10} \approx 2.24$, while $\varv_\infty$ only increases by a factor of $1.48$. Therefore, the models with higher $D_\infty$ posses a higher $\dot{M}_\text{t}$, despite larger terminal velocities reducing its value.

\section{Estimating effective temperatures in stellar structure models}
  \label{sec:t23estimate}	

%--------- Figure ---------------------------------------------------
\begin{figure}
  \includegraphics[width=\columnwidth]{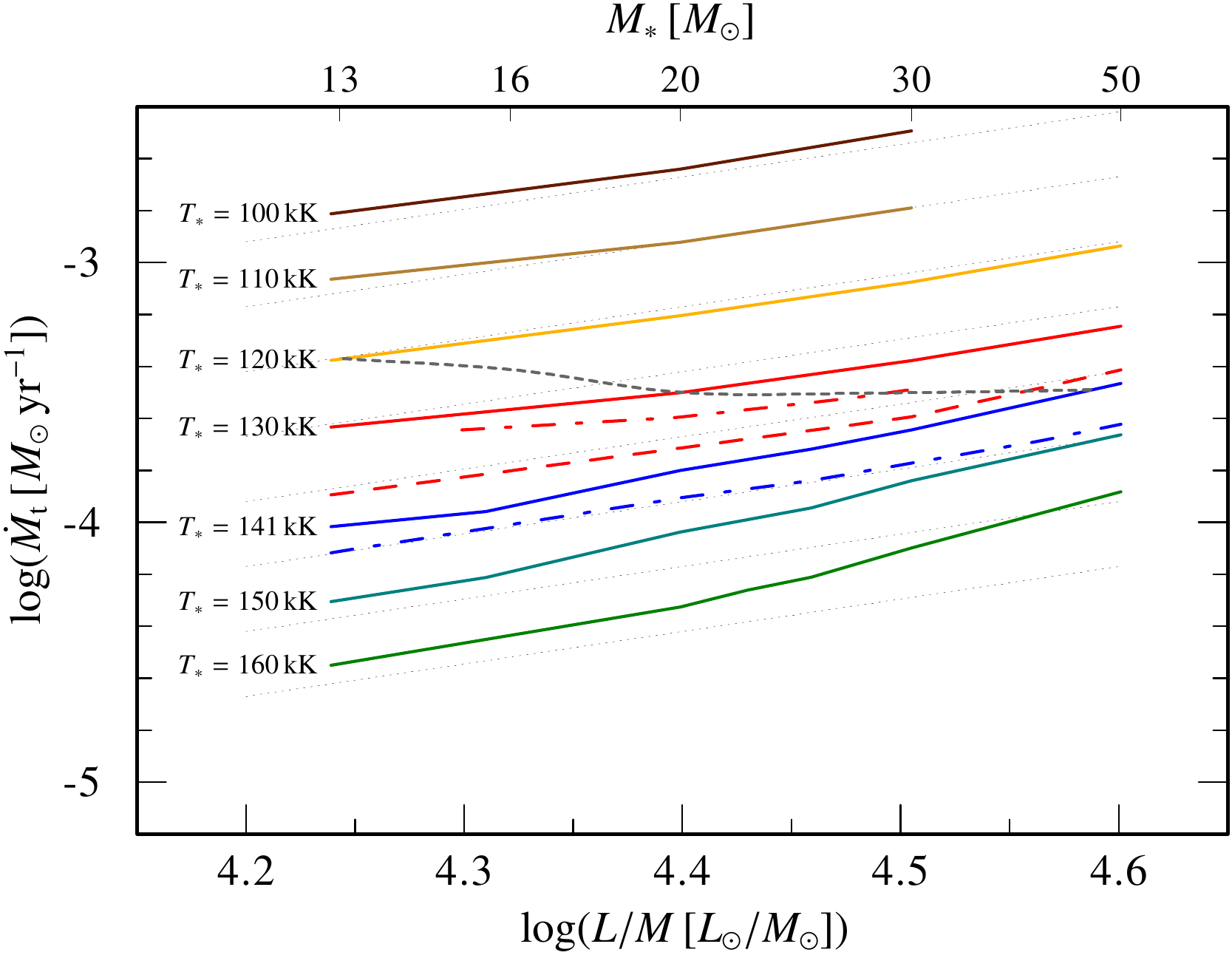}
  \caption{Transformed mass-loss rate $\dot{M}_\text{t}$ as a function of $L/M$ for different sequences varying in $T_\ast$. All models use $X_\text{H} = 0.2$ except the dashed-dotted sequences having $X_\text{H} = 0$ for comparison. Apart from the red, 
  dashed sequence (using $D_\infty = 10$), all sequences employ $D_\infty = 50$. The gray, dotted line is an interpolation of the solid sequences along the theoretical temperatures for the He ZAMS from \citet{Grassitelli+2018} and \citet{Langer1989}. The light dotted linear curves in the background indicate the slope of $1.25$ which is used in Eqs.\,\eqref{eq:mdott-recipe} and onward.}
  \label{fig:mdott-ldm}
\end{figure}
%--------------------------------------------

For stellar structure models, the occurrence of optically thick winds usually spoils the straight-forward prediction of the observable effective temperature $T_{2/3}$ \citep[see, e.g.,][for a more detailed discussion]{Groh+2014}.
In the previous Sect.\,\ref{sec:scaling}, we obtained that for a given clumping factor $D_\infty$, our model sequences collapse almost perfectly to a single line in the $\dot{M}_\text{t}$-$T_{2/3}$-plane, yielding
\begin{equation}
  \label{eq:t23-mdtr-prop}
 T_{2/3} \propto \dot{M}_\text{t}^{-1/2}
\end{equation}
for $\log\,(\dot{M}_\text{t}\,[M_\odot\,\mathrm{yr}^{-1}]) > -4.5$, thereby providing us with a potential path to predict the observable effective temperature. The relation (\ref{eq:t23-mdtr-prop}) seems to be approximately unaffected by abundance ($X_i$) changes, but there is a clear offset for different choices of $D_\infty$. In our calculations, there is a difference of $\Delta \log\,(T_{2/3}\,[\mathrm{K}]) \approx 0.08\dots0.1$ between $D_\infty = 4$ and $D_\infty = 10$ and $\Delta \log\,(T_{2/3}\,[\mathrm{K}]) \approx 0.1\dots0.15$ between $D_\infty = 10$ and $D_\infty = 50$, but the current amount of data along the $D_\infty$ plane is insufficient to provide a robust mathematical formula that could enable a scaling with $D_\infty$. 

A different slope of $\approx -2/3$ was obtained in Sect.\,\ref{sec:scaling}, when considering the sample of \citet{SanderVink2020} instead of our new model sequences. The origin of the difference in the slopes must be rooted in the different nature of the sequences: In the new sequences calculated for this work, $L$ and $M$ are constant. For higher mass-loss rates $\dot{M}$ we then obtain lower values of $\varv_\infty$ along a sequence. In the sequences from \citet{SanderVink2020}, where we proceed to higher $L/M$-ratios along each dataset, such a trend between $\dot{M}$ and $\varv_\infty$ is only reached in the optically thin part, while we obtained $\dot{M} \propto \varv_\infty$ in the dense wind regime. As a consequence, models from the two different sources with approximately the same value of $\dot{M}$ will differ in their $\varv_\infty$. When comparing the $\varv_\infty$-values, the models from the $T_\ast$-sequences in this work will have lower terminal velocities and thus their $\dot{M}_\text{t}$ will be higher. Arguing that $\dot{M}$ is the major factor in setting the $T_{2/3}$ value, we can thus conclude that this difference in the $\varv_\infty$ trends leads to the steeper slope for the sequences along the $L/M$-domain.

Given the focus of this work on the temperature trends and the fact that the steep linear trends in Fig.\,\ref{fig:zrange-t23-mdott} do not provide a good description around the transition region of $\log\,(\dot{M}_\text{t}\,[M_\odot\,\mathrm{yr}^{-1}]) \approx -4.5$, we therefore suggest the more shallow formula
\begin{equation}
  \label{eq:t23estim}
  \log\,(T_{2/3}\,[\mathrm{K}]) = 2.9 - 0.5\,\log\,(\dot{M}_\text{t}\,[M_\odot\,\mathrm{yr}^{-1}])
\end{equation}
as a first attempt to approximate the observable effective temperature $T_{2/3}$ of a WR star in stellar structure models, which is essentially a rounded version of Eq.\,\eqref{eq:t23mdtrfit}. This formula implicitly assumes $D_\infty = 50$ and is recommended for $\dot{M}_\text{t} > 10^{-4.5}\,M_\odot\,\mathrm{yr}^{-1}$. For lower estimates of $D_\infty$, the offset value of $2.9$ would need to be reduced by about $0.1 \dots 0.2$. We emphasize that Eq.\,\eqref{eq:t23estim} is a first approach that needs to be tested and likely refined in future studies.

With Eq.\,\eqref{eq:t23estim} given, only the transformed mass-loss rate $\dot{M}_\text{t}$ needs to be known to determine $T_{2/3}$. In \citet{SanderVink2020}, we could show that for $T_\ast = 140\,$kK the quantity $\dot{M}_\text{t}$ is practically independent of metallicity in the limit of optically thick winds (``pure WR regime''). There, $\dot{M}_\text{t}$ can be expressed as a linear function of $L/M$ with a possible deviation only occurring for He stars with current masses above $50\,M_\odot$.
To check whether this conclusion is independent of $T_\ast$, we calculate a small set of models sequences with different $L/M$ values for different $T_\text{eff,crit} \approx T_\ast$. The resulting trends are shown in Fig.\,\ref{fig:mdott-ldm}. It is clear from Fig.\,\ref{fig:mdott-ldm} that there is some uncertainty in the slopes as well as a potential dependence of the slopes on $T_\ast$ itself, but in general an approximately linear behavior is obtained for each choice of $T_\text{eff,crit} \approx T_\ast$. Hence, we obtain a viable prediction method for stellar evolution models. 
This method is particularly elegant as it does not require any further assumptions about the flux-weighted mean opacity or the shape of the velocity field as for example necessary in the current wind-corrected temperatures in the GENEC models \citep[see, e.g.,][]{Groh+2014}. 

To get a formula for $\dot{M}_\text{t}$ that only depends on quantities which can be obtained from stellar structure calculations, we can use the result derived in Sect.\,\ref{sec:vinf-mdott}. Considering that in the new model sequences calculated for this work both $L$ and $D_\infty$ are constant within one sequence, we can conclude that $\dot{M}_\text{t} \propto \dot{M}/\varv_\infty$ and obtain
\begin{equation}
  \label{eq:mdott-tcrit-scale}
  \log\,(\dot{M}_\text{t}\,[M_\odot\,\mathrm{yr}^{-1}]) = -7.2 \cdot \log\,(T_\text{eff}(\tau_\text{crit})\,[\mathrm{K}]) + \text{offset}
\end{equation}
or $\dot{M}_\text{t} \propto R_\text{crit}^{3.6}$ in the regime of optically thick winds, i.e.\ for $\log\,(\dot{M}_\text{t}\,[M_\odot\,\mathrm{yr}^{-1}]) > -4.5)$.

In a second step, we then merge Eq.\,\eqref{eq:mdott-tcrit-scale}, which has been determined for sequences of constant $L/M$, with the $L/M$-dependence obtained in \citet{SanderVink2020}. Together, we synthesize the formula
\begin{equation}
   \label{eq:mdott-recipe}
   \log \frac{\dot{M}_\text{t}}{M_\odot\,\mathrm{yr}^{-1}} = 1.25 \log \frac{L/M}{L_\odot/M_\odot} + 3.6 \log \frac{R_\text{crit}}{R_\odot} + \dot{M}_\text{t,off}(X_i, D_\infty)\text{.}
\end{equation}
Again, it might be more convenient to replace $R_\text{crit}$ with $T_\text{eff,crit}$ and gauge this with the $20\,M_\odot$ model at $141\,$kK. This then yields
\begin{equation}
   \label{eq:mdott-recipe-gauged}
   \log \frac{\dot{M}_\text{t}}{M_\odot\,\mathrm{yr}^{-1}} = 1.25 \log \frac{L/M}{L_\odot/M_\odot} - 7.2 \log \frac{T_\text{eff,crit}}{141\,\mathrm{kK}} - 9.39 \text{.}
\end{equation}
In Eq.\,\eqref{eq:mdott-recipe-gauged}, we also dropped the offet $\dot{M}_\text{t,off}(X_i, D_\infty)$ which contains further, uncertain dependencies. These can alter the resulting values of $\dot{M}_\text{t}$, e.g.\ by $\approx -0.2\,$dex when changing from $D_\infty = 50$ to $10$. Inserting Eq.\,\eqref{eq:mdott-recipe-gauged} into Eq.\,\eqref{eq:t23estim}, we obtain the final  formula for estimating $T_{2/3}$:
\begin{equation}
  \label{eq:t23-from-basic}
  \log\,\left(\frac{T_{2/3}}{\mathrm{K}}\right) = 7.595 - 0.625 \log \frac{L/M}{L_\odot/M_\odot} + 3.6 \log \frac{T_\text{eff,crit}}{141\,\mathrm{kK}}\text{,}
\end{equation}
This formula is only valid for hydrogen-free WN stars as we have considerable offsets for other chemical compositions in  $\dot{M}_\text{t}(L/M)$ (cf.\,Fig.\,\ref{fig:mdott-ldm}). While the conversion between $\dot{M}_\text{t}$ and $T_{2/3}$ is unaffected by chemical composition (cf.\,Sect.\,\ref{sec:scaling}), the resulting radiative acceleration is not. For example, the additional acceleration from free electrons in partially stripped stars with remaining surface hydrogen leads to higher mass-loss rates than in H-free stars of the same $L/M$-ratio (cf.\,Fig.\,\ref{fig:mdot-teffcrit-all}), thereby substantially shifting the balance between $\dot{M}$ and $\varv_\infty$ and the resulting $\dot{M}_\text{t}$-values. In a future study, we thus plan to extend Eq.\,\eqref{eq:t23-from-basic} by a hydrogen-dependent term. While various uncertainties, e.g., of the precise slopes in $\dot{M}_\text{t}(L/M)$ and $T_{2/3}(\dot{M}_\text{t})$ limit the accuracy of Eq.\,\eqref{eq:t23-from-basic}, it is sufficient enough to tell whether observed effective temperatures of predicted objects are in the range of e.g.\ $100$, $50$, or only $20\,$kK. Depending on the scientific context, such differences in $T_{2/3}$ can have a big impact. While we do not have enough data to draw larger conclusions, the thick gray dotted line Fig.\,\ref{fig:mdott-ldm} illustrates that on the He\,ZAMS, compact radii of the stars at the critical point might even outweigh an expected increase in $\dot{M}_\text{t}$ due to a larger $L/M$. Nonetheless, we can present the first estimate of $T_{2/3}$ for WN stars derived from fundamental principles which can be readily applied in stellar evolution models and population synthesis. The validity of the assumptions made here will have to be tested within dedicated test calculations in stellar evolution models and benchmarked with WR observations analyzed with traditional as well as dynamically consistent models.

\section{Additional Figures}
  \label{sec:addfigures}

%--------- Figure   -----------------------------------------------
\begin{figure}
  \includegraphics[width=\columnwidth]{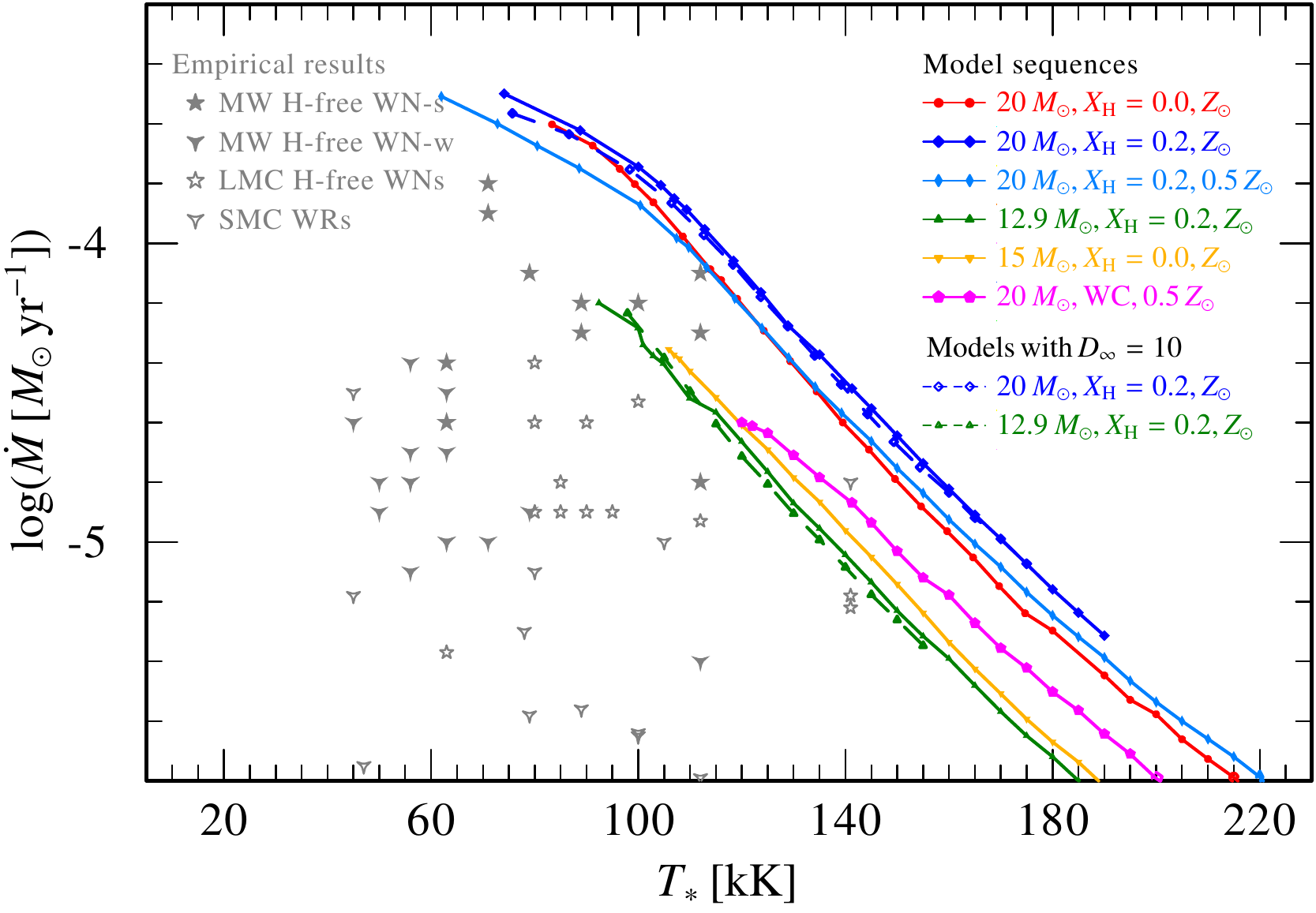}
  \caption{Analogous plot to Fig.\,\ref{fig:mdot-teffcrit-all}, but now showing the mass-loss rate $\dot{M}$ as a function of $T_\ast$ for our model sequences.}
  \label{fig:mdot-tstar-all}
\end{figure}
%--------- end Figure ---------------------------------------------

\begin{figure}
  \includegraphics[width=\columnwidth]{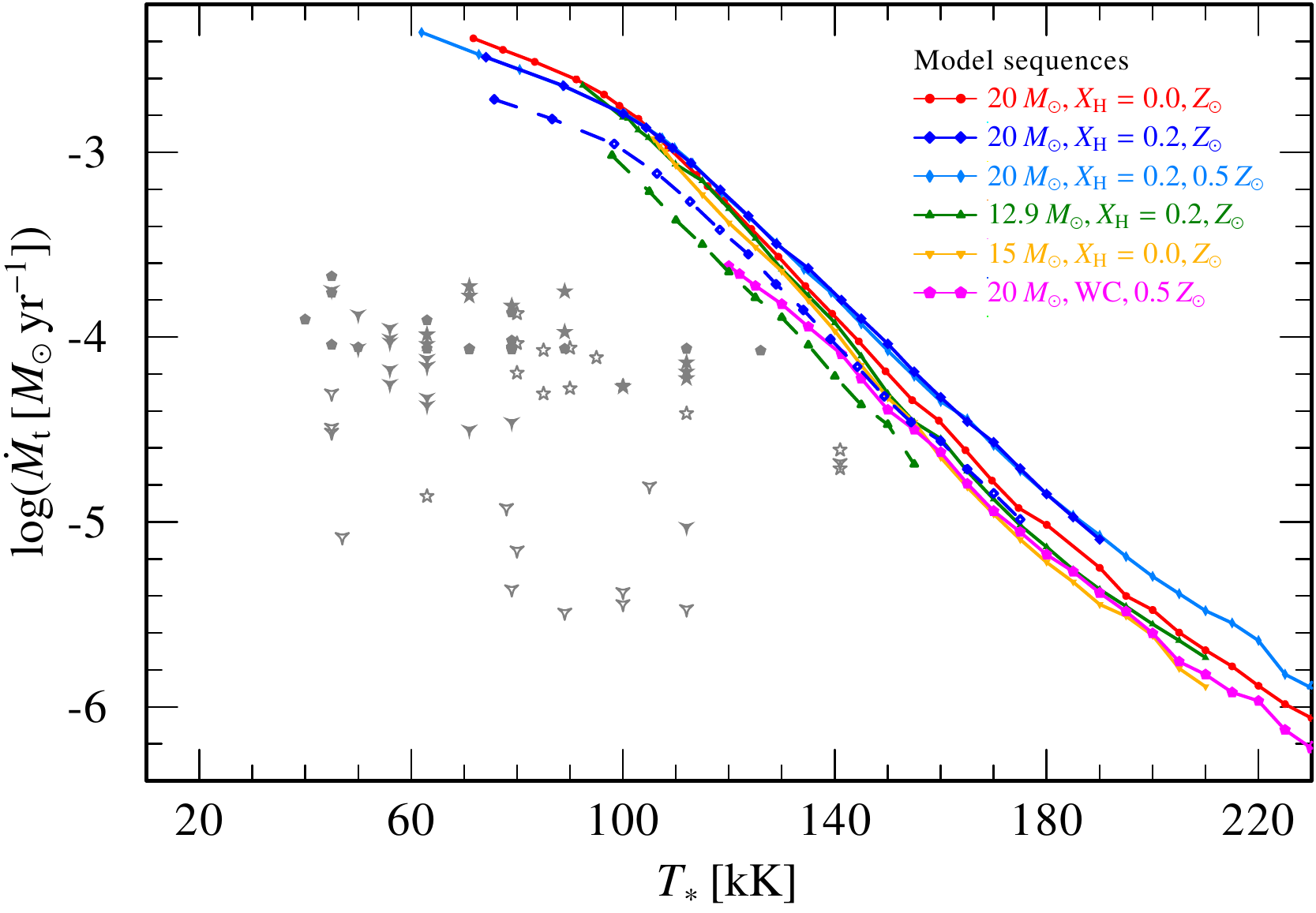}
  \caption{
  Analogous plot to Fig.\,\ref{fig:mdot-tstar-all}, but now showing the transformed mass-loss rate $\dot{M}_\text{t}$ instead of the normal $\dot{M}$.}
  \label{fig:mdott-tstar-all}
\end{figure}
%--------- end Figure ----------------------------------------

%--------- Figure   -----------------------------------------
\begin{figure}
  \includegraphics[width=\columnwidth]{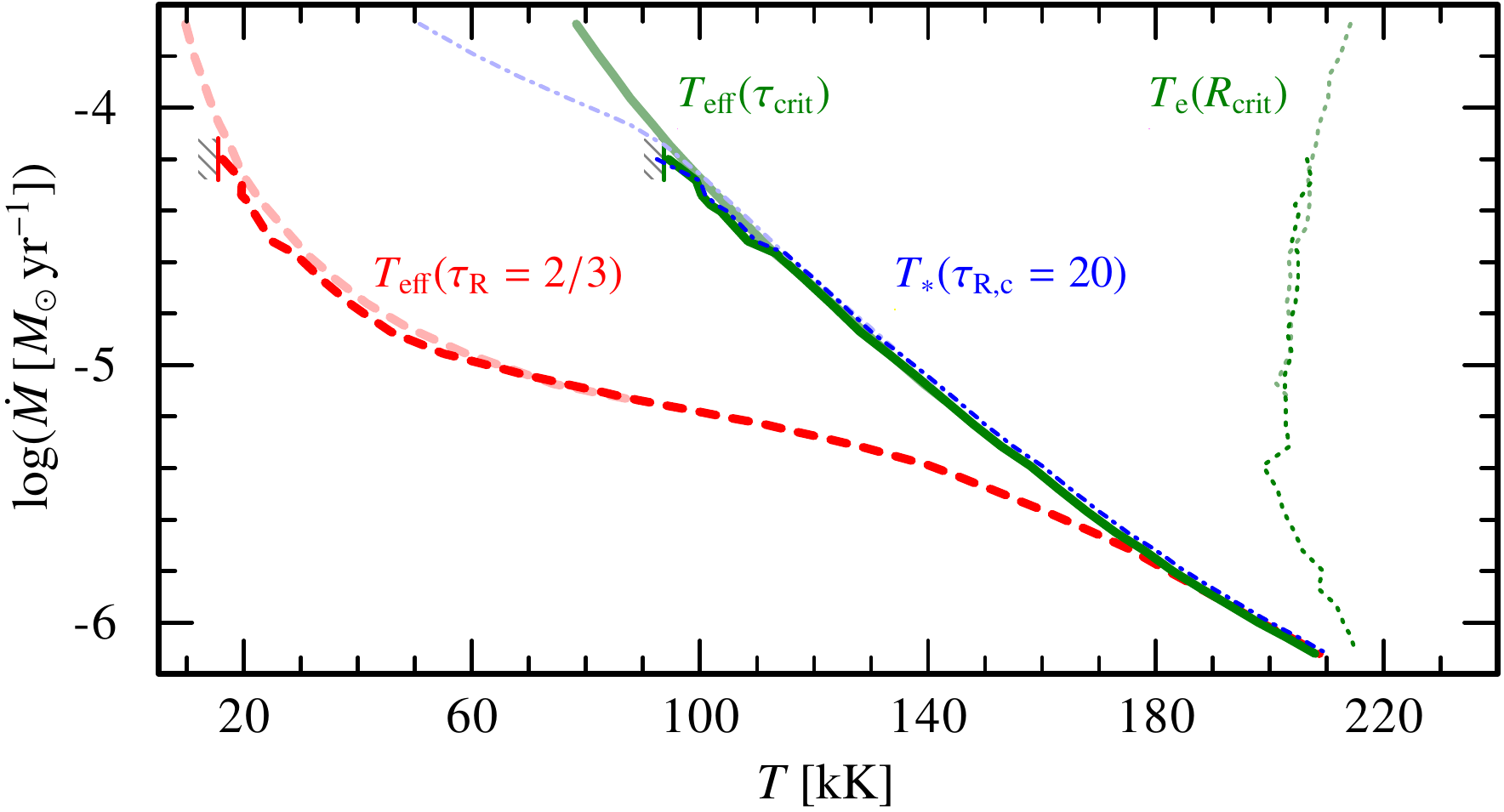}
  \caption{Mass-loss rates as a function of different temperature scales for a series of dynamically consistent atmosphere models  with $\log L/L_\odot = 5.35$, $M = 12.9\,M_\odot$, and $X_\text{H} = 0.2$: The thick red dashed line denoted the classical effective temperatures defined at a Rosseland optical depth of $\tau_\text{R} = 2/3$, while the green solid line and the blue dashed-dotted lines denotes the effective temperatures referring to $\tau_\text{crit}$ and $\tau_\text{R,cont} = 20$ respectively. The green dotted line on the right denotes the (electron) temperature at the critical point. Curves in lighter colors reflect models using the simple integration treatment suppressing negative velocity gradients.}
  \label{fig:mdot-tscales-m12p9-hydrogen}
\end{figure}
%--------- end Figure ----------------------------------------

%--------- Figure   -----------------------------------------
\begin{figure}
  \includegraphics[width=\columnwidth]{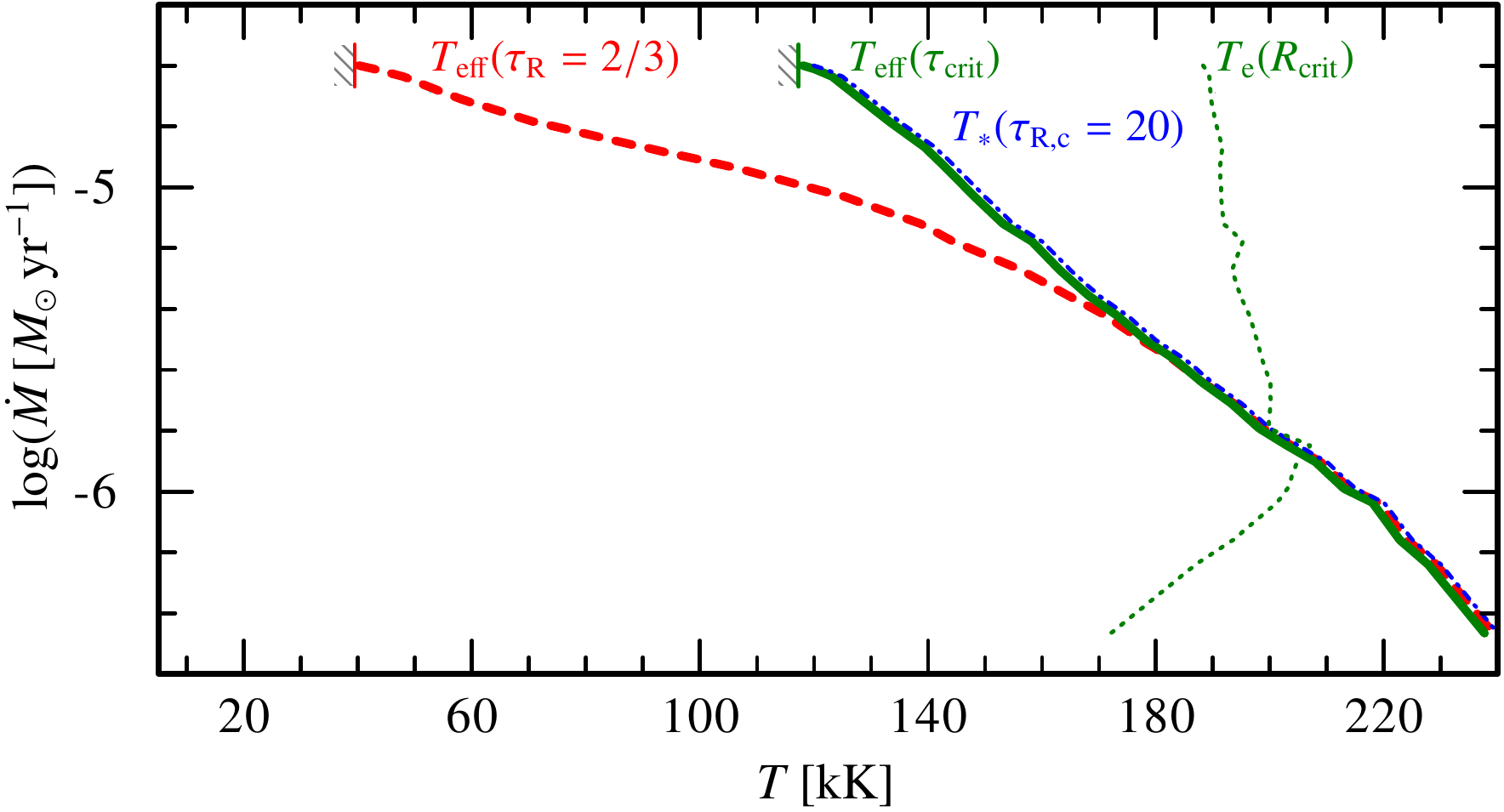}
  \caption{Analogous plot to Fig.\,\ref{fig:mdot-tscales-m12p9-hydrogen}, but now for the WC model series with $\log L/L_\odot = 5.7$, $M = 20\,M_\odot$, and $0.5\,Z_\odot$.}
  \label{fig:mdot-tscales-m20-wc}
\end{figure}
%--------- end Figure ----------------------------------------

In addition to Fig.\,\ref{fig:mdot-teffcrit-all} discussed in Sect.\,\ref{sec:temperatures}, we plot the mass-loss rate $\dot{M}$ as a function of $T_\ast$ in Fig.\,\ref{fig:mdot-tstar-all}. For very high mass-loss rates, we see a bending of the curves toward lower values of $T_\ast$. This is a consequence of the deeper wind launching, which in this regime happens further in than the defining optical depth for $T_\ast$ (i.e.\ at $\tau_\text{R,cont} > 20$). Numerically, these models use a higher optical depth as their inner boundary and we then determine $T_\ast(\tau_{R,cont} = 20)$ for a better comparison with the rest of the model calculations. In this regime, $T_\ast$ is no longer a good approximation for $T_\text{eff,crit}$.

To eliminate the effect of different clumping factors and any remaining distance uncertainties, it is helpful to consider the transformed mass-loss rate $\dot{M}_\text{t}$ instead of $\dot{M}$. In Fig.\,\ref{fig:mdott-tstar-all}, we show the analogous plot to Fig.\,\ref{fig:mdot-tstar-all} with $\dot{M}$ being replaced by $\dot{M}_\text{t}$. While the vertical spread in the observations is slightly reduced, the general temperature mismatch between the empirical $T_\ast$ and our model sequences remains, highlighting once more the ``Wolf-Rayet radius problem'' seen in traditional atmosphere analyses.

In an extend to Fig.\,\ref{fig:mdot-tscales-m20} and Fig.\,\ref{fig:mdot-tscales-m15}, we show similar plots for the model sequences with $12.9\,M_\odot$, $X_\text{H} = 0.2$ and $Z = Z_\odot$ in Fig.\,\ref{fig:mdot-tscales-m12p9-hydrogen} and the WC model sequence with $20\,M_\odot$ and $Z = 0.5\,Z_\odot$ in Fig.\,\ref{fig:mdot-tscales-m20-wc}. Similar to the result obtained for the $15\,M_\odot$-sequence discussed in Sect.\,\ref{sec:breakdown}, there is a lower minimum temperature for obtaining wind solutions driven by the hot iron bump.

\end{appendix}

\end{document}